\pdfoutput=1 
\documentclass[12pt]{elsarticle}

\usepackage{graphicx}
\usepackage{amssymb}
\usepackage{amsthm, amsmath}

\usepackage{multirow}
\usepackage{caption}
\usepackage{subcaption}
\usepackage{xcolor}

\newcommand{\darea}{\alpha}

\begin{document}

\begin{frontmatter}


\title{Ripple-like instability in the simulated gel phase of finite size phosphocholine bilayers}



\author[kcl]{Vivien Walter\corref{cor1}}
\author[ics]{C\'eline Ruscher}
\author[ics]{Adrien Gola}
\author[ics]{Carlos M. Marques}
\author[ics]{Olivier Benzerara}
\author[ics]{Fabrice Thalmann\corref{cor2}}

\cortext[cor1]{\ead{vivien.walter@kcl.ac.uk}}
\cortext[cor2]{\ead{fabrice.thalmann@ics-cnrs.unistra.fr}}

\address[kcl]{Department of Chemistry, King’s College London, Britannia House, 7 Trinity Street, SE1~1DB, London, United Kingdom}
\address[ics]{Institut Charles Sadron, CNRS and University of Strasbourg, 23 rue du Loess, F-67034 Strasbourg, France}

\begin{abstract}

Atomistic molecular dynamics simulations have reached a degree of maturity that makes it possible to investigate the lipid polymorphism of model bilayers over a wide range of temperatures. However if both the fluid $L_{\alpha}$ and tilted gel $L_{\beta'}$ states are routinely obtained, the $P_{\beta'}$ ripple phase of phosphatidylcholine lipid bilayers is still unsatifactorily described. Performing simulations of lipid bilayers made of different numbers of DPPC  (1,2-dipalmitoylphosphatidylcholine) molecules ranging from 32 to 512, we demonstrate that the tilted gel phase $L_{\beta'}$ expected below the pre-transition cannot be obtained for large systems ($>$ 94 DPPC molecules) through common simulations settings or temperature treatments. Large systems are instead found in a disordered gel phase which display configurations, topography and energies reminiscent from the ripple phase $P_{\beta'}$ observed between the pretransition and the main melting transition. We show how the state of the bilayers below the pretransition can be controlled and depends on thermal history and conditions of preparations. A mechanism for the observed topographic instability is suggested. 
\end{abstract}


\end{frontmatter}


\section{Introduction}
\label{S:1}

Lipid membranes are fundamental components of living organisms, for their pivotal role in the structure and  the biochemistry of the cells. Their properties are for a large part the consequence of the organisation of the phospholipid molecules that compose them to a large extent.  This organisation is best revealed by experimenting with artificial lipid bilayers~\cite{Mouritsen_MatterOfFat, Dimova_Marques_TheGiantVesicleBook}. 
Through a perfect control of the molecular composition, an extended range of geometries and the possibility to insert membrane proteins and all sort of molecules, artificial bilayers have become a standard tool in modern biophysics~\cite{Marsh_HandbookLipidBilayers2, Cevc_Marsh_PhospholipidBilayers}.

A striking property of pure phospholipids bilayers is to exhibit a number of thermodynamic transitions upon temperature changes, meaning that they can be found in several phases including a dense structured $L_{\beta'}$ tilted gel phase at low temperature or a disordered fluid phase at higher temperature~$L_{\alpha}$~\cite{Marsh_HandbookLipidBilayers2, Cevc_Marsh_PhospholipidBilayers, 1976_Mabrey_Sturtevant, Heimburg_BiophysicsMembrane}. Decades ago, a new phase have been identified in the most common sort of phospholipid, the phosphocholines (PC lipids)~\cite{tardieu_reman_1973,1996_Sun_Nagle,Marsh_HandbookLipidBilayers2}. 
This new phase, specific to the PC lipids and called the ripple phase $P_{\beta'}$, is characterised by important corrugation along the bilayer and the alternation between interdigitated leaflets and non-interdigitated ones. The ripple phase is observed experimentally in pure bilayers of 1,2-dipalmitoylphosphatidylcholine (DMPC) between 16 and 24$^{\circ}$C, 1,2-dipalmitoylphosphatidylcholine (DPPC) between 34 and 41$^{\circ}$C
and 1,2-distearoylphosphatidylcholine (DSPC) between 51 and 55$^{\circ}$C~\cite{1987_Lewis_Mcelhaney,Marsh_HandbookLipidBilayers2}.


This structure is sometimes presented as resulting from an alternation of gel and fluid lipid configurations~\cite{heimburg_2000}. There is no consensus regarding the cause\cite{1979_Doniach, 1993_Lubensky_Mackintosh,1998_Fournier,1998_Misbah_Houchmandzadeh,2001_Sengupta_Hatwalne}, and the experimental structure is still subject to detailed investigations~\cite{akabori_nagle_2015}. The interest in the ripple phase has grown over the years, and intense research in its nature are made jointly in experimental and numerical sciences~\cite{2005_deVries_Marrink,lenz_schmid_2007, debnath_ayappa_2014}.

In a recent paper, Khakbaz and Klauda investigated the phase transition of DMPC and  DPPC~\cite{2018_Khakbaz_Klauda}. They reported formation of a structure resembling the $P_{\beta'}$ phase for DMPC bilayers in a range of temperature, while for DPPC only a transition from the tilted  to the fluid phase was observed in bilayers composed of 70 lipids. More recently, we investigated the phase transition of DPPC bilayers using Machine Learning algorithms~\cite{walter_thalmann_2020}. In membranes composed of 212 lipids, we observed a transition at 315~K from a fluid phase to a condensed disordered gel phase similar to the ripple phase that persists well below the pre-transition temperature of 307K and the $L_{\beta'}$ tilted phase was never obtained directly from a quench of the $L_{\alpha}$ state. 

In this article, we provide a detailed analysis of the nature of the phase of DPPC simulated with the CHARMM36 force field below its melting temperature.  We investigate the state of DPPC bilayers at 288~K, for different system sizes and different thermal routes. While the $L_{\beta'}$ phase was observed for small systems, we found that a disordered gel phase, reminiscent from the $P_{\beta'}$ phase,  occurs in larger systems. In particular, we noticed the formation of corrugations whose amplitude increases with the system sizes investigated. By applying different thermal treatments, we characterize the metastability of DPPC and show that disordered gel is the preferred phase of simulated DPPC at 288K, in the thermodynamic limit.  


\section{Material \& Methods}

\subsection{System Description}

DPPC bilayers were obtained using the CHARMM-GUI online Membrane Builder~\cite{jo_im_2007, jo_im_2008, jo_im_2009, wu_im_2014}. The size of the membrane was controlled \textit{via} the number of lipids, ranging from 32 to 512 and with equal amounts of lipids in both leaflets. The bilayers were hydrated with water blocks of 10~nm on each side to prevent any interaction of the leaflets through the PBCs along the Z-axis. The exact composition of each system used is given in the Supplementary Informations.

The constructed system were minimized in energy using a steepest descent algorithm and equilibrated by running two NVT simulations at 288~K for 10~ps, with respectively a 0.001~ps and a 0.002~ps step; and two NPT simulations at 288~K and 1~bar, with a 0.002~ps step for respectively 100~ps and 1~ns.

\subsection{Simulation Runs}

Unless specified, the following conditions and parameters have been used for all simulations. All simulations were performed using GROMACS 2016.4~\cite{berendsen_vandrunen_1995,abraham_lindahl_2015} along with the CHARMM-36 all-atom force-field~\cite{best_mackerell_2012} (June 2015 version). The force field parameters for the lipid molecules were provided directly by CHARMM-GUI~\cite{brooks_karplus_2009, lee_im_2016}.

All the molecular dynamics simulations used the leap-frog integration algorithm~\cite{hockney_eastwood_1974} with a time step set to 2~fs. The temperature was controlled during the simulation using a Nos\'e-Hoover thermostat~\cite{nose_1984, hoover_1985} with a correlation time of $\tau_T = 0.4$~ps, and a Parrinello-Rahman semi-isotropic barostat~\cite{nose_klein_1983, parrinello_rahman_1998} set to 1~bar in all directions was applied to the system (correlation time  $\tau_P = 2.0$~ps, compressibility $4.5 \times 10^{-5}$~bar$^{-1}$).

Lipid and water molecules were separately coupled to the thermostat. Following GROMACS recommendations for the CHARMM-36 all-atom force field, a Verlet cut-off scheme on grid cells was used with a distance of 1.2~nm, and non-bonded interactions cut-offs (Van der Waals and Coulombic) were also set to 1.2~nm. Fast smooth Particle-Mesh Ewald electrostatics was selected for handling the Coulombic interactions, with a grid spacing of 4~nm. A standard cut-off scheme with a force-switch smooth modifier at 1.0~nm was applied to the Van der Waals interactions. We did not account for long range energy and pressure corrections, and constrained all the hydrogen bonds of the system using the LINCS algorithm.

Molecular dynamics production were run for a minimum of 50~ns. When temperature treatments occurred, the systems were simulated for another 50~ns to allow for the bilayer to reach equilibrium. Regardless of the length of the simulation, the analysis were always performed on the last 25~ns of the simulations.

\subsection{Analysis}

The areas per lipid of the systems were measured using two different methods: (i) measure by projection of the bilayer in the XY plane of the box, noted $A_H^p$, and (ii) measure by meshing of the water-lipid interfaces of each leaflet of the bilayer, written $A_H^m$. The projected area per lipid was measured directly in Gromacs by measuring the area of the simulation box in the XY plane and by dividing it by the number of lipids per leaflet. The meshed area per lipid was computed using Ovito 2.9. To do so, the surface of the water blocks at the water-lipid interface was meshed using a probe sphere radius of 6 and a smoothing level of 30 after removing the lipid molecules. The area of the meshing was then divided by the total number of lipids in the bilayer.

The amplitude and the period of the corrugations were measured using the meshing of the water-lipid interface collected for the measurement of the area per lipid. The position of the vertices of the meshing were interpolated using SciPy~\cite{virtanen_vanMulbregt_2020} on a (200,200) uniform grid of the size of the simulation box, and the height of these 200$\times$200 points were analysed. This grid was directly used to generate the height maps. The amplitude of the corrugations, represented by the root mean square (RMS) height of the points on the grid, were calculated by subtracting the mean and then measuring the standard deviation of the height $\sigma(h)$. The period of the corrugations were calculated by computing the azimuthal power spectrum density (PSD) of the points of the grid and determining the frequency of the highest peak of the PSD.

To measure the local chain tilt direction, the positions of the atoms of the tails of the lipids were extracted from the simulation files using MDAnalysis~\cite{michaud_beckstein_2011, gowers_beckstein_2016}. The vectors from the first carbon atom of each tail to all the carbon atom of their respective tails were computed, before calculating the mean vector. The vector field was then created by binning the membrane in $n \times m$ squares and averaging over time the tilt direction vectors found in each square $(X_i, Y_i)$. As a consequence, the vector field account for the average direction of the tails found at this location on the membrane over time and not to the average direction of a lipid over time.

The enthalpy of the systems were extracted from the simulation using directly the tool provided with Gromacs, \textit{gmx energy}. The systems used strictly had the exact same number of atoms, for both lipids and water molecules, to prevent changes due to the system composition. The (tilted) gel - fluid phase transition enthalpy was measured using the method from~\cite{2020_deOliveira_Marques}. Briefly, the total enthalpy of a system simulated at different temperatures over a wide range, here 283 to 358~K, was collected. The effects of the temperature increase on the enthalpy, besides any phase transition, were removed from the measurement by subtracting the affine baseline measured in each phase. The values were then divided by the total number of lipids in the system, and the transition enthalpy was calculated by integrating the difference between the fitted gel and fluid baseline over the range of temperature at which the transition occurs (here 308 to 318~K, accounting for Gromacs accuracy in setting a temperature). For the tilted to disordered gel phase transition, the enthalpy was measured by bringing the two systems at the exact same temperature. The difference in enthalpy measured between these two systems was then divided by the total number of lipids in the system.


\section{Results}

\subsection{Corrugation formation and characterization}

\begin{figure}[ht!]
\centering
\includegraphics[width=.9\linewidth]{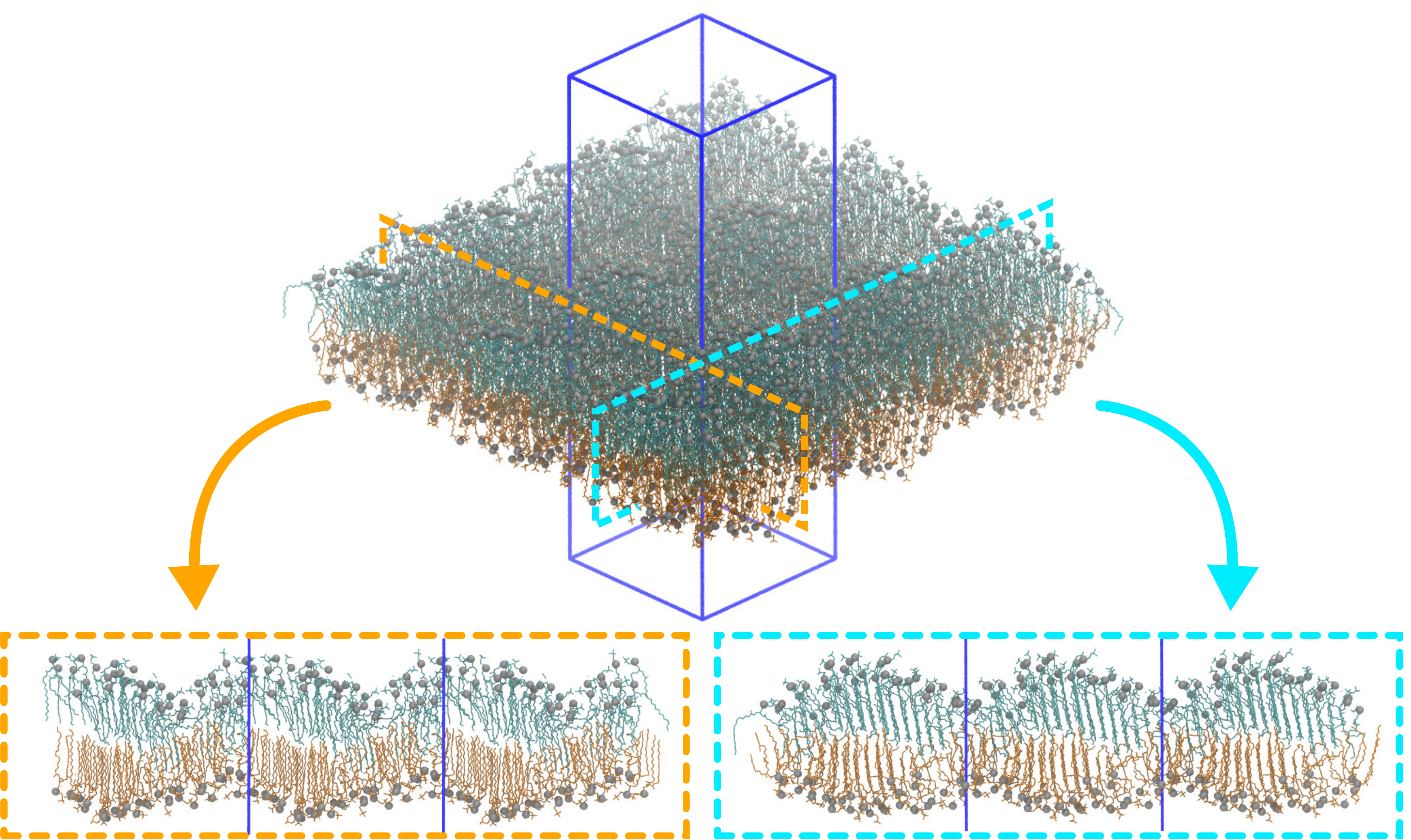}
\caption{DPPC bilayer made of 212 lipids simulated at 288~K right after construction and equilibration. 2.5~nm thick slices of the system along each axis are shown on the bottom, highlighting the corrugations of the membrane. Top and bottom lipid leaflets are coloured respectively in cyan and orange while the phosphorus atoms are shown as silver beads. Hydrogen atoms were removed from the screenshot to improve readability. The PBC box is displayed using the blue plain lines.}
\label{fig_corrugation}
\end{figure}

We considered DPPC bilayers prepared with CHARMM-GUI at temperature $T=288$K whose size is ranging from 32 to 512 lipids. For small bilayers made of 64 lipids or less, we observed the formation of a smooth $L_{\beta'}$ tilted gel phase, in agreement with the numerous experiments and simulations reported in the literature~\cite{cunningham_brain_1998}. The geometry of these bilayers remained stable over time, even after a 50~ns simulation. When larger systems $(>94)$ lipids were concerned, we noticed the formation of a corrugation deforming significantly the leaftlets  (\textit{\textit{cf}} Figure~\ref{fig_corrugation}).

\begin{figure}[ht!]
\centering
\begin{tabular}[t]{cc}

\begin{subfigure}{0.5\textwidth}
\centering
\smallskip
\includegraphics[width=\linewidth]{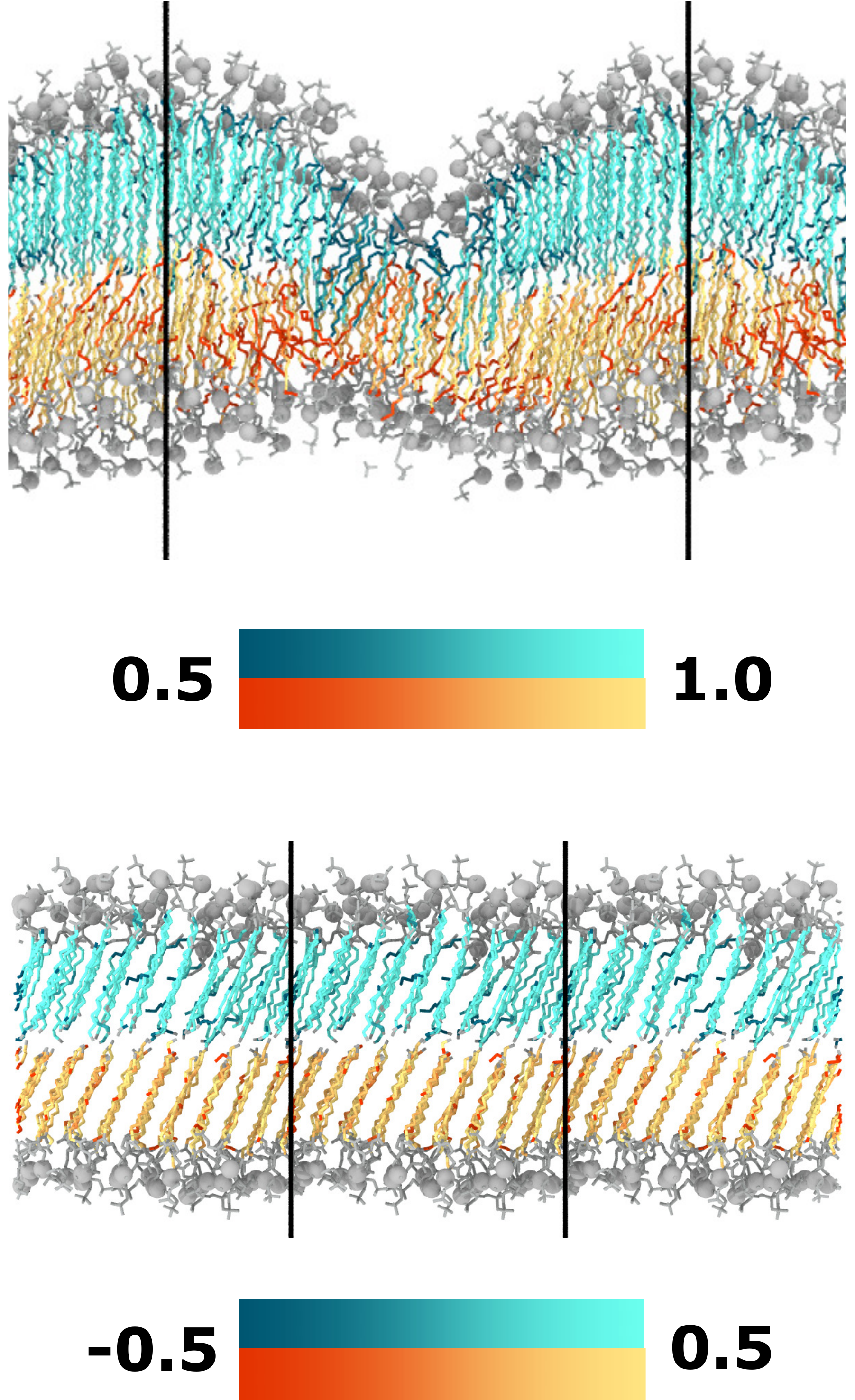}
\caption{} 
\end{subfigure}
&
\begin{subfigure}{0.35\textwidth}
\centering
\smallskip
\includegraphics[width=\linewidth]{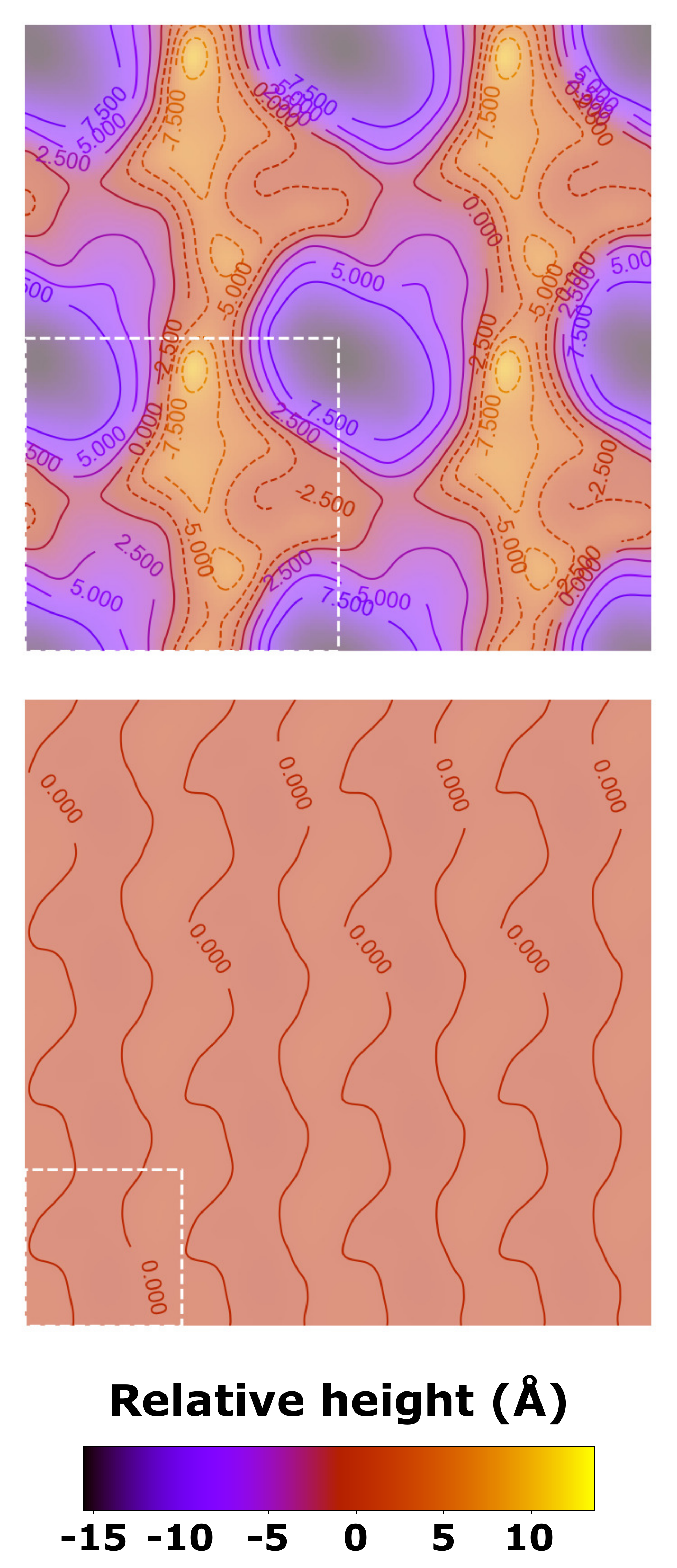}
\caption{} 
\end{subfigure}
\end{tabular}

\caption{(Left) Topography of the upper leaflet of (Top) 256 and (Bottom) 64 lipid bilayers, highlighting the corrugations observed in the membrane. Both contour plot shares the same color code and contour line scale, in \AA. The white dashed line show the system before replication. (Right) Representation of slices of the lipid membranes on the left, with a color code on the tail atoms corresponding to their order parameters.}
\label{fig_height_2d}
\end{figure}

The corrugations takes place along both simulations box axis ($x$ and $y$). As already observed and investigated in details for DMPC by Khakbaz and Klauda~\cite{2018_Khakbaz_Klauda}, the molecular configurations of the lipids do not appear uniform along the corrugations, and the typical stretched tails of the gel phase seem to turn into the typical disordered tail configurations of the fluid phase in the thin portions of the corrugations where interdigitation happens. These variations in configuration can be highlighted by computing and mapping the local segment order parameter $S_{\text{mol}}$ of each atom of each lipid (\textit{\textit{cf}}~Figure~\ref{fig_height_2d}(a)).

\begin{figure}[ht!]
\centering
\begin{tabular}[t]{cc}
\begin{subfigure}{0.445\textwidth}
\centering
\smallskip
\includegraphics[width=\linewidth]{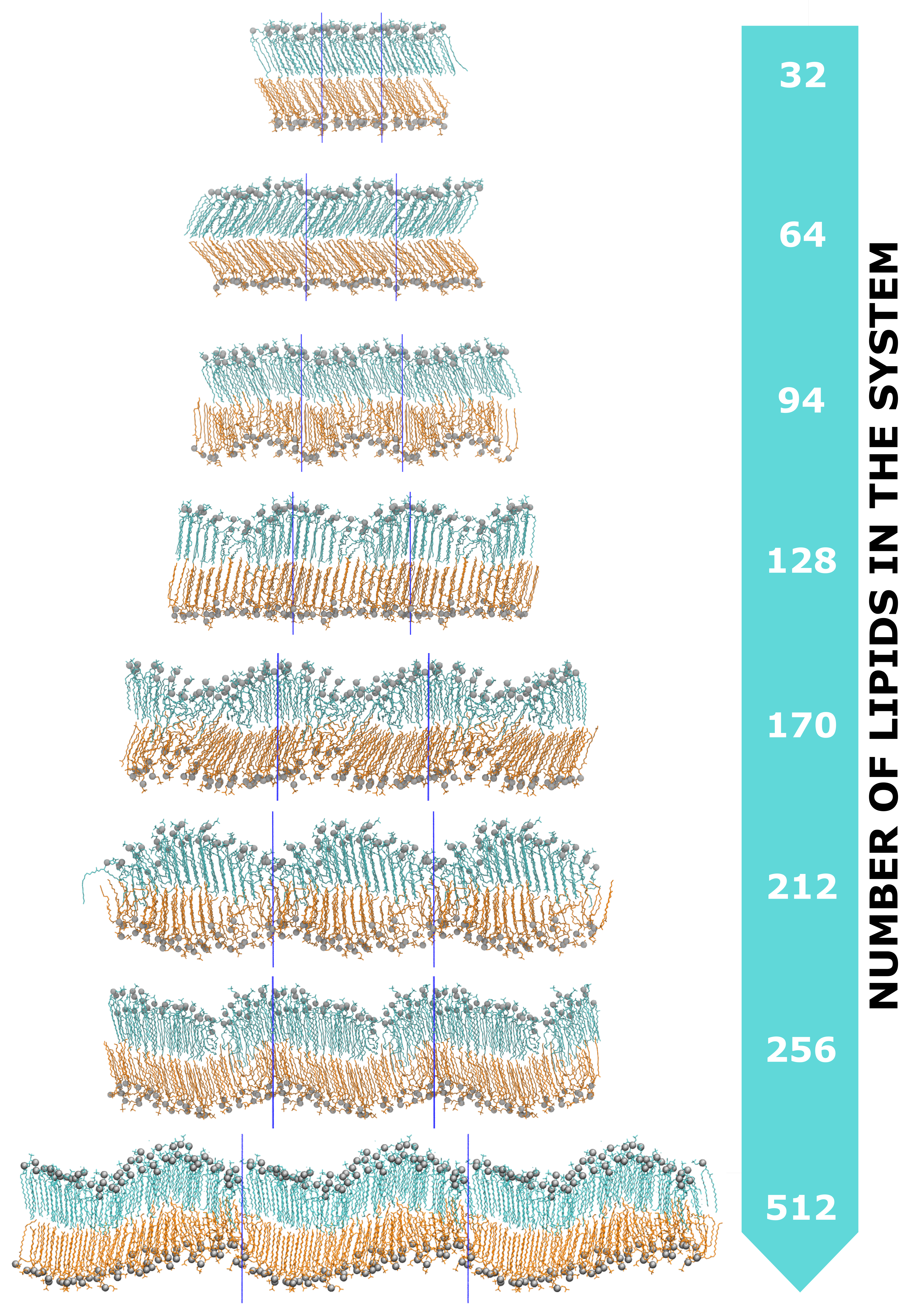}
\caption{} 
\end{subfigure}
&
\begin{tabular}{c}
\smallskip
\begin{subfigure}[t]{0.45\textwidth}
\centering
\includegraphics[width=\textwidth]{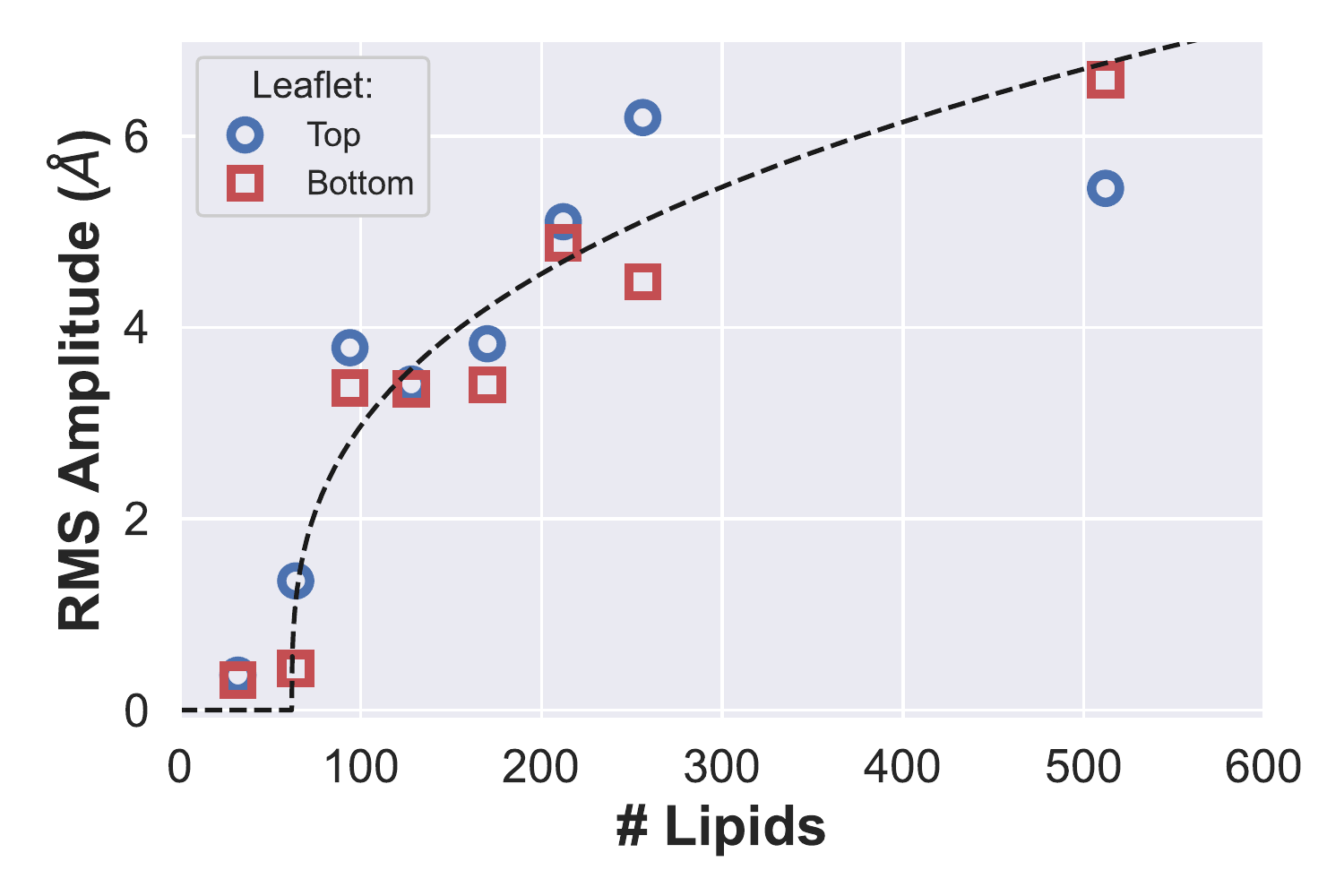}
\caption{}
\end{subfigure}\\
\begin{subfigure}[t]{0.45\textwidth}
\centering
\includegraphics[width=\textwidth]{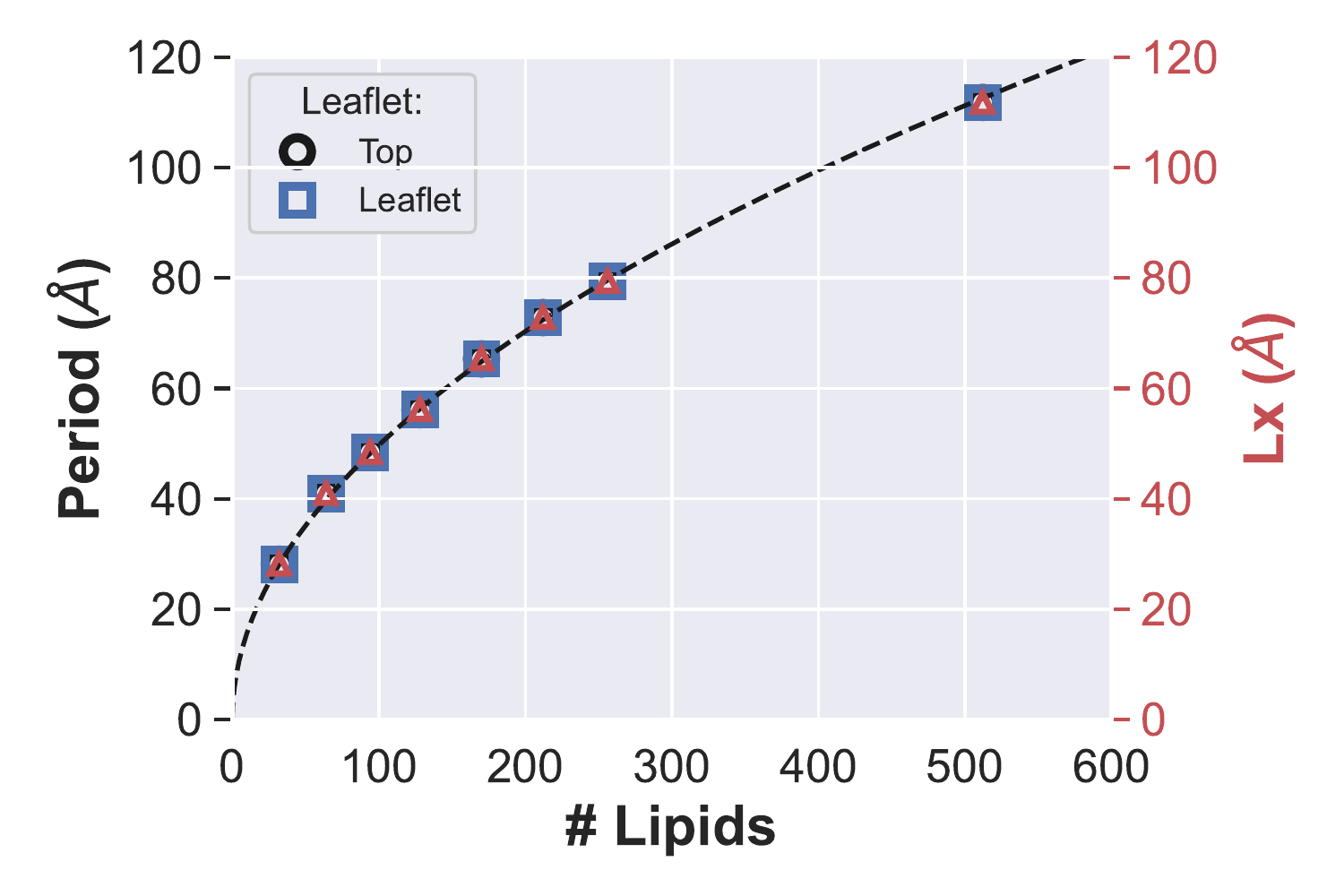}
\caption{}
\end{subfigure}
\end{tabular} \\
\end{tabular}

\caption{Effect of the size of the simulation box on the geometry of bilayers. (a) 2.5~nm thick slices of the systems made with different amount of lipid molecules and simulated at 288~K. Measurement of the (b) amplitude, or $h_{\text{RMS}}$ and (c) period of the corrugations observed in different systems versus the number of lipids. The evolution of the size of the box is shown on top of the period. Dashed lines are respectively the cubic root and square root fits of the data.}
\label{fig_system_size}
\end{figure}

We systematically characterized the corrugation and probed its evolution with system size. To quantify it in a more specific way, we investigated the topographic elevation function for each leaflet (see Figure~\ref{fig_height_2d}(b)) from which we defined the amplitude of the corrugation as the root mean square (RMS) height $h_{\text{RMS}}$ of the two water-lipid interfaces of the bilayers. The heights of the interfaces were obtained by meshing the water surface and removing the mean height of each leaflet. In these circumstances, $h_{\text{RMS}}$ is equal to the standard deviation of the heights $\sigma(h)$. The leaflet corrugation amplitudes are identical for both leaflets even though cross section pictures might suggest otherwise (Figure~\ref{fig_system_size}(a)). The corrugation amplitude increases with system size but saturates for large systems (Figure~\ref{fig_system_size}(b)). The longitudinal period of the corrugations always coincide with the periodic boundary conditions (PBC, see Figure~\ref{fig_system_size}(c)). The power spectrum densities (PSD) used to measure the periods are given in the Supplementary Materials.

\begin{figure}[ht!]
\centering
\includegraphics[width=.45\linewidth]{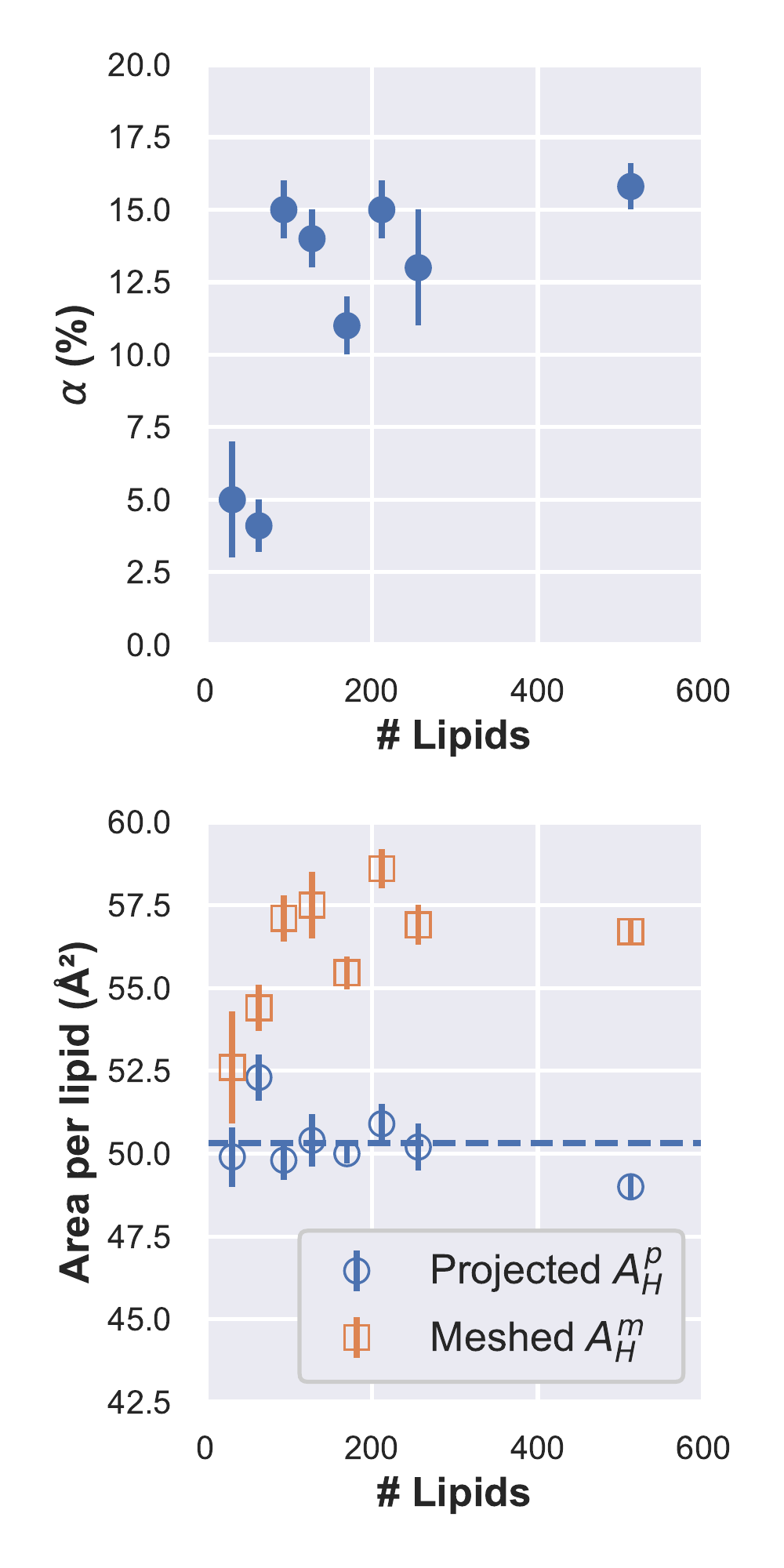}
\caption{(Top) Differences between the area per lipid measured by surface meshing and by projection in the XY plane $\darea$ (in \%) and (Bottom) evolution of each type of area per lipid for different system sizes. Blue dashed line on the bottom graph is the average projected area per lipid.}
\label{fig_area_size}
\end{figure}

To better quantify the nature of the corrugation, we investigated the area per lipid using two methods: i) \emph{projection} in the $(XY)$ plane of the box, i.e the area $A_H^p$ of the box divided by the number of lipid; ii) \emph{meshing} of the water lipid interface that enables the determination of the interfacial area $A_H^m$. We found that the difference between interfacial and projected area per lipid 
\begin{align}
    \darea = \frac{A_H^m - A_H^p}{A_H^p}
\label{new_order_parameter}
\end{align}
provides an excellent characterization of the corrugated phase (Figure~\ref{fig_area_size}). Indeed, $\darea$ is three times larger for corrugated systems ($\darea = 14\pm 2 \%$) than for tilted ones ($\darea = 5\pm 2 \%$). The differentiation using usual methods such as the area per lipid or the tail order parameter achieved less significant results (\textit{\textit{cf}} Supplementary materials).

\begin{figure}[ht!]
\centering
\begin{tabular}[t]{cc}
\begin{subfigure}{0.3\textwidth}
\centering
\includegraphics[width=\textwidth]{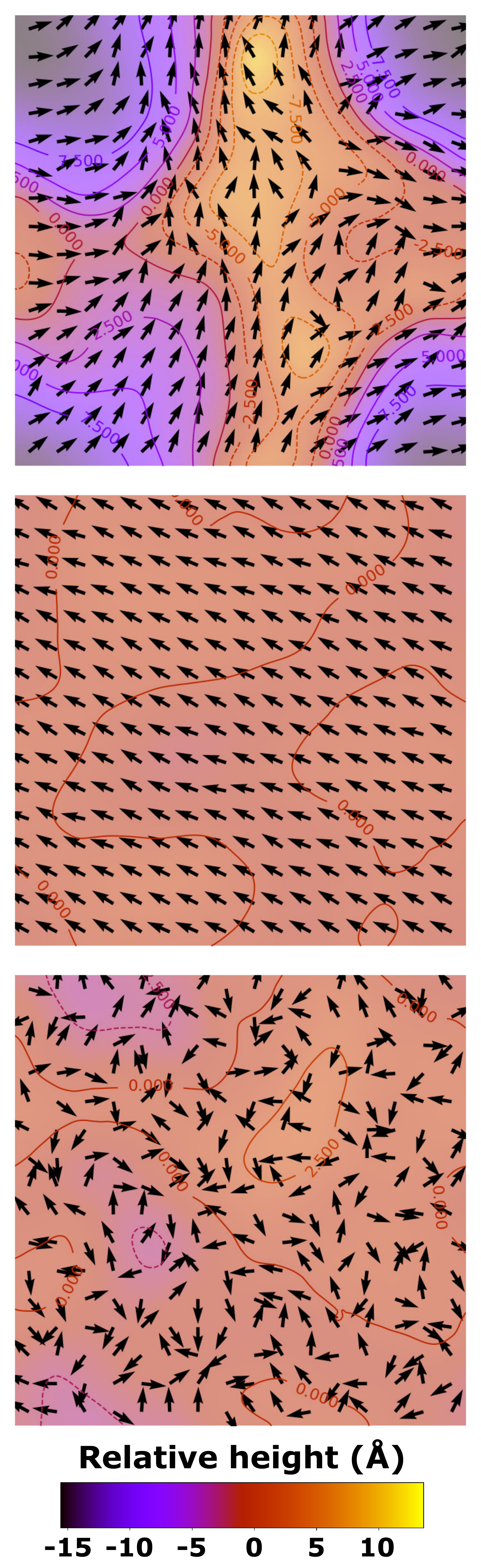}
\caption{}
\end{subfigure}
&
\begin{subfigure}{0.3\textwidth}
\centering
\includegraphics[width=\linewidth]{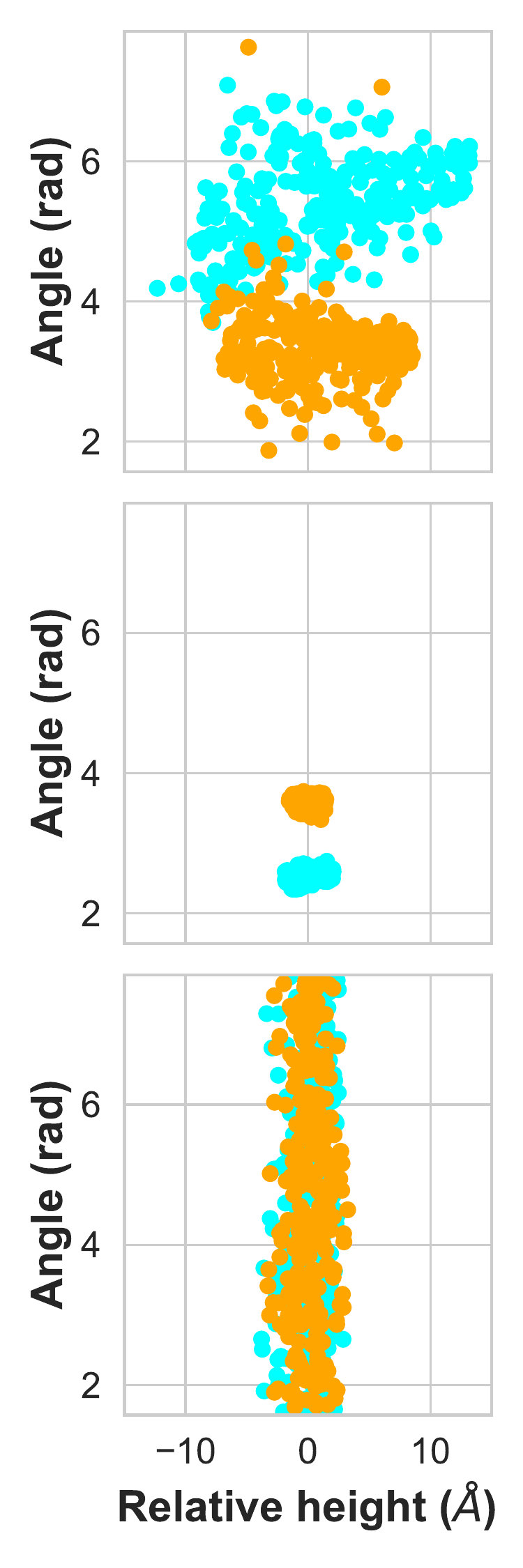}
\caption{}
\end{subfigure}
\end{tabular}

\caption{(a) Orientation in the XY plane of the tails of each of the 256 lipids forming a DPPC bilayer (Top) in the disordered gel phase, (Middle) in the tilted gel phase and (Bottom) in the fluid phase. Only one leaflet of the membrane is shown for each system. (b) Respective scatter distribution of the angles between the tails of the lipids and the X-axis of the membrane, as a function of the height on the leaflet. The measurement were performed on both top and bottom leaflets, respectively colored in cyan and in orange.}
\label{fig_tail_vectors}
\end{figure}

The local chain tilt direction in the XY plane was also determined and mapped (Figure~\ref{fig_tail_vectors}(a)). As expected, tilts are uniform in the $L_{\beta'}$ state and present a random short-range correlation in the fluid $L_\alpha$ state.  This was confirmed by the distribution of the angles with respect to the X-axis, as a function of the height on the leaflet. In the corrugated systems, the local chain tilts displays long-range variations. The angle distribution of the corrugated systems shows that small heights have random orientation similar to the fluid, while small heights are subject to less variations.

\begin{figure}[ht!]
\centering
\begin{tabular}[t]{c}

\begin{subfigure}{0.7\textwidth}
\centering
\includegraphics[width=\textwidth]{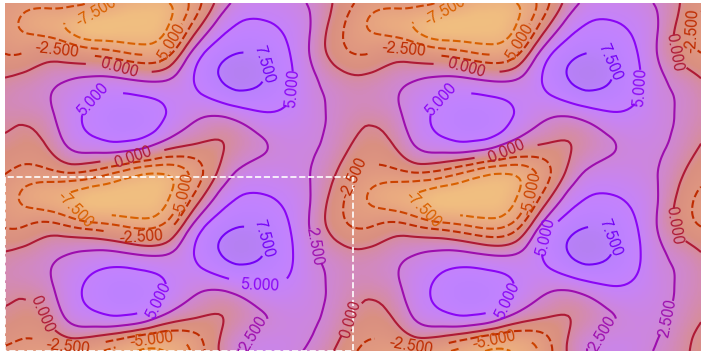}
\caption{}
\end{subfigure}
\\
\begin{subfigure}{0.7\textwidth}
\centering
\includegraphics[width=\textwidth]{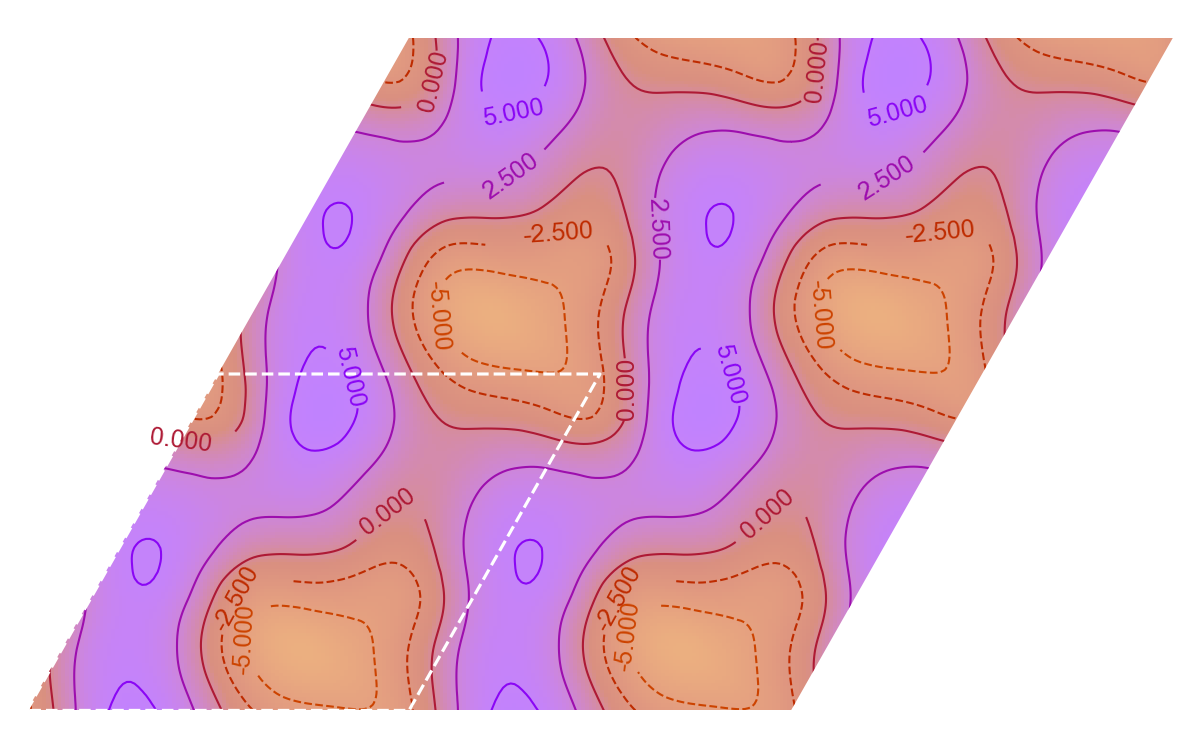}
\caption{}
\end{subfigure}
\end{tabular}

\caption{Topography of bilayers (a) made of 188 DPPC molecules in a simulation box with $L_x = 2L_y$ and (b) made of 128 DPPC molecules in an hexagonal simulation box, both simulated at 288~K.}
\label{fig_non_square}
\end{figure}

We also probed the sensitivity of the corrugation to modifications of space group. A simple change from cubic to rectangular simulation boxes with $L_x = 2L_y$ shows commensurate corrugations periods. Moreover we carry out simulations in hexagonal and monoclinic simulation boxes counting 128 lipids. Both of these systems were found to be corrugated, with modulation vectors directed along the PBC/crystallographic directions (Figure~\ref{fig_non_square}).

\begin{figure}[ht!]
\centering
\includegraphics[width=.8\linewidth]{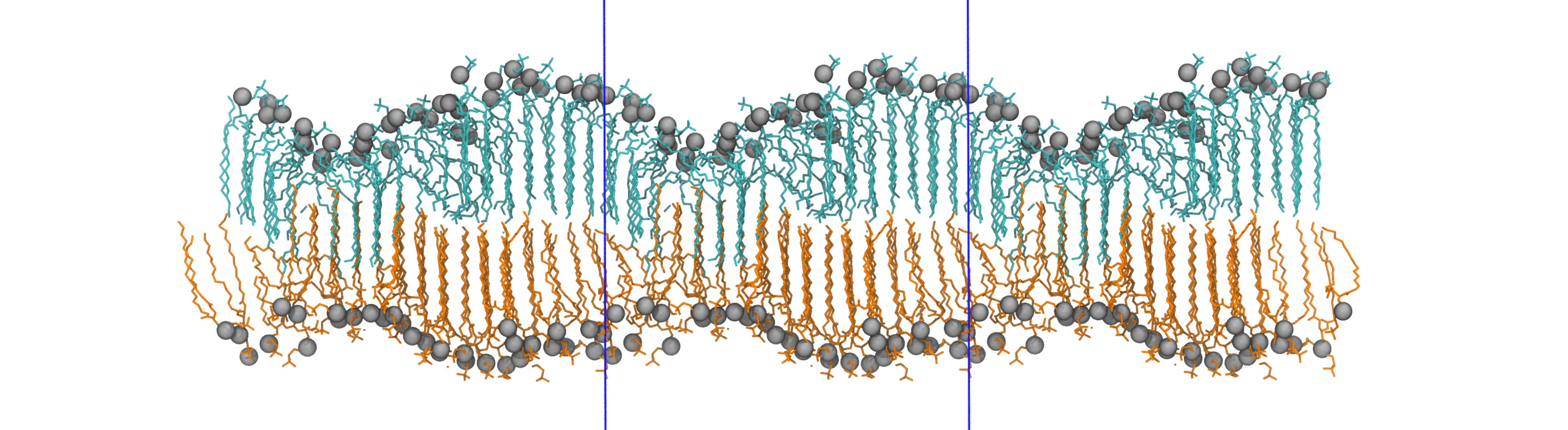}
\caption{Slice of the 212 DPPC molecule system obtained after being simulated at 288~K for 50~ns with an anisotropic barostat.}
\label{fig_stress_anisotropic}
\end{figure}

To release residual stress on the bilayer configuration that could have been brought by the semi-isotropic barostat, an anisotropic barostat was applied to the systems, with 1~bar and a $4.5 \times 10^{-5}$~bar$^{-1}$ compressibility set along each axes, and the pressure crossed terms set to 0 for both pressure and compressibility. We found that releasing residual anisotropic stress did not modify the structure of the bilayers. After this simulation run, the system was still found in a corrugated state, as shown in Figure~\ref{fig_stress_anisotropic}, with a relative area increase $\darea$ of 11.9 $\pm$ 0.8\% and a RMS corrugation amplitude of 4.89~\AA\, comparable to the amplitude before the run (5.18~\AA). We therefore conclude from this result that both tilted gel and corrugated states behave as a cohesive, solid state on the simulated time scales. They also display significant residual static stresses of 0.8 and 1.0~bars respectively in the x and the y directions. Based on these characterizations we refer to the corrugated state as the "disordered gel state", noted here $L^d_{\beta}$, to make a clear distinction between these configurations and the ripple phase $P_{\beta'}$ that has only be experimentally reported above the pre-transition temperature.

\begin{figure}[ht!]
\centering
\includegraphics[width=.9\linewidth]{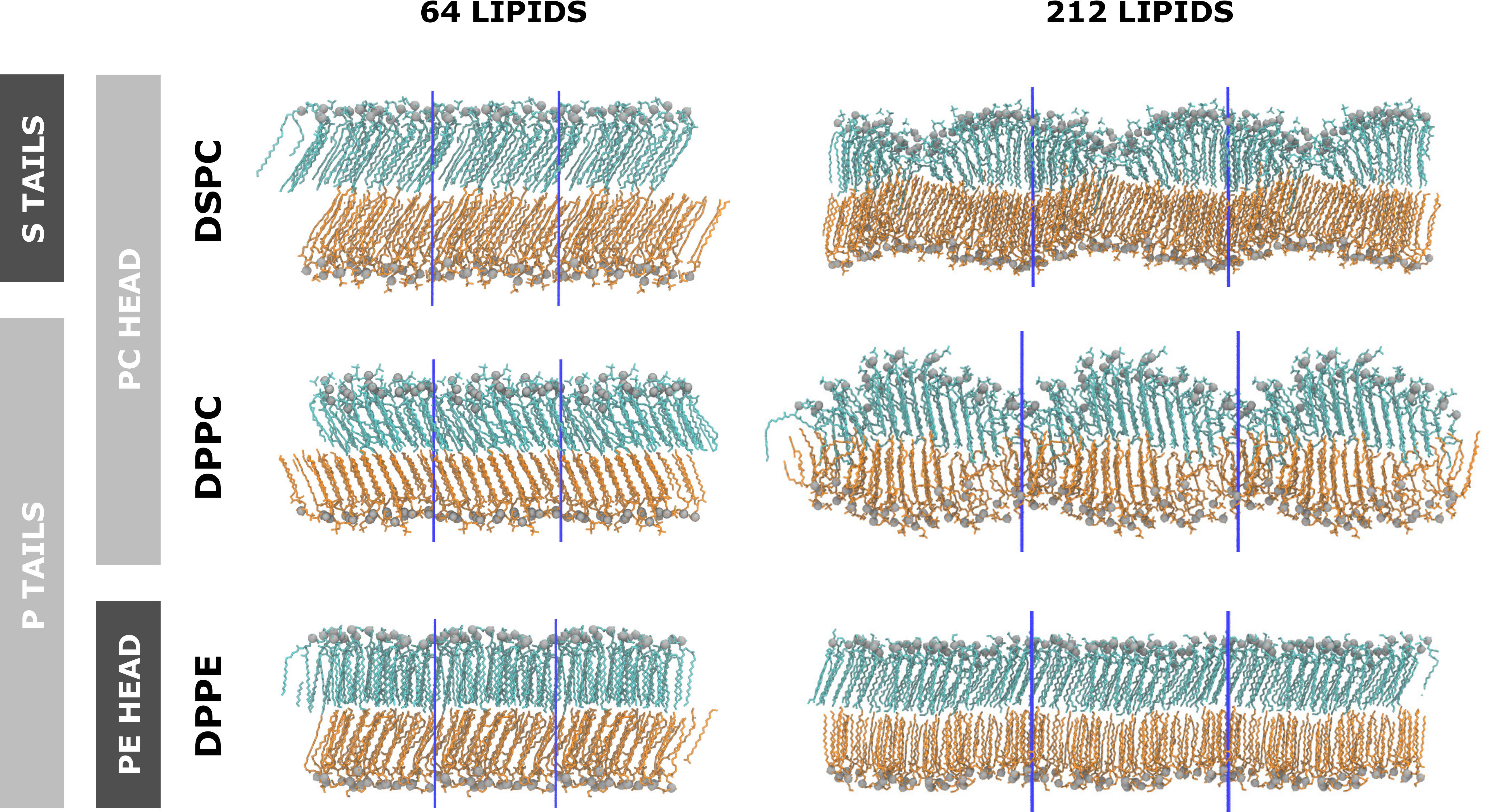}
\caption{Profiles of bilayers made of 64 or 212 lipids, either with DPPC, DSPC or DPPE, after a 50~ns simulation at 288~K. Only the DPPE membrane remained in the tilted gel phase when constructed with a large number of lipids.}
\label{fig_lipid_type}
\end{figure}

Finally one can wonder whether the appearance of the corrugation is restricted to DPPC or DMPC. To answer this question, we probed the effect of the tail and the head groups by considering the longer-tailed DSPC and the ethanolamine-based DPPE. As shown in Figure~\ref{fig_lipid_type}, the large DSPC systems were found in the disordered gel state while the DPPE systems remained in the expected homogeneous tilted gel state. The respective area differences of these systems are 12 $\pm$ 1 and 5 $\pm$ 1 \%.

\subsection{Influence of thermal history}

The previous observations suggest that disordered $L^d_{\beta}$ and  tilted $L_{\beta'}$ gel states are two competing states whose appearance seem to be correlated to the system size for freshly thermalised systems generated with CHARMM-GUI. In what follows we probe the sensitivity of the stability of these phases to different routes of thermal treatments.  

We focus our analysis on two system sizes: 64 lipids, which has been found in the $L_{\beta'}$ phase, and 256 lipids which has been found in the $L^d_{\beta}$. Both systems were subjected to the following thermal treatment: starting from $T=288K$, systems were annealed at $T=358$K in the fluid phase. These systems are respectively named pc64-A and pc256-A, for annealing. They were then cooled down to $T=288$K in two different ways: either with a brutal fast cooling, named here quenched, or with a slow gradual cooling of $1$K/ns that we denote gentle cooling. The quenched systems are noted pc64-AQ and pc256-AQ (annealing-quenching), and the cooled systems called pc64-AC and pc256-AC (annealing-cooling).

\begin{figure}[ht!]
\centering
\includegraphics[width=.9\linewidth]{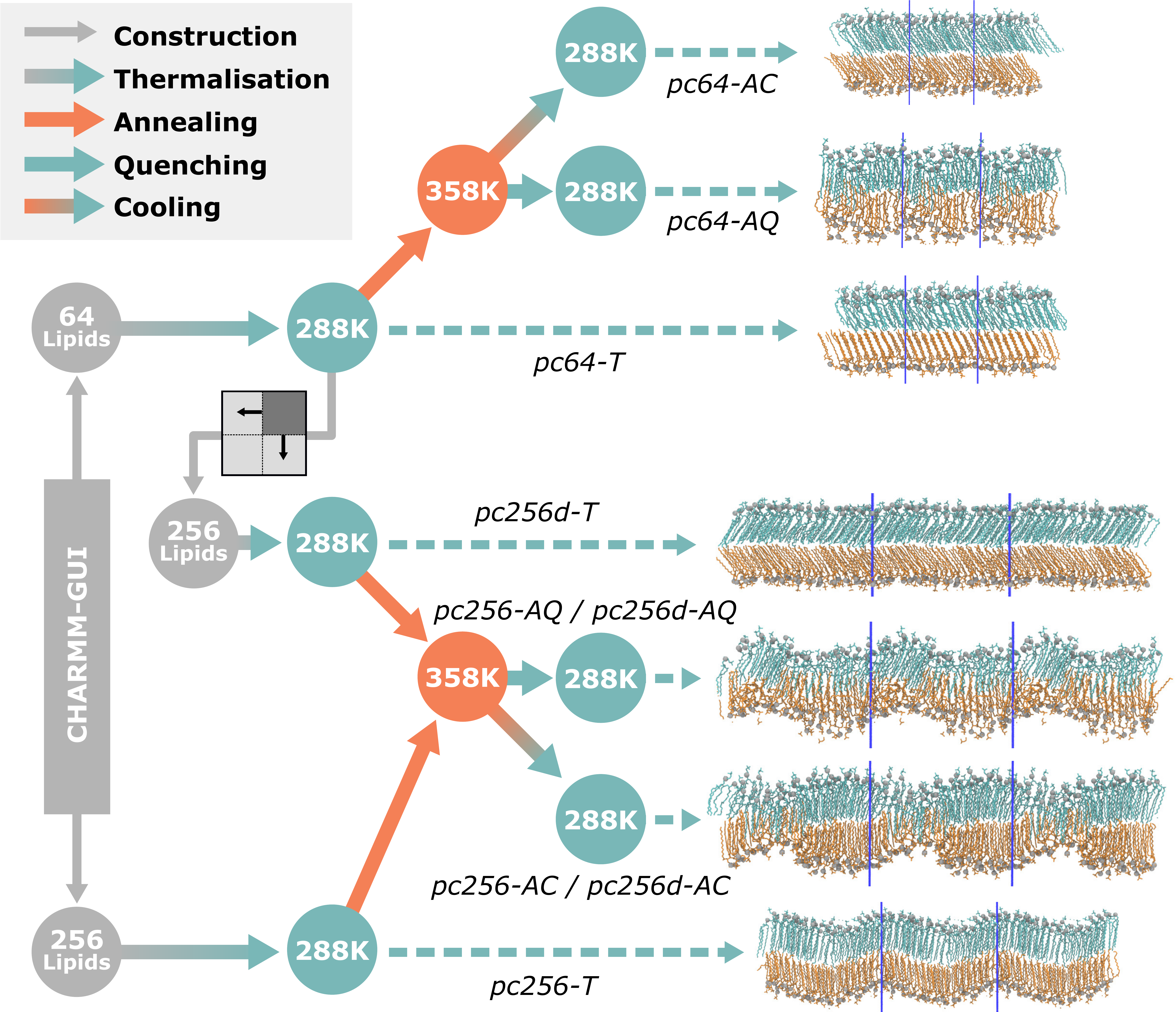}
\caption{Synopsis of the different thermal trajectories, showing the systems obtained for different cooling rates as well as the systems they originated from. Only the results from the 256 DPPC system obtained by replication of the 64 lipid system are shown for the quenching and cooling experiments. Results for the 256 DPPC system constructed by CHARMM-GUI, as well as the intermediate systems, are shown in the Supplementary Materials.}
\label{fig_temperature}
\end{figure}

We first notice that whatever the thermal history, large systems were always found in the disordered gel state as shown in Figure~\ref{fig_temperature}. Gentle cooling of small systems allowed them to recover the $L_{\beta'}$ phase while quenching lead to $L^d_{\beta}$ phase. As in this latter case, either phases could be obtained depending on the thermal history, we conclude that the disordered gel is metastable with respect to the tilted gel for small systems. Furthermore, we can note that both the gently cooled and the quenched systems have a final projected area per lipid $A_H^p$ (respectively 50.1 $\pm$0.7 and 50.8 $\pm$ 0.8~\AA$^2$) close to the average projected area of 50~\AA$^2$ found for systems of all sizes, hence correcting the odd value of 52.3~\AA$^2$ found before temperature treatment (\textit{cf}~Figure~\ref{fig_area_size}).

As large systems of 256 lipids were never spontaneously found in the $L_{\beta'}$ state, we decided to force them into this state by duplicating along both $x$ and $y$ directions the 64 $L_{\beta'}$ system. The flat tilted duplicated system was found to be stable on the simulation time scale, suggesting that $L_{\beta'}$ could be either stable or metastable in large systems too. The nature of the relevant stable thermodynamic phase for large systems remains an open question, while our simulations clearly favor the disordered gel state $L^d_{\beta}$. Measurements of the difference in area $\darea$ can be found in the Supplementary Materials.

\subsection{Thermodynamics of tilted and disordered states}

Since the small systems made of 64 lipids can be prepared and controlled to reach all the observed phases, we used them to compare the energetic properties of the respective gel phase. Having in mind the idea of a complex underlying potential energy landscape composed of several minima located at different energy levels, we probe how far energetically the disordered gel phase stand from the tilted gel phase minimum. Therefore we performed an energy minimization using conjugate gradient to remove thermal fluctuations on the tilted and disordered gel configurations. We found $\Delta E_P=E_P(\text{disordered})-E_P(\text{tilted})=64$kJ/mol, meaning that if both states are metastable, the tilted system is the most stable one. In addition, we also computed the difference in enthalpy between the disordered and the tilted thermalised gel phases. To this aim, we used the same initial states, after quenching or gentle cooling but without energy minimisation. Since the two systems share the same atomic compositions (same number of lipid, water molecules and constraints), the difference in enthalpy (kinetic and potential energy according to the force field, plus the $PV$ contribution)  at a given temperature should be characteristic of the enthalpy difference between the two states. We found an enthalpy difference of $12 \pm 4$kJ/mol at $288$K. This value is higher than the one reported experimentally for the calorimetric gel to ripple phase pretransition, 4.6~kJ/mol~\cite{Cevc_Marsh_PhospholipidBilayers} but obtained at a higher temperature. We therefore repeated the measure after rising both systems at a temperature close to the pretransition temperature, namely 305~K, and then found a difference in enthalpy of $3.9 \pm 0.4$ kJ/mol now compatible with experimental measurements. It is essential to note here that, in order to obtain a disordered system configuration at 305~K, the system had first to be annealed at 375~K instead of 358~K, meaning that the difference in temperature required for quenching should be at least of 70~K in order to obtain the $L^d_{\beta}$ phase (see Supplementary materials).

The enthalpy of the $L^d_{\beta}$  $ \longrightarrow L_\alpha$ transition was also measured in our small systems. This was performed by simulating the 64 DPPC system at temperatures ranging from 283 to 358~K and by removing the change in enthalpy due to the change in temperature (data shown in Supplementary Materials). The transition enthalpy was found equal to 27.3~kJ/mol, which is also comparable with the experimental reported values  \textit{circa} 32.2~kJ/mol~\cite{Cevc_Marsh_PhospholipidBilayers}. We can conclude that despite the approximations, and the absence of quantum corrections to the bond vibrations contributions, the CHARMM36 force field thermodynamic predictions seem in good agreement with experimental observations.

\section{Discussion}

A careful inspection of the corrugations shows that the topographic modulation is imposed by the periodicity of the simulation box. The corrugation shows finite system size dependence, as its amplitude increases up to sizes of the order of $L_x= 8$~nm where the modulation saturates to a value of 7 $\pm$ 1~\AA. 
Assuming a sine-like corrugation, this maximum RMS amplitude measured can be converted into a peak-to-peak amplitude of 2.0 $\pm$ 0.2~nm similar to the previously published values of 2.4~nm for the DPPC or even 1.8~nm in DMPC bilayers~\cite{deVries_marrink_2005, akabori_nagle_2015}. The saturation of the corrugation amplitude is expected, given that it can only reach a fraction of the total membrane thickness. The associated lateral length 8~nm can therefore be considered as a lower bound of the instability characteristic longitudinal length scale.

We believe that the relative difference between the interfacial area and the projected one, $\darea$, introduced in equation (\ref{new_order_parameter}) can be taken as a relevant order parameter for the transition between the $L_{\beta'}$ and the disordered gel phases $L^d_{\beta}$. The latter being reminiscent from $P_{\beta'}$ ripple phase, $\darea$ could be seen as a critical parameter to investigate the existence of the ripple phase and discriminate it from the tilted gel or from the fluid phase.

However, unlike experiments, the numerical instability occurs along two orthogonal directions, or along the hexagonal axes. Non-square boxes fail to select only one modulation direction. We nevertheless think that the numerical corrugation instability is related to the experimental ripple instability, as also suggested by the dependence of the presence of corrugation to the chemical nature of the heads and tails of the lipids. Indeed, experiments have shown that the ripple phase is specific from the phosphocholine (PC) lipids~\cite{cunningham_brain_1998, katsaras_nagle_2000}. 

Another striking observation is the insensitivity of large systems (256 lipids) to thermal treatment which have been systematically ended in the disordered gel state. On the opposite small systems can alternate between both phases. However, when thermalised at low temperatures from CHARMM-GUI or slowly quenched, they end up preferentially into the tilted gel state, suggesting that in the range of temperatures investigated the $L_{\beta'}$ state is thermodynamically favored. By contrast the $L_{\beta'}$ state  is never selected spontaneously by larger systems. For the latter, MD suggests that the disordered gel state $L^d_{\beta}$ is the most preferred state for all temperatures below melting. Metastability and kinetic effects are certainly significant and may hide to true nature of the stable phase. 

Assuming that the difference between tilted and disordered gels has something to do with the pretransition, we found an enthalpy difference 3 times larger than the premelting latent heat at 288K, and of the same order of magnitude at 305K, close to the observed experimental transition. This points towards the relevance of disordered gel state $L_{\beta}^d$ as a ripple state analogue.  Moreover the higher enthalpy of the disordered gel with respect to the tilted gel is consistent with a $L_{\beta'} \longrightarrow$ $L^d_{\beta}$  $ \longrightarrow L_{\alpha}$ sequence of transitions as temperature is increased.

The reason of the outcome of a ripple instability below the melting temperature is nothing but obvious. Our simulations point out to a competition between an homogeneous tilted state and an inhomogeneous corrugated state. The transition between these states is discontinuous. The corrugated state is not very tilted, and partially melted, or disordered, and interdigitated. We suggest now a possible mechanism explaining the observed situation. The tilted phase can be understood as the result of a frustration between lipid headgroups which try to increase their exposure to water in the interface region, lipid chains which try to reach an optimal packing density as a result of cohesive forces, and chain stiffness for which the introduction of \textit{gauche} dihedral angles is unfavorable at low temperatures. Tilt allows lipid to optimize simultaneously those three constraints. On the other hand, melting a lipid chain enables the release of the constraint acting on the chain stiffness, and makes it possible to increase the hydration free-energy by reducing the membrane thickness. Below melting temperature, thermodynamics makes it unfavorable to melt all the lipid molecules. However, some local disordering of the lipids may still be favorable, increasing the hydration of the headgroups without need of spending too much energy in melting the chains. 

\begin{figure}
    \centering
    \resizebox{0.6\textwidth}{!}{\includegraphics{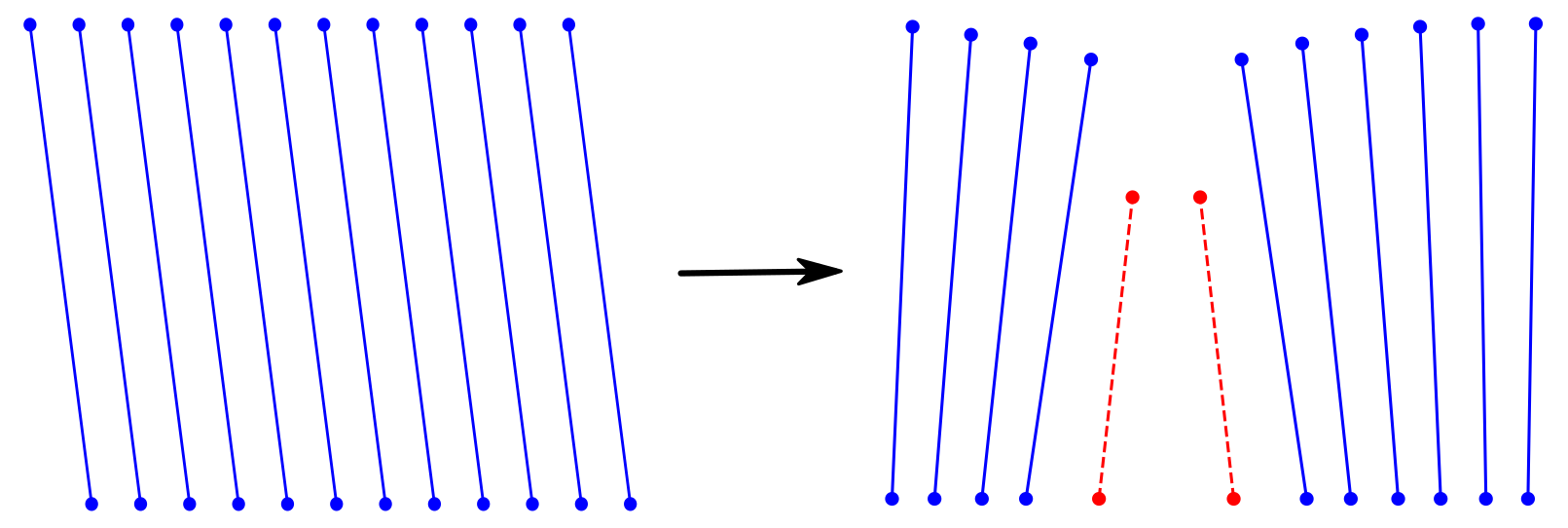}}
    \caption{Schematic representation of the mechanical instability occurring in a simple one dimensional lipid chain model, where the tilted state has a higher energy than a locally disordered state. The gain in energy originates mostly from the hydration term. Blue solid rods stand for gel state, red dashed rods for fluid state.}
    \label{fig_corrugationInstability1D}
\end{figure}

Based upon those considerations, we designed a simple one dimensional lipid chain model that supports the idea that in an temperature range just below melting, the homogeneous tilted state energetically unstable with respect to a local corrugation of the bilayer, see Figure~\ref{fig_corrugationInstability1D}. Details of the parameterisation of the model and further results are presented in the Supplementary Information section. We conclude that the thickness modulation might indeed be caused by a subtle interplay between headgroup hydration, hydrophobic chain packing, \textit{trans-gauche} isomerisation and tilt elasticity energy terms.  

\section{Conclusion}

We have successfully demonstrated in this work how the size of the simulation box influences the ripple-like instability in a PC membrane simulated with the Charmm36 force field at low temperature, where it is usually expected to be in the tilted gel $L_{\beta'}$ phase. This unexpected organisation, which we called the disordered gel phase $L^d_{\beta}$, does not appear in small systems, which is consistent with the results from Khakbaz and Klauda~\cite{2018_Khakbaz_Klauda} as well as with our results in a previously published paper~\cite{walter_thalmann_2020}. The energy and geometry analyses demonstrated that this disordered gel phase has a lot in common with the $P_{\beta'}$ phase. Furthermore, this instability was not observed with PE lipids, in agreement with experimental findings. For small systems, we found ways of preparing the system in either tilted gel or disordered gel states by acting on the thermal treatment. More work is needed to determine whether the Charmm36 force-field can describe a one-dimensional $P_{\beta'}$ spatial thickness modulation, with the right periodicity, and at which temperatures.  Our work suggests that simulations will have to be guided to the desired structure. Finally, we conclude that the ripple instability looks like a generic mechanism adopted by the phosphatidylcholines lipids to increase the headgroup hydration while still satisfying the packing  constraints, at the expense of a mild cost in disordering/melting a small fraction of the chains.   






\section*{Acknowledgements}

V. W. warmly thanks T.E. de Oliveira for helping with simulations set-up. A.G. acknowledges partial support from the Investissements d’Avenir program “D\'{e}veloppement de l’\'{E}conomie Num\'{e}rique” through the SMICE project.
The authors gratefully acknowledge support from the high performance cluster (HPC) Equip@Meso from the University of Strasbourg, through grants no.~ G2019A131C and G2020A140C.

\bibliographystyle{elsarticle-num-names}
\bibliography{references.bib}







\end{document}


\begin{frontmatter}

\title{Ripple-like instability in the simulated gel phase of finite size phosphocholine bilayers \\ -Supplementary Materials-}



\author[kcl]{Vivien Walter}
\author[ics]{C\'eline Ruscher}
\author[ics]{Adrien Gola}
\author[ics]{Carlos M. Marques}
\author[ics]{Olivier Benzerara}
\author[ics]{Fabrice Thalmann}

\address[kcl]{Department of Chemistry, King’s College London, Britannia House, 7 Trinity Street, SE1~1DB, London, United Kingdom}
\address[ics]{Institut Charles Sadron, CNRS and University of Strasbourg, 23 rue du Loess, F-67034 Strasbourg, France}

\end{frontmatter}

\newpage
\appendix

\section{System Composition}

The atomic composition of all systems used in this work is given in the Table~\ref{si_table_composition} below. Each DPPC molecule is made of 130 atoms, DSPC of 142 atoms and DPPE of 121. The average number of water molecules per lipid is 163 $\pm$ 8.

\begin{table}[!ht]
\centering
\caption{
{\bf Atomic composition and size of the different simulation systems used in this work.}}
\begin{tabular}{c c||c|c|c|c}
{\bf Type} & {\bf System} & {\bf \# Lipids}  & {\bf \# Water} & {\bf L\textsubscript{x}, L\textsubscript{y}\textsuperscript{a} (nm)} & {\bf L\textsubscript{z}\textsuperscript{a} (nm)} \\ \hline
\textbf{DPPC} & 32 Lipids & 4,160 & 15,570 & 2.91 & 22.55 \\
& 64 Lipids & 7,020 & 31,869 & 4.24 & 21.57 \\
& 94 Lipids & 12,220 & 46,590 & 5.12 & 21.65 \\
& 128 Lipids & 16,640 & 63,897 & 5.98 & 21.71 \\
& 170 Lipids & 22,100 & 84,315 & 6.87 & 21.77 \\
& 188 Lipids\textsuperscript{b} & 24,440 & 93,180 & (L\textsubscript{x}) 11.09 & 19.74 \\
& & & & (L\textsubscript{y}) 5.55 & \\
& 212 Lipids & 27,560 & 89,478 & 7.69 & 19.01 \\
& 256 Lipids & 33,280 & 127,926 & 8,46 & 21.71 \\
& 256 Lipids\textsuperscript{c} & 33,280 & 127,476 & 8.13 & 23.26 \\
& 512 Lipids & 66,560 & 255,963 & 11.97 & 21.71 \\
\hline
\textbf{DSPC} & 64 Lipids & 9,088 & 32,526 & 4.29 & 21.74 \\
& 212 Lipids & 30,104 & 66,447 & 7.82 & 15.18 \\ \hline
\textbf{DPPE} & 64 Lipids & 7,744 & 30,582 & 4.01 & 23.07 \\
& 212 Lipids & 25,652 & 62,925 & 7.44 & 15.64 \\
\end{tabular}
\begin{flushleft} \textsuperscript{a} Values measured on the last frame of the equilibrated system before simulation. \\
\textsuperscript{b} System constructed by replication of the system made of 94 lipids along the X axis for the $L_x = 2 L_y$ system. \\
\textsuperscript{c} System constructed by replication of the system made of 64 lipids along both X and Y axes.
\end{flushleft}
\label{si_table_composition} 
\end{table}

\newpage
\section{Corrugation formation and characterization}

\begin{figure}[ht!]
\centering
\begin{tabular}[t]{ccc}

\begin{subfigure}{0.31\textwidth}
\centering
\smallskip
\includegraphics[width=\linewidth]{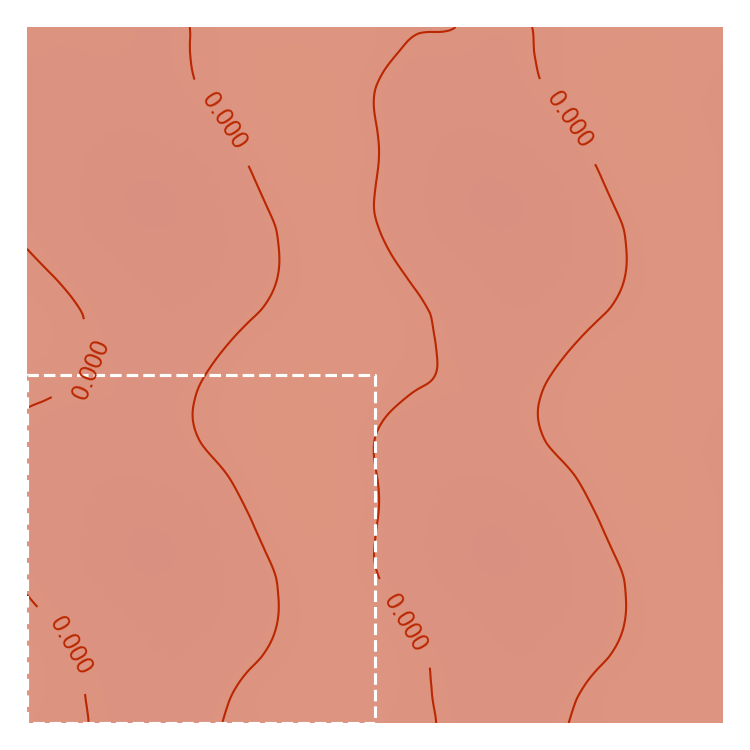}
\caption{32 Lipids}
\end{subfigure}
&
\begin{subfigure}{0.31\textwidth}
\centering
\smallskip
\includegraphics[width=\linewidth]{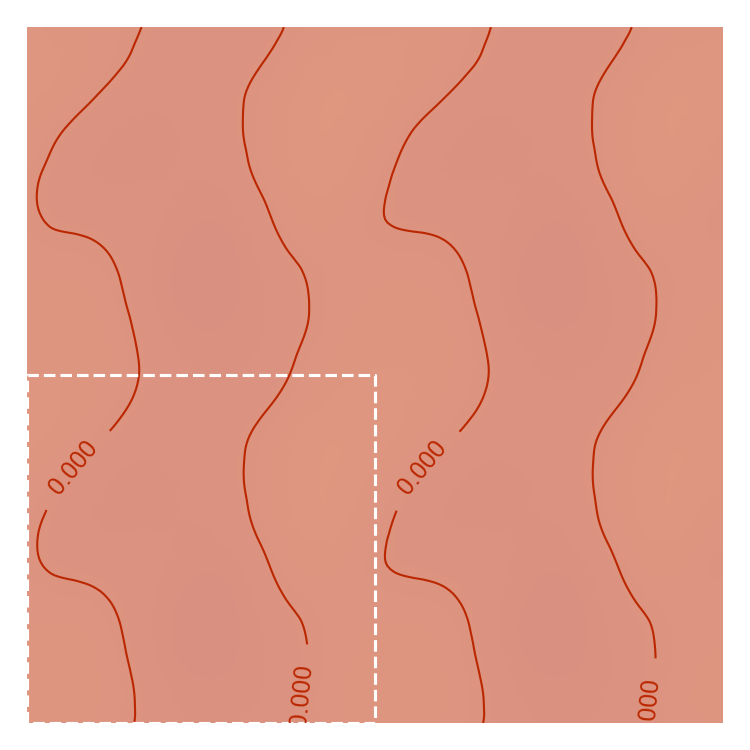}
\caption{64 Lipids}
\end{subfigure}
&
\begin{subfigure}{0.31\textwidth}
\centering
\smallskip
\includegraphics[width=\linewidth]{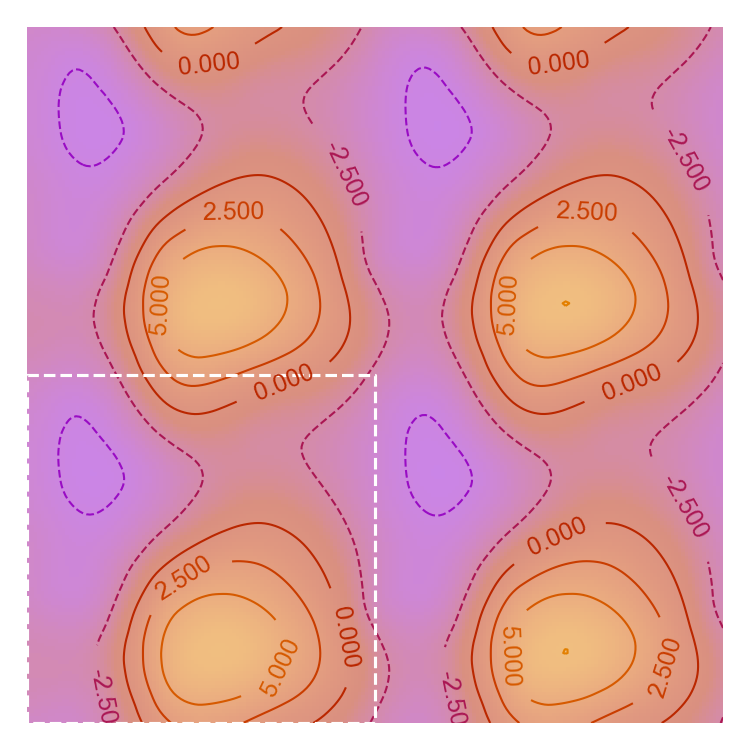}
\caption{94 Lipids}
\end{subfigure}
\\
\begin{subfigure}{0.31\textwidth}
\centering
\smallskip
\includegraphics[width=\linewidth]{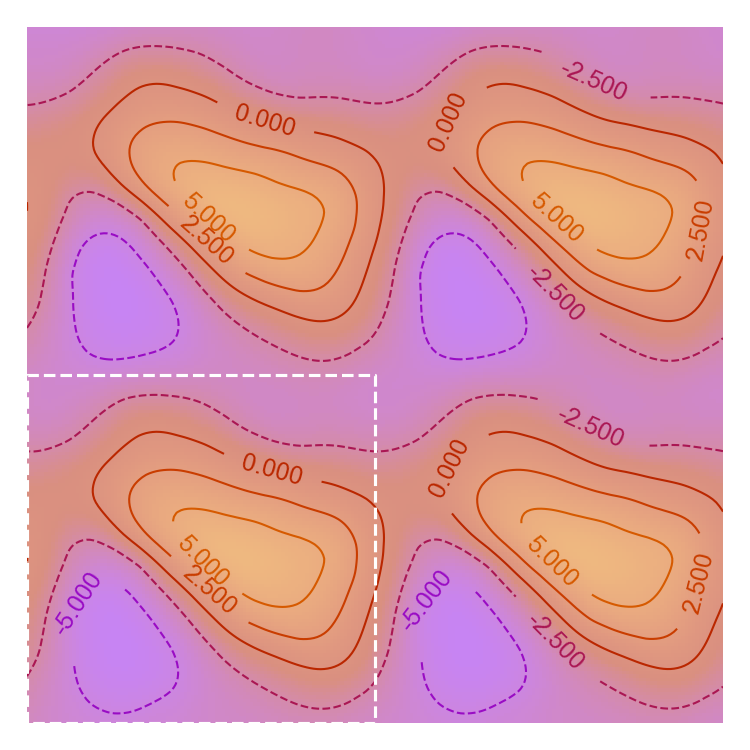}
\caption{128 Lipids}
\end{subfigure}
&
\begin{subfigure}{0.31\textwidth}
\centering
\smallskip
\includegraphics[width=\linewidth]{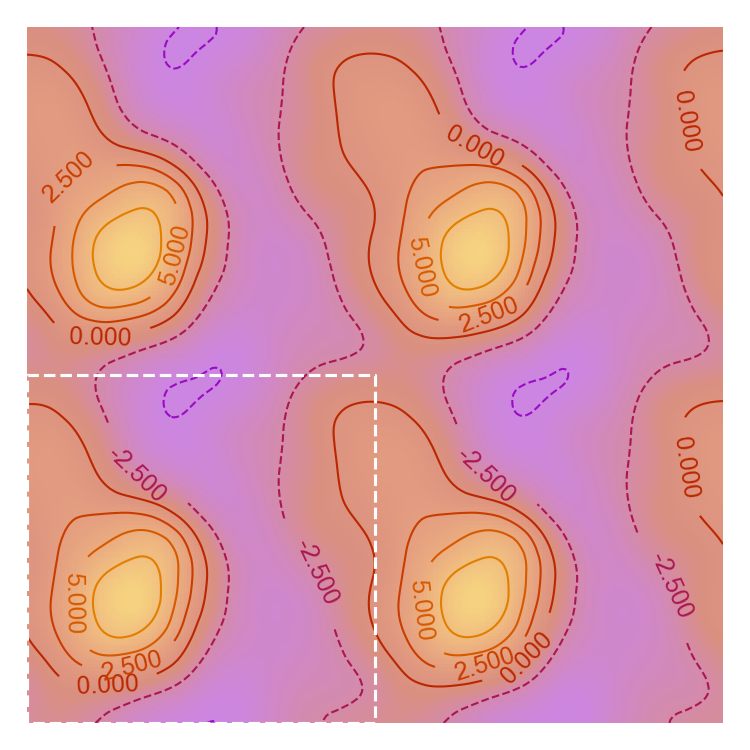}
\caption{170 Lipids}
\end{subfigure}
&
\begin{subfigure}{0.31\textwidth}
\centering
\smallskip
\includegraphics[width=\linewidth]{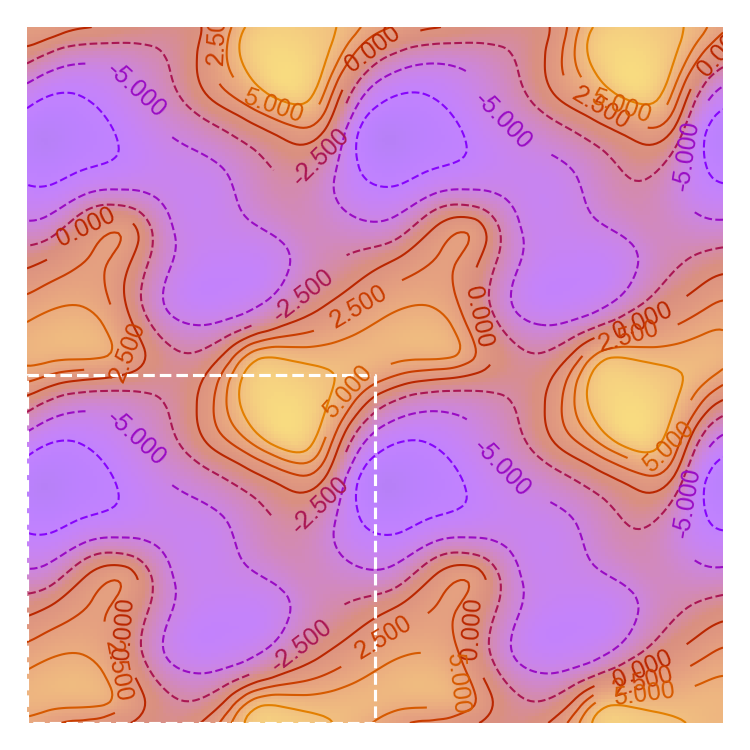}
\caption{212 Lipids}
\end{subfigure}
\\
\begin{subfigure}{0.31\textwidth}
\centering
\smallskip
\includegraphics[width=\linewidth]{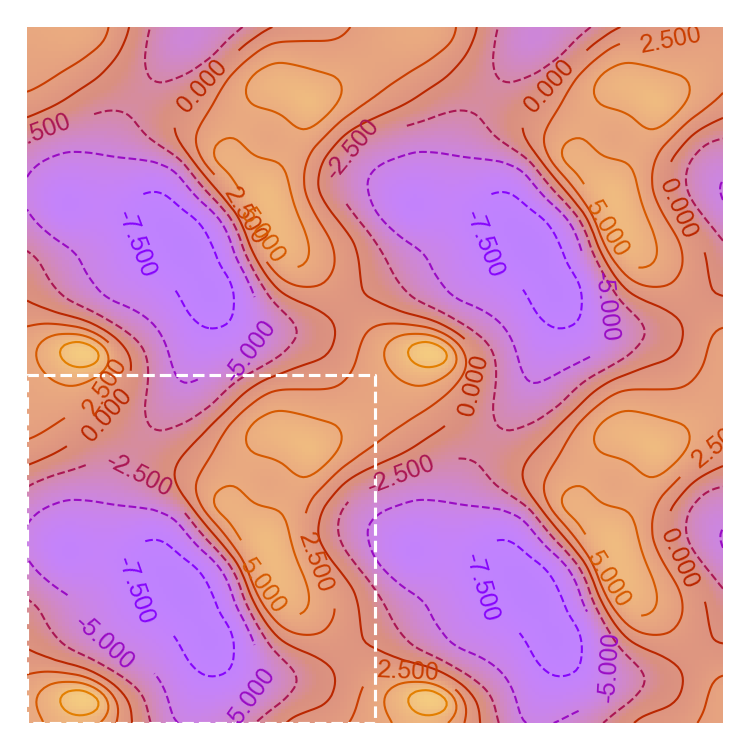}
\caption{256 Lipids}
\end{subfigure}
&
\begin{subfigure}{0.31\textwidth}
\centering
\smallskip
\includegraphics[width=\linewidth]{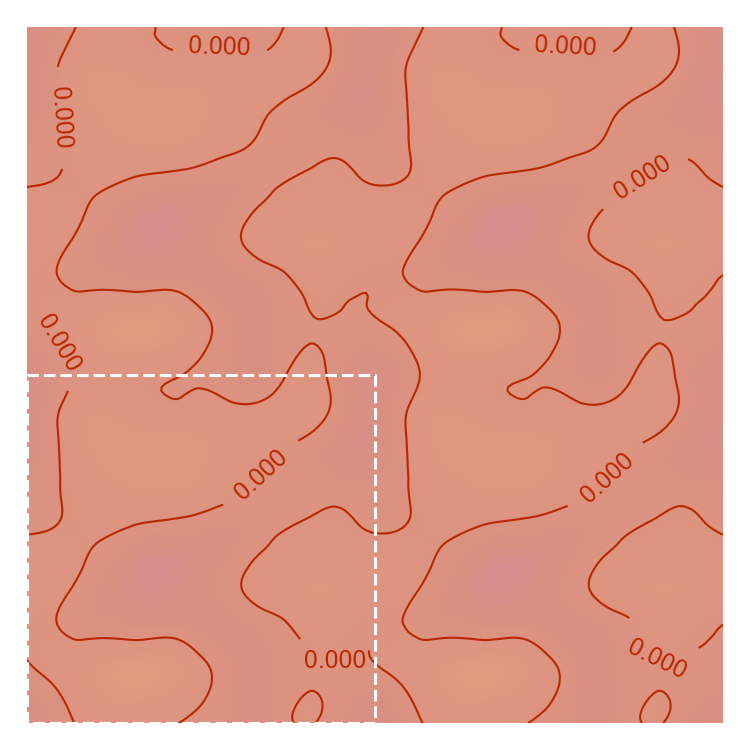}
\caption{256 Lipids (replicated)}
\end{subfigure}
&
\begin{subfigure}{0.31\textwidth}
\centering
\smallskip
\includegraphics[width=\linewidth]{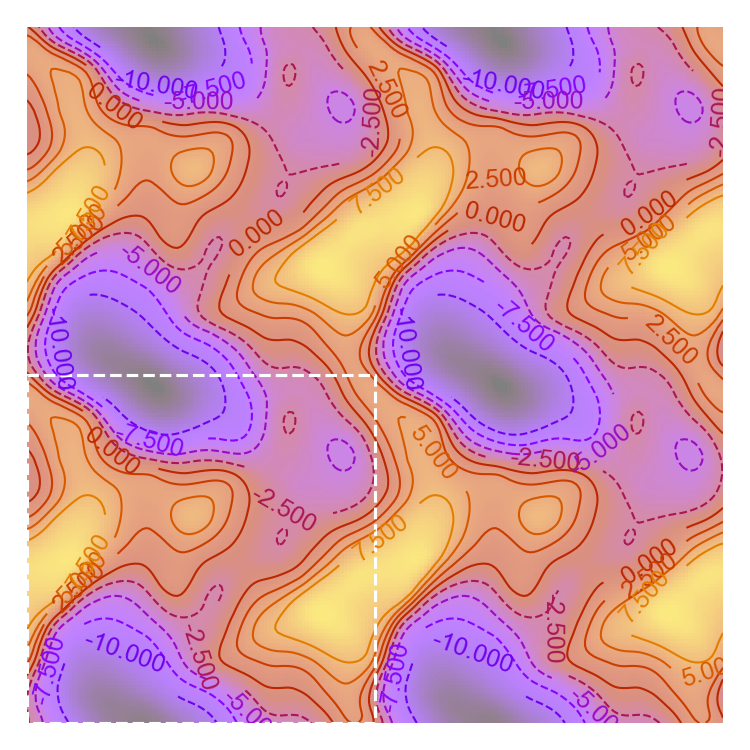}
\caption{512 Lipids}
\end{subfigure}

\end{tabular}

\caption{Bottom leaflets.}
\label{si_fig_view_bottom}
\end{figure}

\begin{figure}[ht!]
\centering
\begin{tabular}[t]{c}

\begin{subfigure}{0.8\textwidth}
\centering
\includegraphics[width=\linewidth]{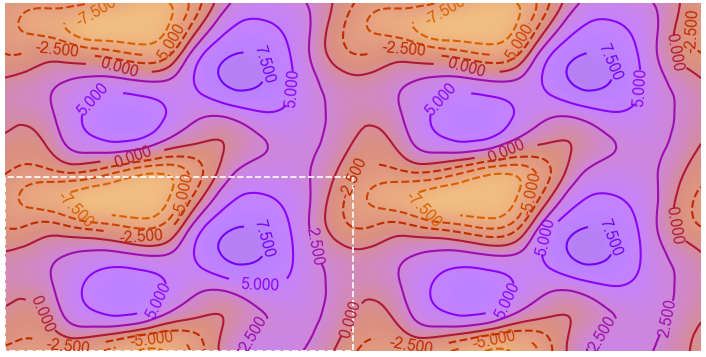}
\caption{} 
\end{subfigure} 
\\
\begin{subfigure}[t]{0.8\textwidth}
\centering
\includegraphics[width=\linewidth]{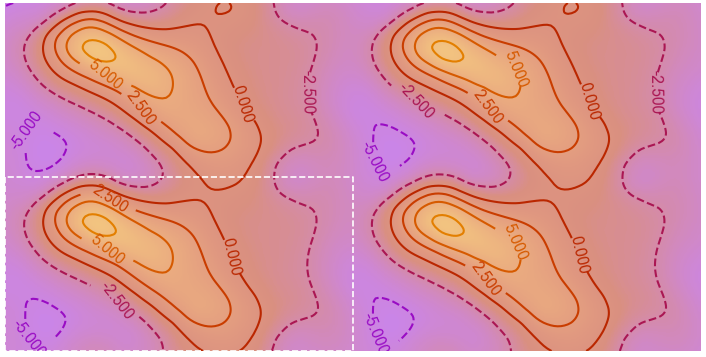}
\caption{}
\end{subfigure}
\end{tabular}

\caption{Corrugation in a system where $L_x = 2L_y$. (a) Top and (b) bottom leaflets are both shown. Color code shows in blue the thick portions of the membrane while red are the thin ones.}
\label{si_fig_non_square}
\end{figure}

\begin{figure}[ht!]
\centering
\includegraphics[width=.9\linewidth]{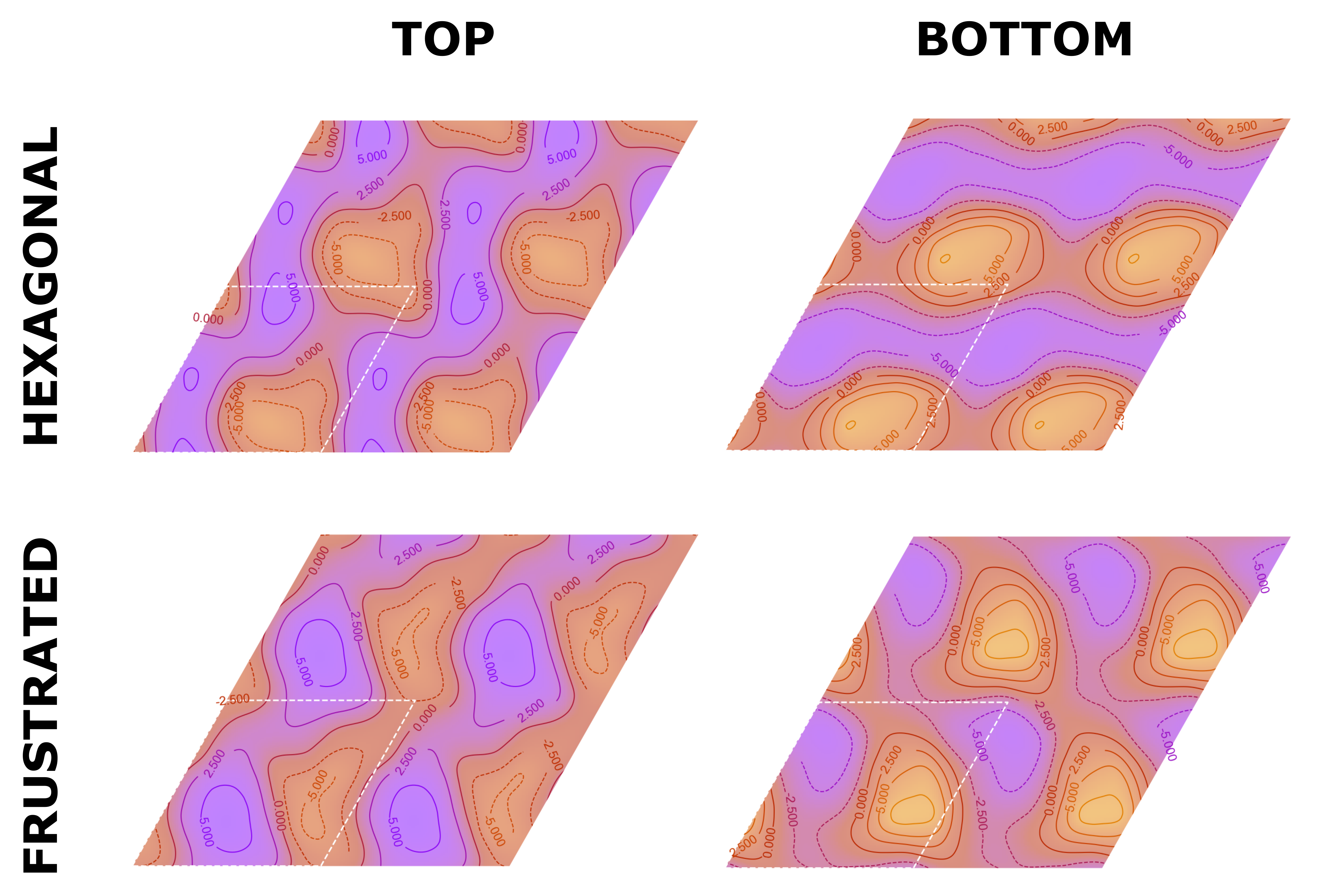}
\caption{Corrugation in a hexagonal system. (a) Top and (b) bottom leaflets are both shown. Color code shows in blue the thick portions of the membrane while red are the thin ones.}
\label{si_fig_hexagonal}
\end{figure}

The topography of all the DPPC bilayers simulated at 288~K right after construction is shown below. The figures are separated into top (Figure~\ref{si_fig_view_top}) and bottom leaflets (Figure~\ref{si_fig_view_bottom}). The complete sets of leaflets for the non-square systems, both rectangular (with $L_x = 2 L_y$, Figure~\ref{si_fig_non_square}) and hexagonal or frustrated (Figure~\ref{si_fig_hexagonal}). To facilitate the comparison between systems, all the color codes and contour lines were scaled with respect to the system with the largest amplitude. The color scale of the height is set as the distance from the mean height of the leaflet (0) toward the center of the bilayer (positive values, yellow/red). Negative height values are height further from the center of the bilayer than the mean height (purple/blue).

\begin{figure}[ht!]
\centering
\begin{tabular}[t]{c}

\begin{subfigure}{0.6\textwidth}
\centering
\smallskip
\includegraphics[width=\linewidth]{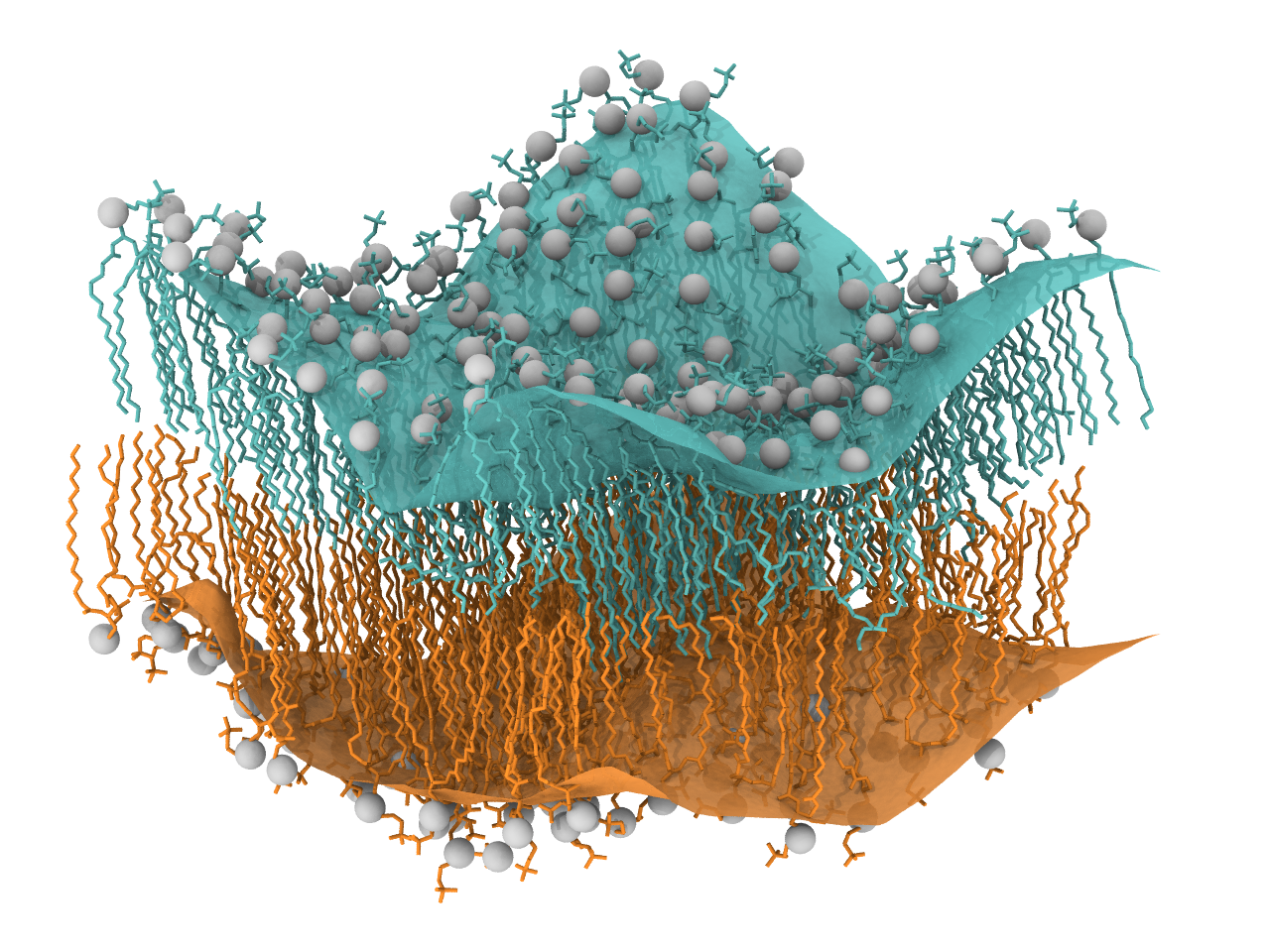}
\caption{}
\end{subfigure}
\\
\begin{subfigure}{0.6\textwidth}
\centering
\smallskip
\includegraphics[width=\linewidth]{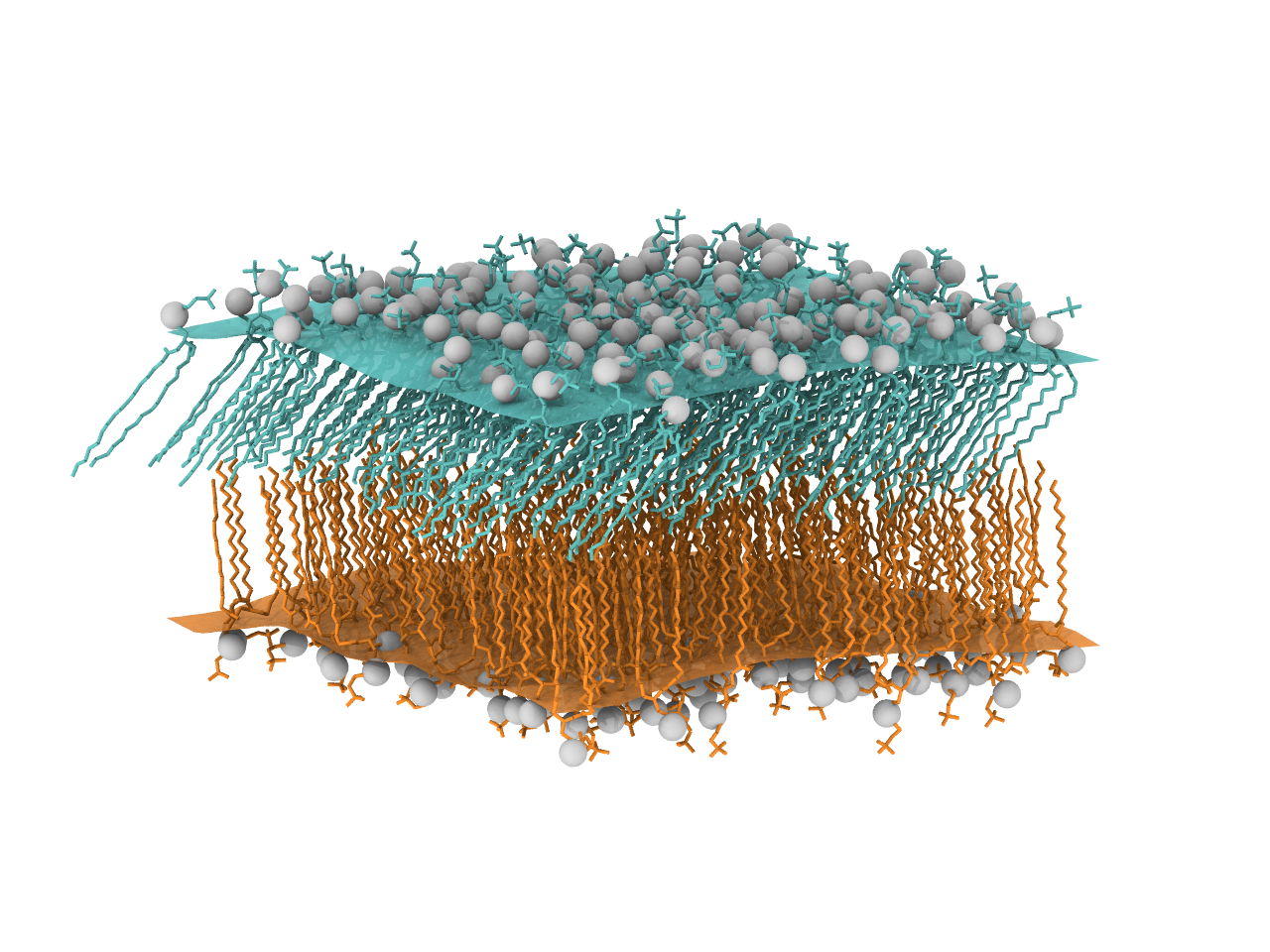}
\caption{}
\end{subfigure}

\end{tabular}

\caption{Rendering of DPPC bilayers made with 256 lipids (a) generated by CHARMM-GUI and simulated at 288~K or (b) produced by replication of a 64 DPPC bilayer found in the tilted gel phase; with the surface meshing generated on the water-membrane interface (using water molecules as the reference) to calculate the meshed area per lipid. The respective area difference of the systems were calculated at 13 $\pm$ 2 and 2 $\pm$ 1\%.}
\label{si_fig_surface_mesh}
\end{figure}

\begin{figure}[ht!]
\centering
\includegraphics[width=.9\linewidth]{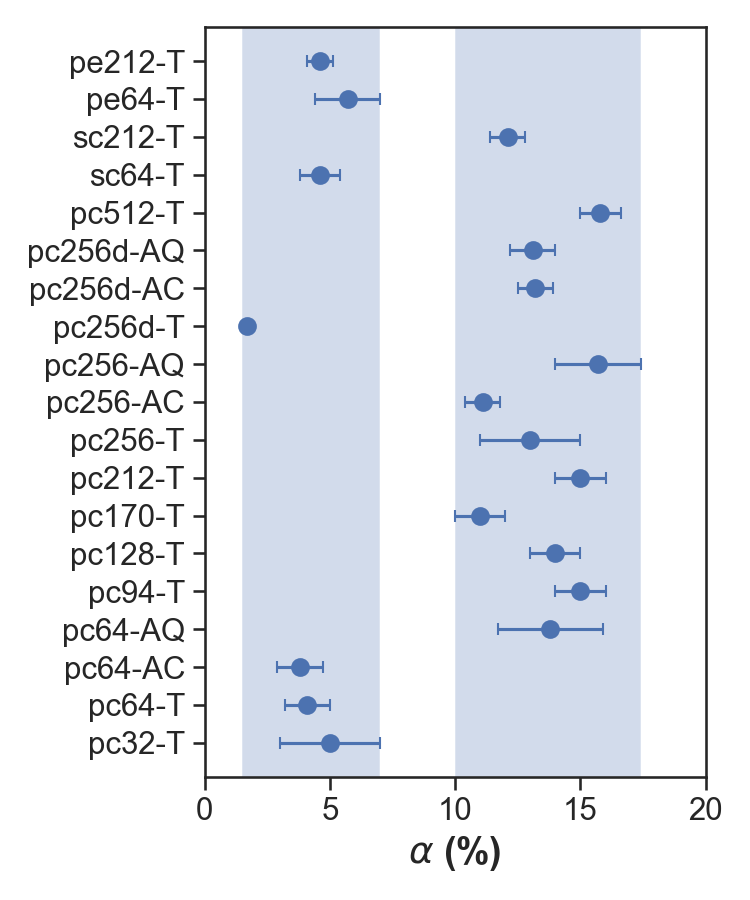}
\caption{Distribution of the mean $\darea$ of all the square systems simulated at 288~K. The blue bands highlights the two ranges of values of $\darea$ in which the tilted and disordered phases were observed. All systems with a $\darea < 10$~\% were visually classified as tilted gel phase, while all systems with $\darea \geq 10$~\% were classified as disordered gel. Error bars are the standard deviations of each distribution.}
\label{si_fig_area_ratio}
\end{figure}

Figure~\ref{si_fig_surface_mesh} is a rendering generated in Ovito 2.9 to illustrate the surface meshing performed to measure the meshed area per lipid $A_H^m$ of the membrane. The mesh was done on the water molecules at the interface with the lipid bilayer, with a probe sphere radius of 6 and a smoothing level of 30. The meshed surface was fed to a two-dimensional Fast-Fourier Trasnform (FFT) routine, and the squared moduli of the Fourier coefficients (power spectrum density PSD) plot as a function of the $x$ and $y$ components of the wave numbers $k_x$, $k_y$. Figures~\ref{si_fig_psd},~\ref{si_fig_psd2} shows that no significant peak exist except but the periodicity of the simulation box. The projected area per lipid $A_H^p$ was simply calculated by multiplying together the size of the box along the X and Y axes and dividing by the number of lipid per leaflet. These areas were measured for all the square systems and gathered in Table~\ref{si_table_area}. The relative increase of area per lipid $\darea$ from $A_H^p$ to $A_H^m$ was also measured for all these and shown in Figure~\ref{si_fig_area_ratio}. For this Figure, the nomenclature used is \textit{\{tail\}\{head\}\{number of lipids\}-\{temperature treatment\}}. For example, pe64-T is a 64 DPPE (pe) bilayer thermalised (T) after construction of the system. The different temperature treatments used are thermalisation (T), annealing (A), quenching (Q) and gentle cooling (C) - AC indicates that the system was obtained first by annealing followed by a quenching.

\begin{figure}[ht!]
\centering
\includegraphics[width=.9\linewidth]{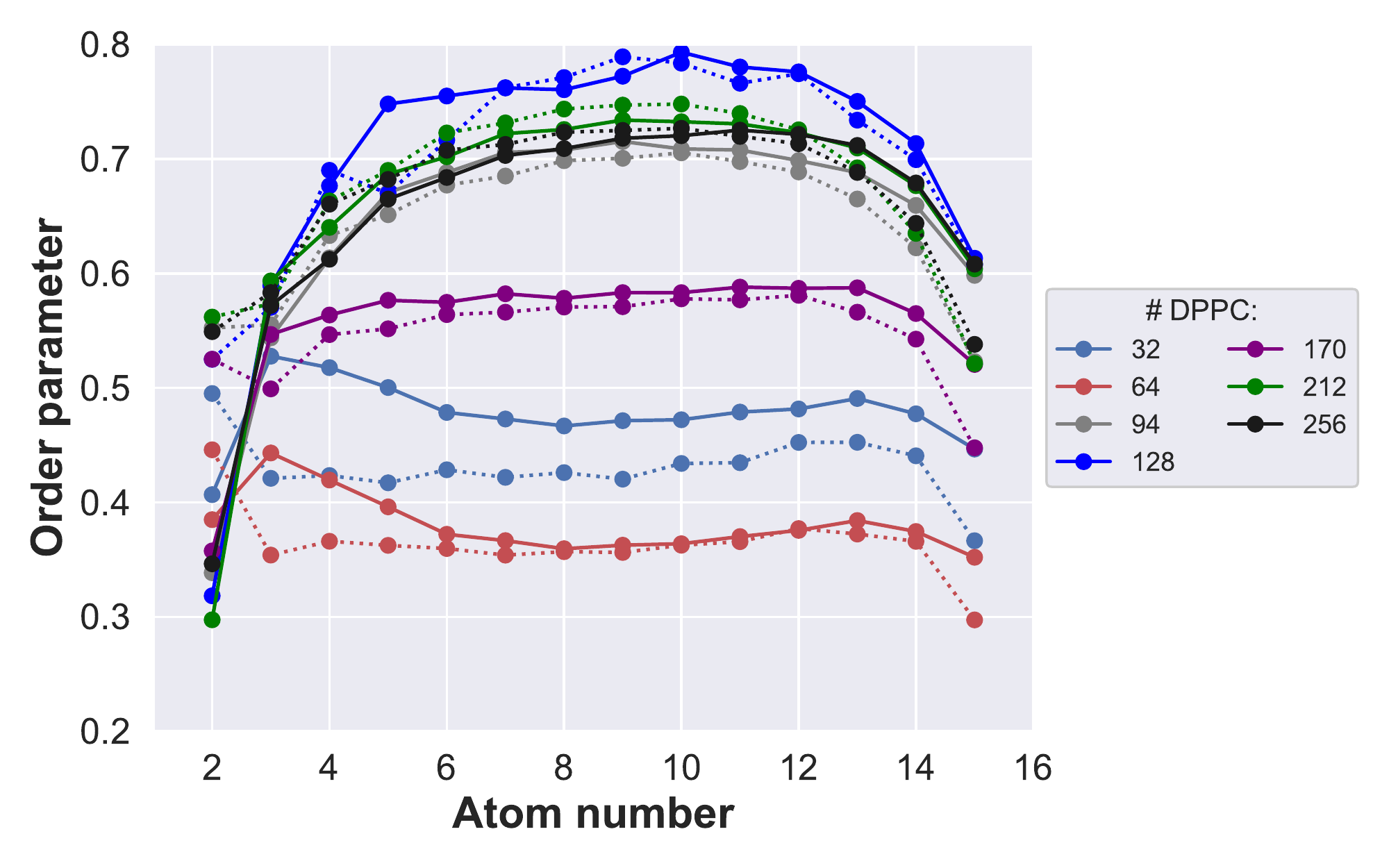}
\caption{Tail order parameters of the carbon atoms from the DPPC tails measured for different systems made of increasing amount of lipids per bilayers. Plain and dashed lines are respectively the tails sn1 and sn2 of the lipids. While generally the system visually classified as disordered have a very order parameter, occasionally some systems were found with an order parameter close to the one of the tilted gel phase - here 170 DPPC.}
\label{si_fig_order_smol}
\end{figure}

\begin{figure}[ht!]
\centering
\includegraphics[width=.9\linewidth]{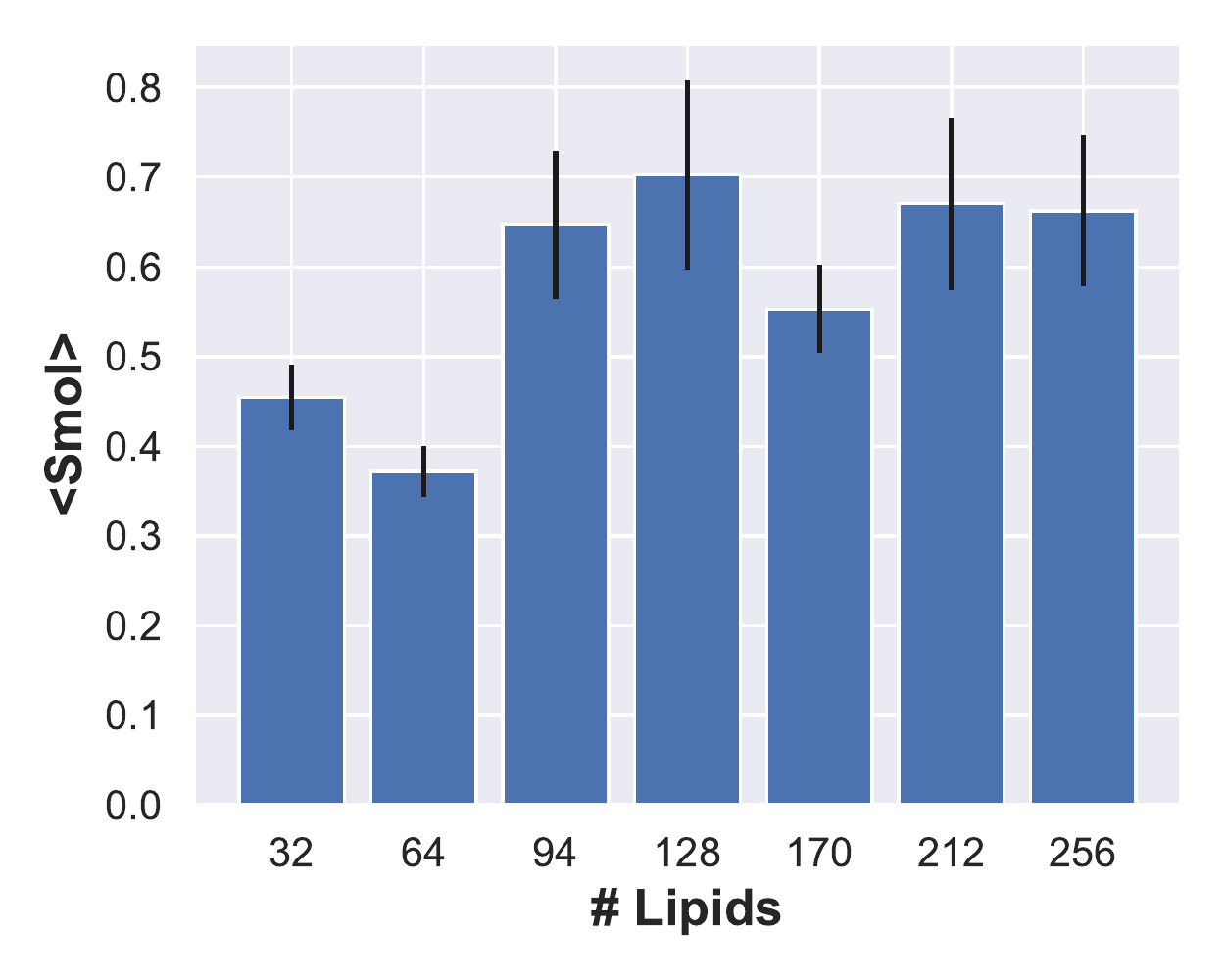}
\caption{Mean order parameters of the carbon atoms from both DPPC tails measured in Figure \ref{si_fig_order_smol} for different systems made of increasing amount of lipids per bilayers. Error bars are the standard deviation over all the atom of both chains.}
\label{si_fig_average_smol}
\end{figure}

The efficiency of the relative increase of area per lipid $\darea$ to be used as an order parameter distinguishing between the tilted and disordered phase has been compared to the usual order parameters of the carbon atom $n$ of the tail $k$ of each lipid, noted $S_{\mathrm{mol}}(k,n)$, used to discriminate between the gel and fluid phases. The parameter $S_{\mathrm{mol}}(k,n)$ was calculated using the following formula:

\begin{equation}
\label{eq_order_parameter}
    S_{\mathrm{mol}}(k,n) = \left< \frac{3 \cos^2 \theta_{k,n} - 1}{2} \right>,
\end{equation}
with $\theta_{k,n}$ the angle formed between the vector generated by the atoms $n-1$ and $n+1$ along the tail $k$ chain and the vector orthogonal to the plane of the bilayer, here Z. The results are illustrated in Figure~\ref{si_fig_order_smol}. While the tail order parameters could discriminate most of the disordered systems (94, 128, 212 and 256) from the tilted ones (32, 64), the system made of 170 DPPC molecules could not be undoubtedly told apart from neither the tilted nor the disordered phases. Same results was observed by measuring the average tail order parameter $\langle S_{\mathrm{mol}}\rangle$ of all atoms of both tails (cf. Figure~\ref{si_fig_average_smol}). In contrast, the Figure~\ref{si_fig_area_ratio} shows that the relative increase of area $\darea$ could easily classify the same system as a system in the disordered phase.

\begin{figure}[ht!]
\centering
\begin{tabular}[t]{ccc}

\begin{subfigure}{0.31\textwidth}
\centering
\smallskip
\includegraphics[width=\linewidth]{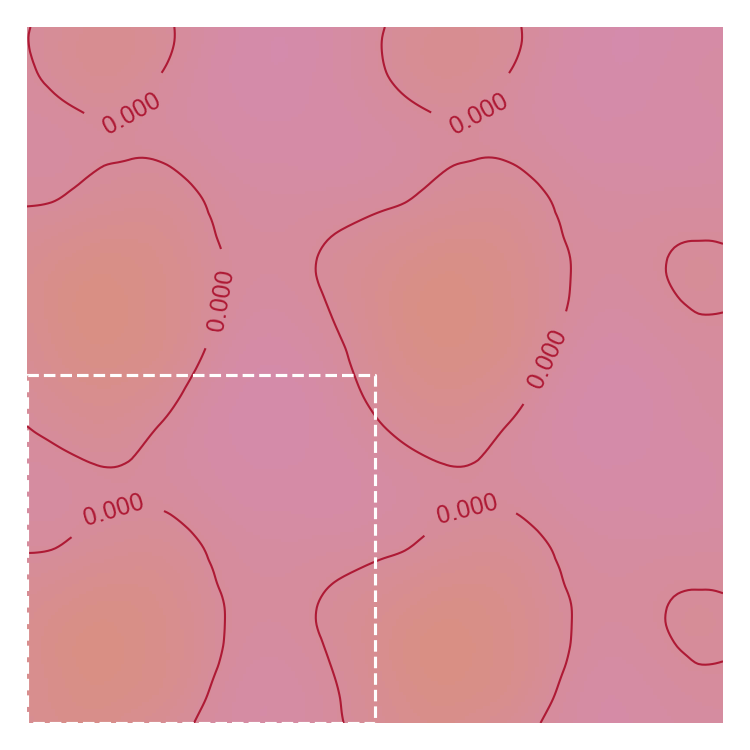}
\caption{32 Lipids}
\end{subfigure}
&
\begin{subfigure}{0.31\textwidth}
\centering
\smallskip
\includegraphics[width=\linewidth]{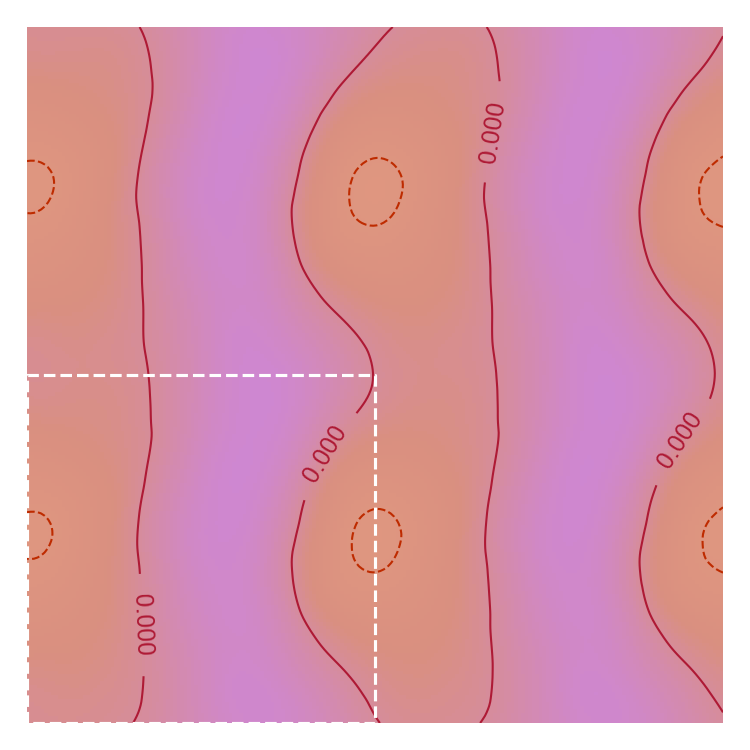}
\caption{64 Lipids}
\end{subfigure}
&
\begin{subfigure}{0.31\textwidth}
\centering
\smallskip
\includegraphics[width=\linewidth]{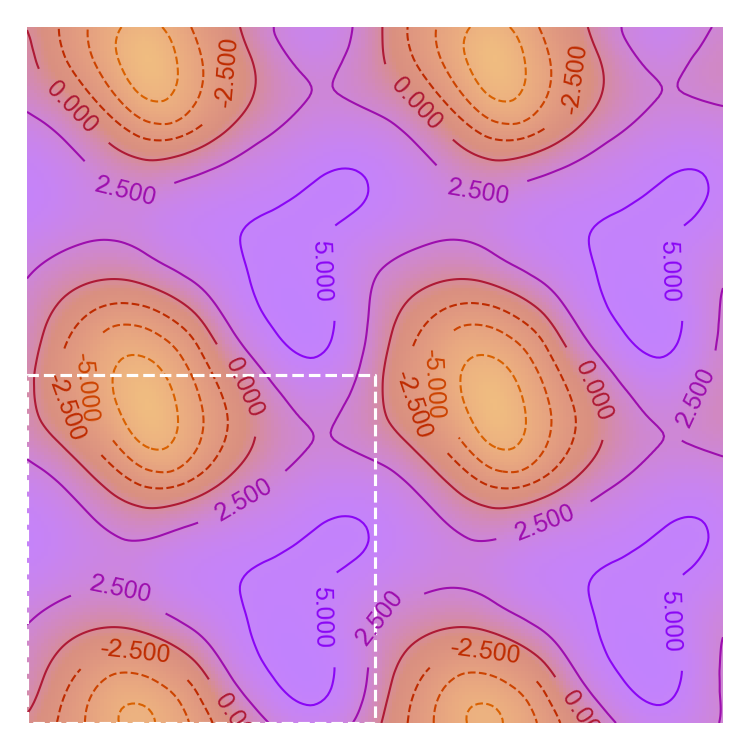}
\caption{94 Lipids}
\end{subfigure}
\\
\begin{subfigure}{0.31\textwidth}
\centering
\smallskip
\includegraphics[width=\linewidth]{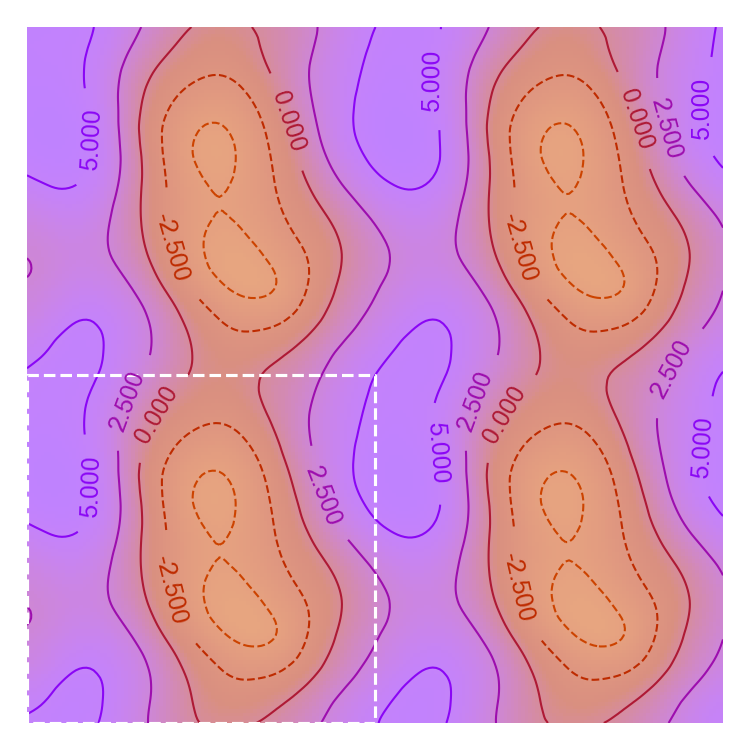}
\caption{128 Lipids}
\end{subfigure}
&
\begin{subfigure}{0.31\textwidth}
\centering
\smallskip
\includegraphics[width=\linewidth]{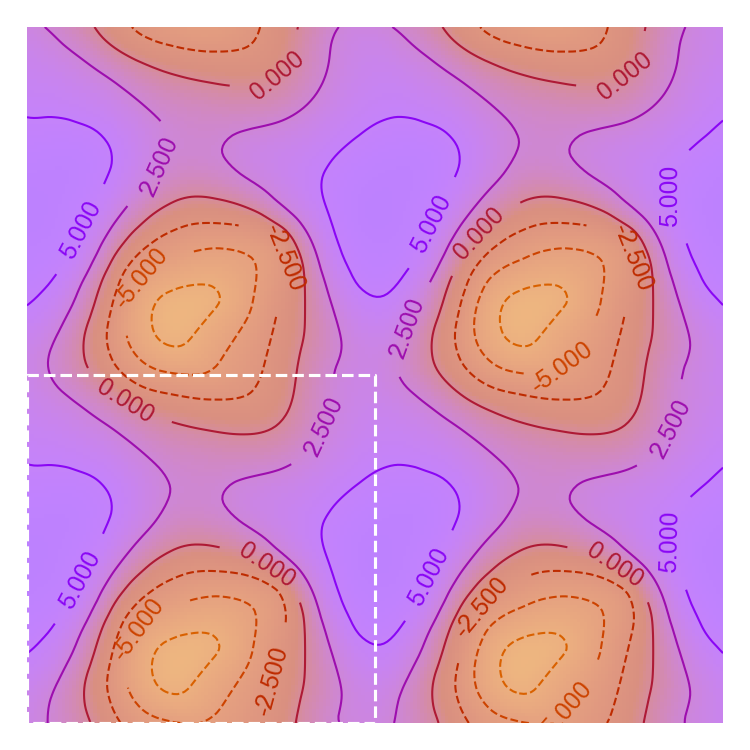}
\caption{170 Lipids}
\end{subfigure}
&
\begin{subfigure}{0.31\textwidth}
\centering
\smallskip
\includegraphics[width=\linewidth]{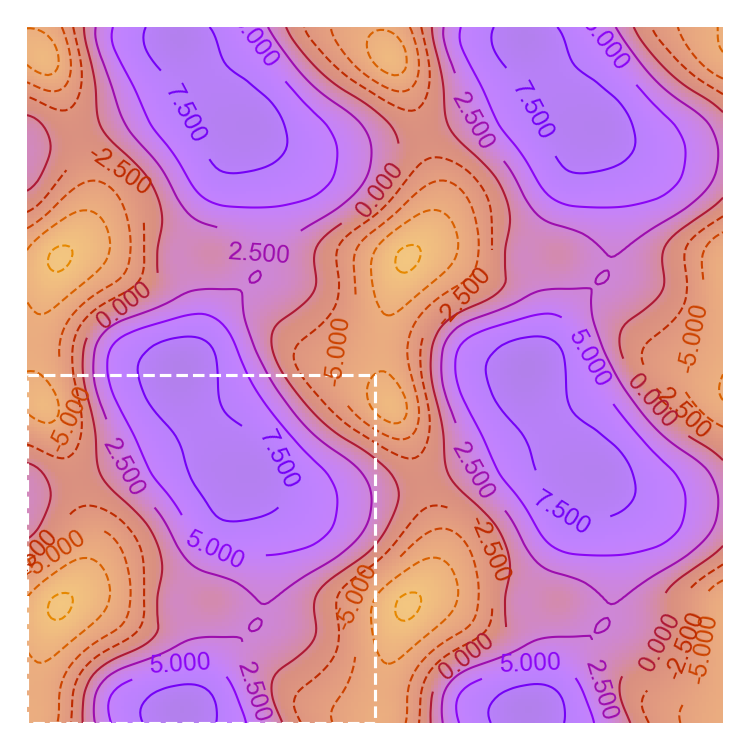}
\caption{212 Lipids}
\end{subfigure}
\\
\begin{subfigure}{0.31\textwidth}
\centering
\smallskip
\includegraphics[width=\linewidth]{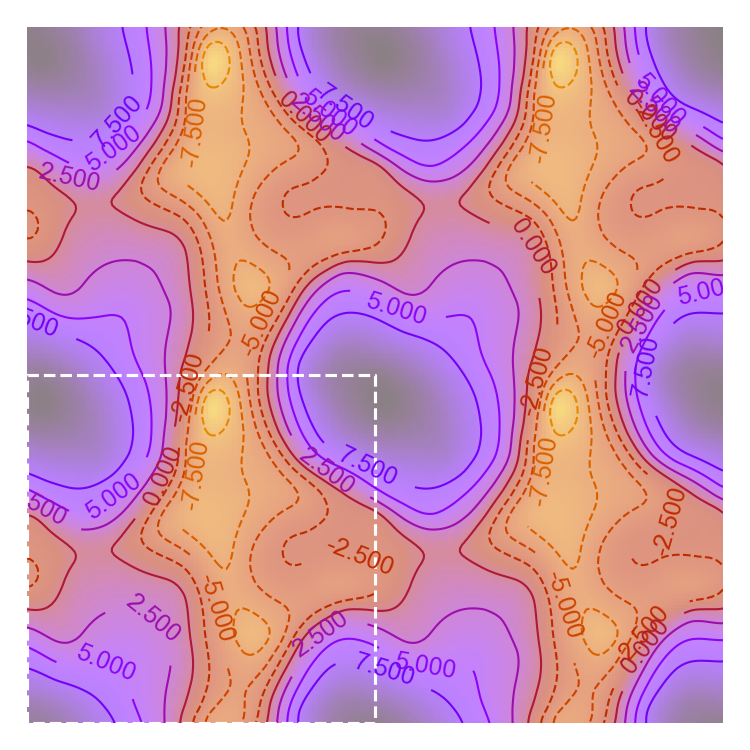}
\caption{256 Lipids}
\end{subfigure}
&
\begin{subfigure}{0.31\textwidth}
\centering
\smallskip
\includegraphics[width=\linewidth]{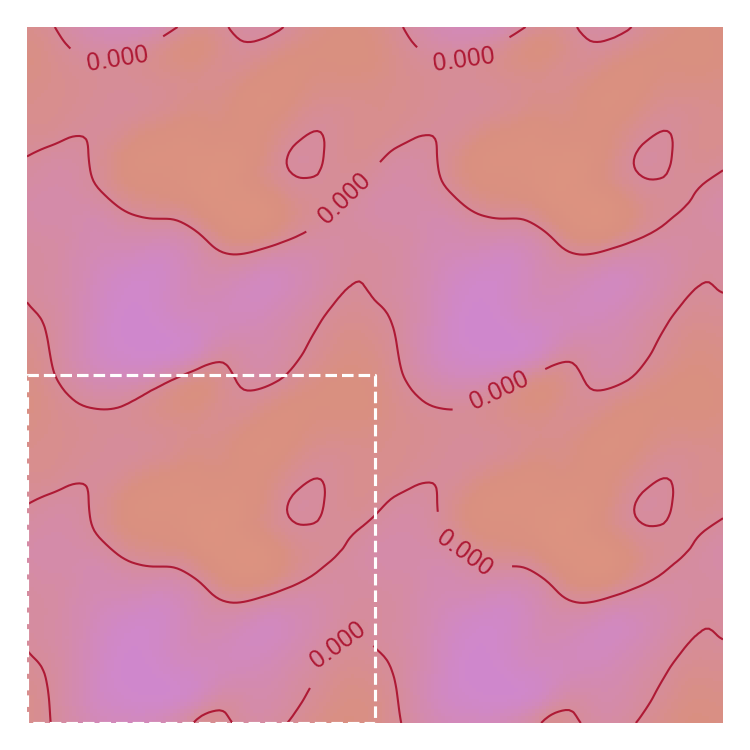}
\caption{256 Lipids (replicated)}
\end{subfigure}
&
\begin{subfigure}{0.31\textwidth}
\centering
\smallskip
\includegraphics[width=\linewidth]{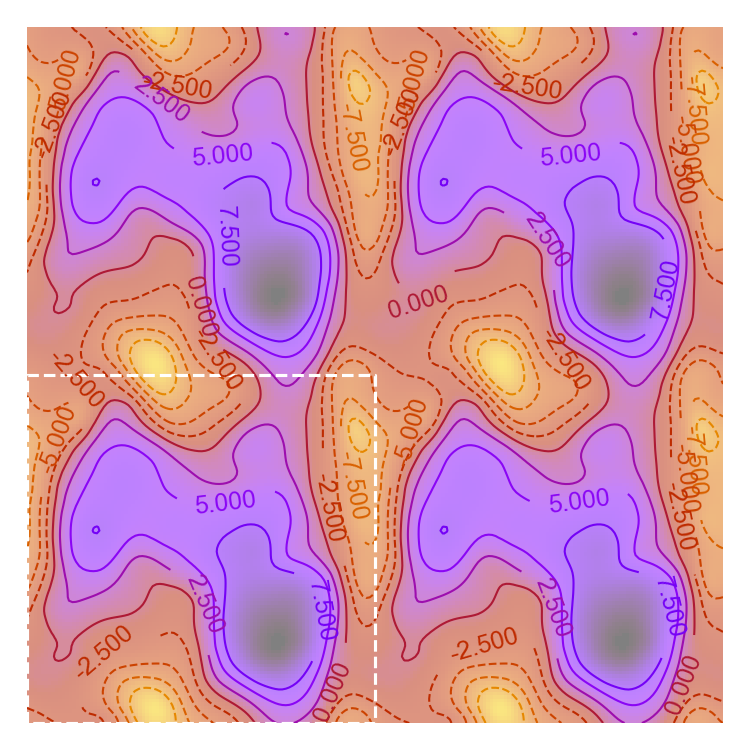}
\caption{512 Lipids}
\end{subfigure}

\end{tabular}

\caption{Top leaflets.}
\label{si_fig_view_top}
\end{figure}

\newpage

\newpage

\newpage

\newpage
\begin{figure}[ht!]
\centering
\begin{tabular}[t]{cc}

\begin{subfigure}{0.45\textwidth}
\centering
\smallskip
\includegraphics[width=\linewidth]{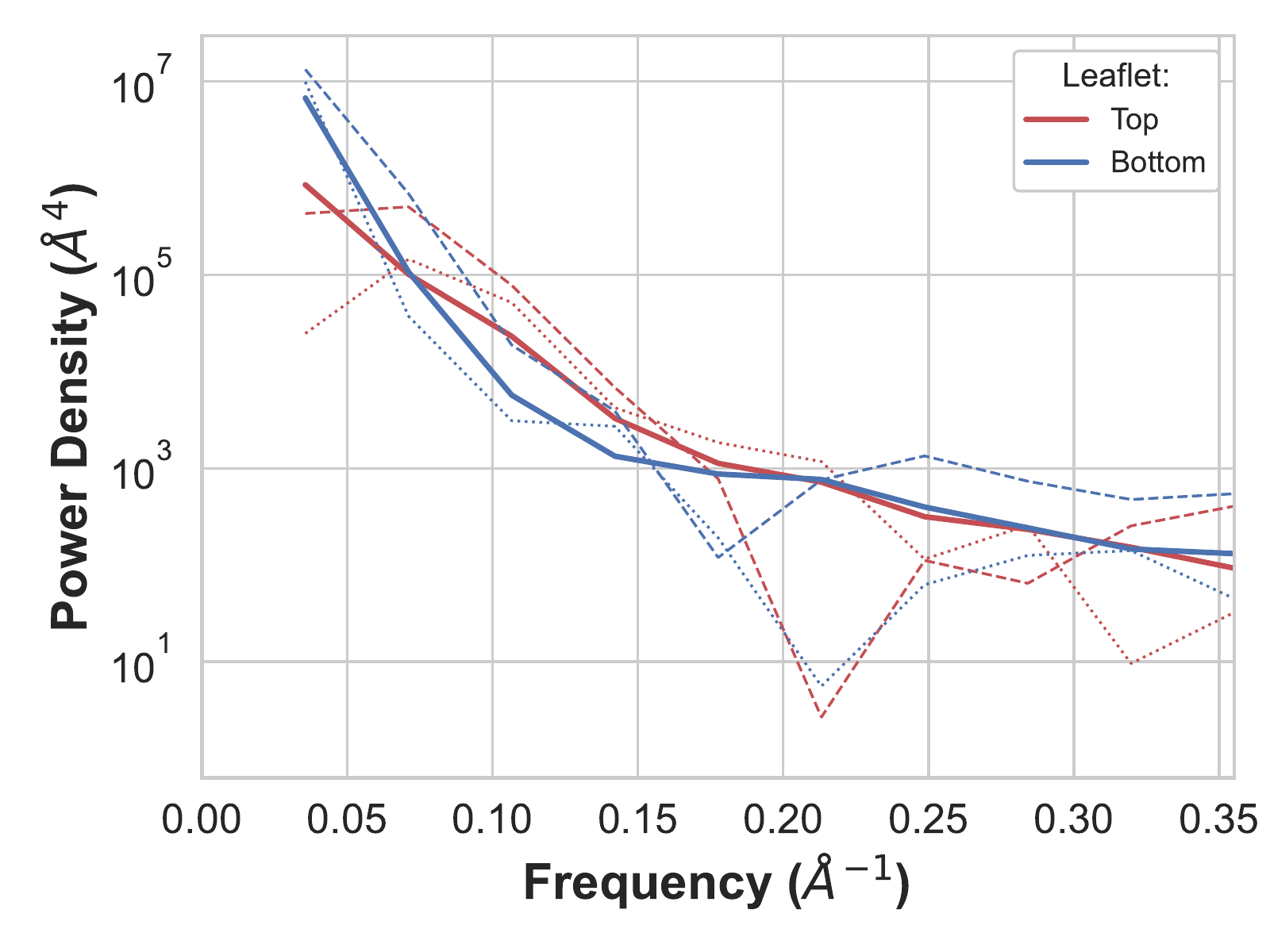}
\caption{32 Lipids}
\end{subfigure}
&
\begin{subfigure}{0.45\textwidth}
\centering
\smallskip
\includegraphics[width=\linewidth]{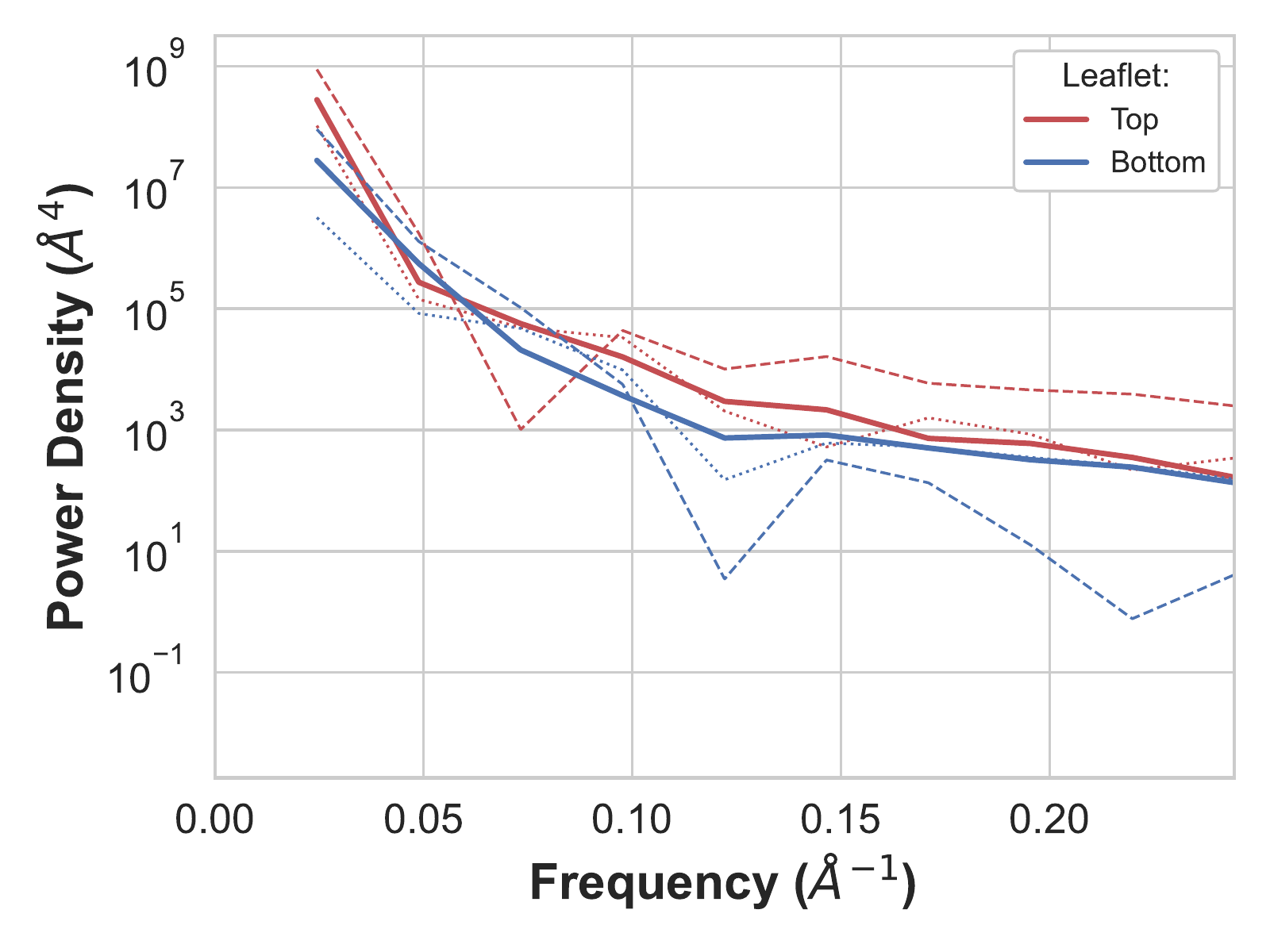}
\caption{64 Lipids}
\end{subfigure}
\\
\begin{subfigure}{0.45\textwidth}
\centering
\smallskip
\includegraphics[width=\linewidth]{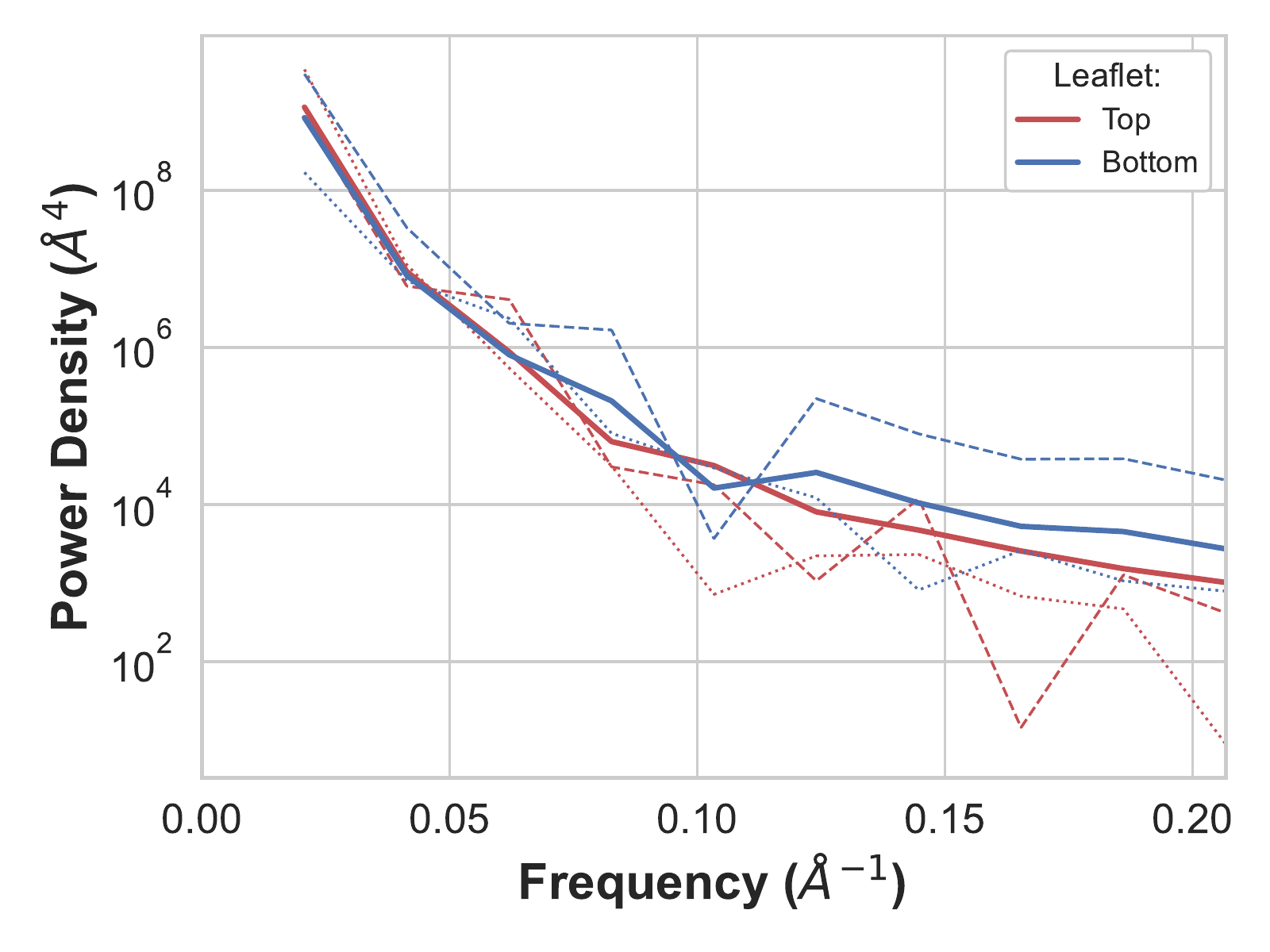}
\caption{94 Lipids}
\end{subfigure}
&
\begin{subfigure}{0.45\textwidth}
\centering
\smallskip
\includegraphics[width=\linewidth]{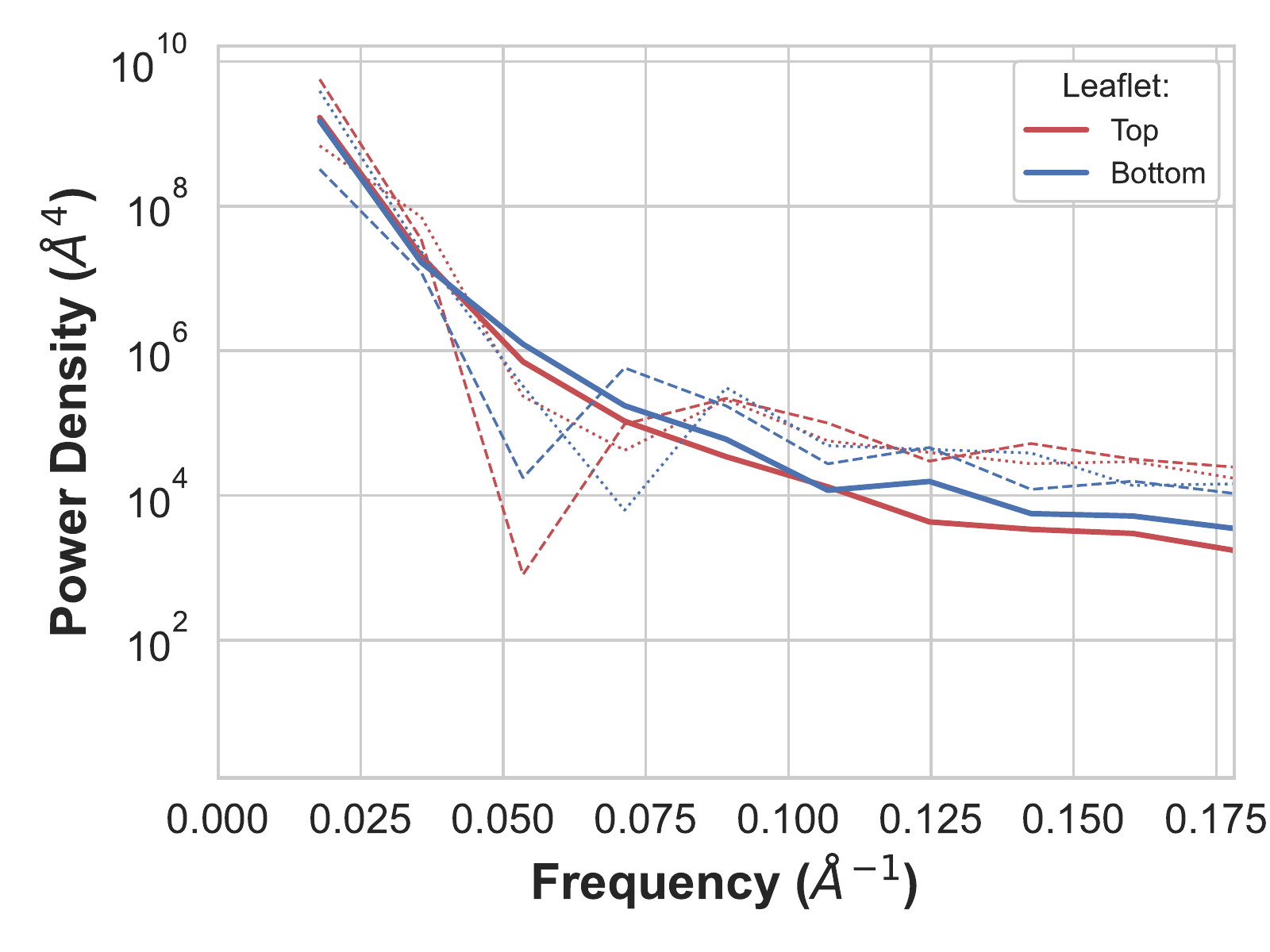}
\caption{128 Lipids}
\end{subfigure}

\end{tabular}

\caption{Power spectrum densities of the topography used to calculate the period of the corrugations. The frequency of the mean background value (0~\AA$^{-1}$) has been removed from the spectrum to improve the clarity. Plain lines are the spectrum along the X axis, while dashed lines are along the Y axis. Plain black lines are the average PSD of the membrane.}
\label{si_fig_psd}
\end{figure}

\newpage
\begin{figure}[ht!]
\centering
\begin{tabular}[t]{cc}

\begin{subfigure}{0.45\textwidth}
\centering
\smallskip
\includegraphics[width=\linewidth]{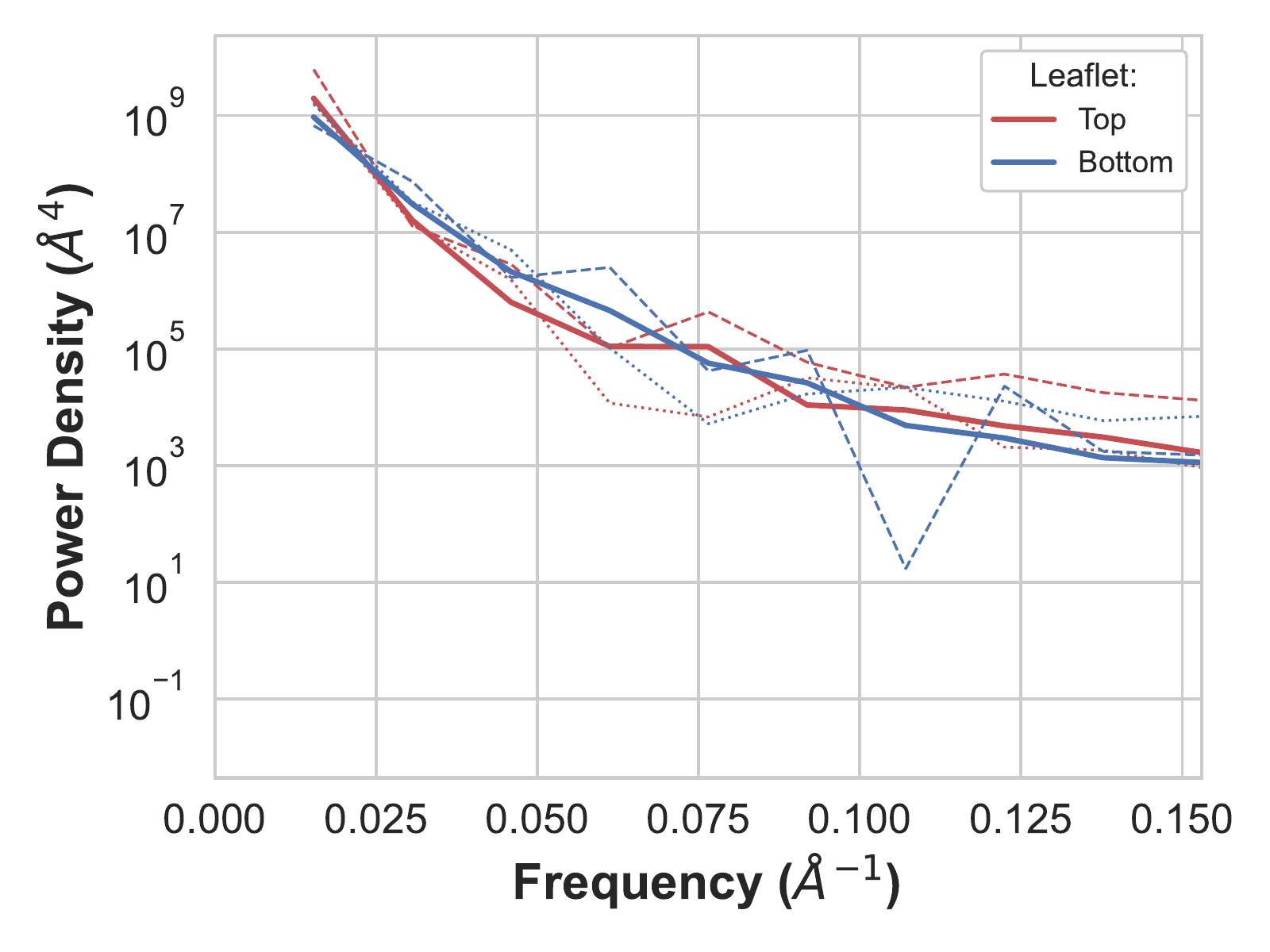}
\caption{170 Lipids}
\end{subfigure}
&
\begin{subfigure}{0.45\textwidth}
\centering
\smallskip
\includegraphics[width=\linewidth]{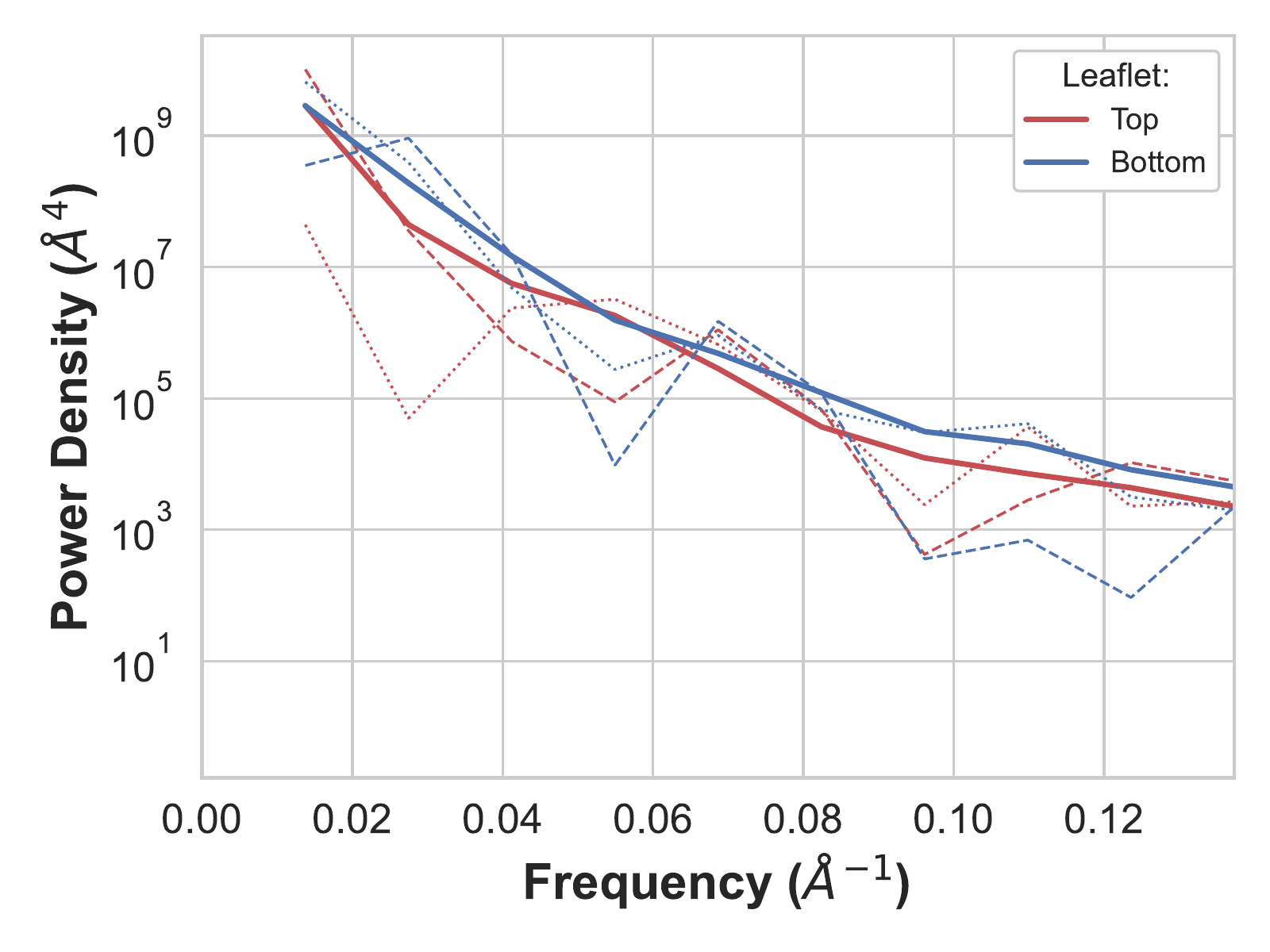}
\caption{212 Lipids}
\end{subfigure}
\\
\begin{subfigure}{0.45\textwidth}
\centering
\smallskip
\includegraphics[width=\linewidth]{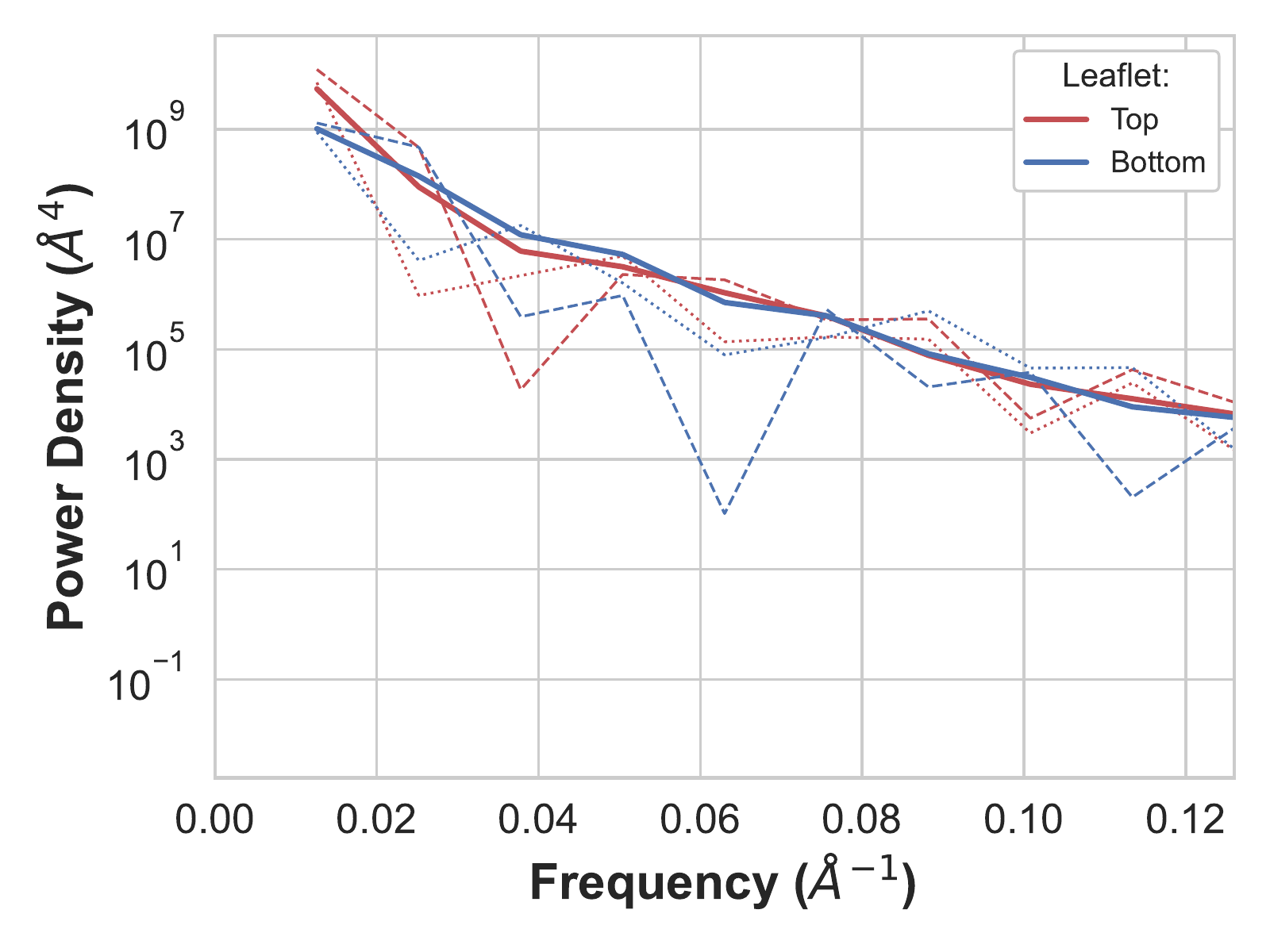}
\caption{256 Lipids}
\end{subfigure}
&
\begin{subfigure}{0.45\textwidth}
\centering
\smallskip
\includegraphics[width=\linewidth]{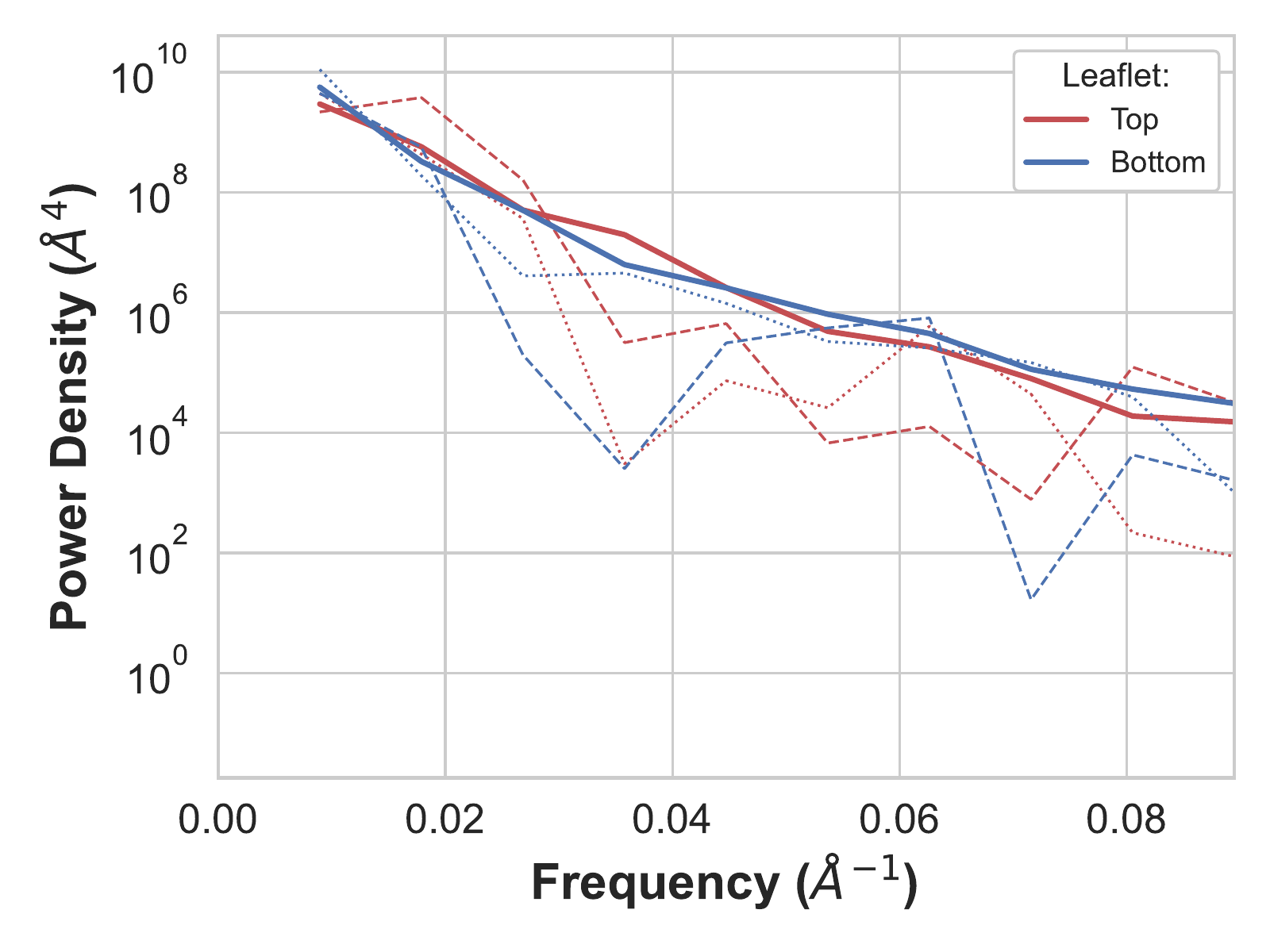}
\caption{512 Lipids}
\end{subfigure}

\end{tabular}

\caption{(Continuation of Figure~\ref{si_fig_psd}).}
\label{si_fig_psd2}
\end{figure}

\newpage

\newpage
\begin{table}[!ht]
\centering
\caption{
{\bf All mean values of area per lipid measured in this work on the systems simulated at 288~K, either by projection of the area in the XY plane of the system box ($A_H^p$) or by meshing of the surface of the bilayer using Ovito ($A_H^m$). Error bars are the standard deviation of the measured values.}
}
\resizebox*{!}{.8\textheight}{
\rotatebox{90}{
\begin{tabular}{c c||c|c||c|c||c|c}
& & \multicolumn{2}{c||}{\bf Thermalised (\textit{T})} & \multicolumn{2}{c||}{\bf Cooling (\textit{AC})} & \multicolumn{2}{c}{\bf Quenching (\textit{AQ})} \\
{\bf Type} & {\bf System} & {\bf A\textsubscript{H}\textsuperscript{p}}  & {\bf A\textsubscript{H}\textsuperscript{m}} & {\bf A\textsubscript{H}\textsuperscript{p}} & {\bf A\textsubscript{H}\textsuperscript{m}} & {\bf A\textsubscript{H}\textsuperscript{p}} & {\bf A\textsubscript{H}\textsuperscript{m}} \\ \hline
\textbf{DPPC} & 32 Lipids & 49.9 $\pm$ 0.9 & 52.6 $\pm$ 1.7 & - & - & - & - \\
(\textit{pc}) & 64 Lipids & 52.3 $\pm$ 0.7 & 54.4 $\pm$ 0.7 & 50.1 $\pm$ 0.7 & 52.0 $\pm$ 0.7 & 50.8 $\pm$ 0.8 & 57.8 $\pm$ 1.5  \\
& 94 Lipids & 49.8 $\pm$ 0.6 & 57.1 $\pm$ 0.7 & - & - & - & - \\
& 128 Lipids & 50.4 $\pm$ 0.8 & 57.5 $\pm$ 1.0 & - & - & - & - \\
& 170 Lipids & 50.0 $\pm$ 0.3 & 55.5 $\pm$ 0.5 & - & - & - & - \\
& 212 Lipids & 50.9 $\pm$ 0.6 & 58.6 $\pm$ 0.6 & - & - & - & - \\
& 256 Lipids & 50.2 $\pm$ 0.7 & 56.9 $\pm$ 0.6 & 50.6 $\pm$ 0.3 & 56.2 $\pm$ 0.4 & 51.4 $\pm$ 0.7 & 59.5 $\pm$ 0.6\\
& 256 Lipids\textsuperscript{a} & 51.3 $\pm$ 0.5 & 52.1 $\pm$ 0.5 & 51.7 $\pm$ 0.3 & 58.5 $\pm$ 0.4 & 52.2 $\pm$ 0.4 & 59.0 $\pm$ 0.6 \\
& 512 Lipids & 49.0 $\pm$ 0.4 & 56.7 $\pm$ 0.4 & - & - & - & - \\\hline
\textbf{DSPC} & 64 Lipids & 48.6 $\pm$ 0.4 & 50.9 $\pm$ 0.5 & - & - & - & - \\
(\textit{sc}) & 212 Lipids & 41.3 $\pm$ 0.6 & 46.3 $\pm$ 0.6 & - & - & - & - \\ \hline
\textbf{DPPE} & 64 Lipids & 45.9 $\pm$ 0.5 & 48.5 $\pm$ 0.6 & - & - & - & - \\
(\textit{pe}) & 212 Lipids & 45.4 $\pm$ 0.3 & 47.5 $\pm$ 0.3 & - & - & - & - \\
\end{tabular}
}}
\begin{flushleft} \textsuperscript{a} System constructed by replication of the system made of 64 lipids.
\end{flushleft}
\label{si_table_area}
\end{table}

\newpage

\newpage

\newpage

\newpage
\section{Influence of the thermal history}

\begin{figure}[ht!]
\centering
\begin{tabular}[t]{cc}

\begin{subfigure}{0.7\textwidth}
\centering
\smallskip
\includegraphics[width=\linewidth]{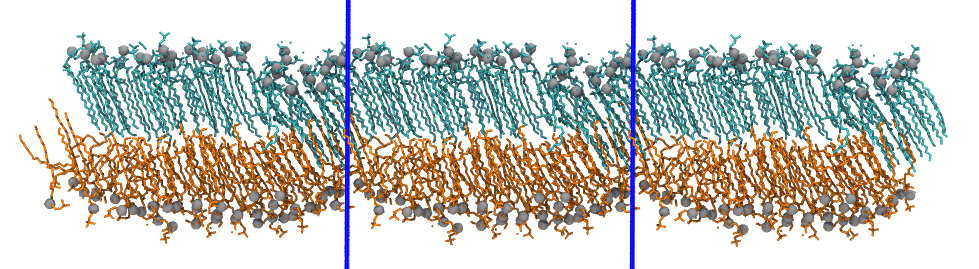}
\caption{Gentle cooling}
\end{subfigure}
\\
\begin{subfigure}{0.7\textwidth}
\centering
\smallskip
\includegraphics[width=\linewidth]{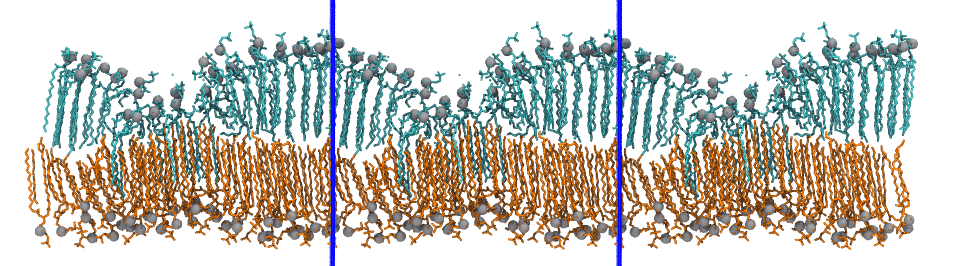}
\caption{Quenching}
\end{subfigure}
\end{tabular}

\caption{Screenshots of the bilayers with 256 DPPC molecules constructed using CHARMM-GUI and simulated at 288~K, followed by an annealing over the gel-fluid melting temperature before cooled down to 288~K, either with a (a) gentle cooling or a (b) quenching.}
\label{si_fig_snapshot_non_replicated}
\end{figure}

\begin{figure}[ht!]
\centering
\begin{tabular}[t]{cc}

\begin{subfigure}{0.35\textwidth}
\centering
\smallskip
\includegraphics[width=\linewidth]{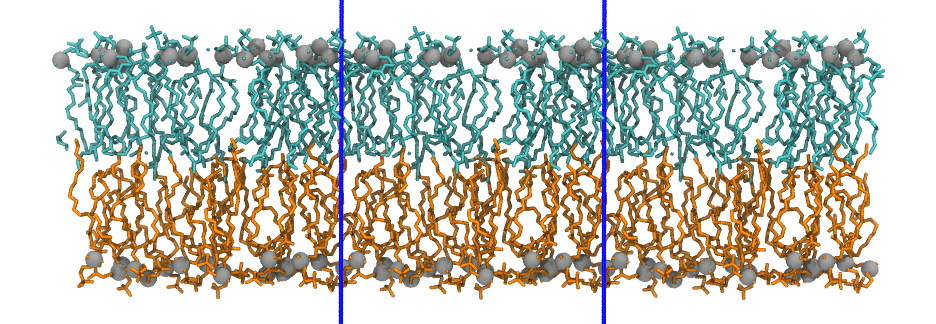}
\caption{DPPC 64 - CHARMM-GUI}
\end{subfigure}
    &
\begin{subfigure}{0.35\textwidth}
\centering
\smallskip
\includegraphics[width=\linewidth]{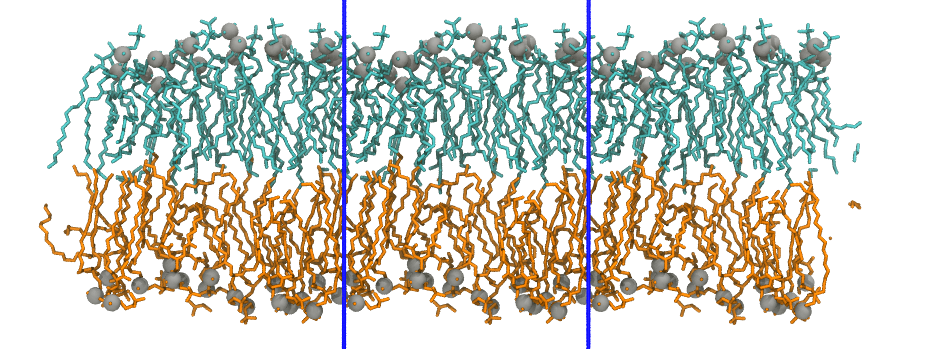}
\caption{DPPC 64 - Equilibration}
\end{subfigure}
\\
\multicolumn{2}{c}{
\begin{subfigure}{0.7\textwidth}
\centering
\smallskip
\includegraphics[width=\linewidth]{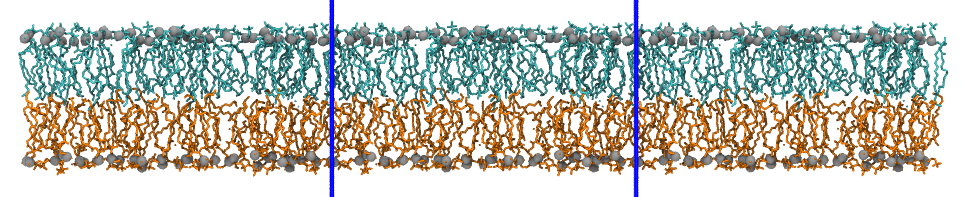}
\caption{DPPC 256 - CHARMM-GUI}
\end{subfigure}
}
\\
\multicolumn{2}{c}{
\begin{subfigure}{0.7\textwidth}
\centering
\smallskip
\includegraphics[width=\linewidth]{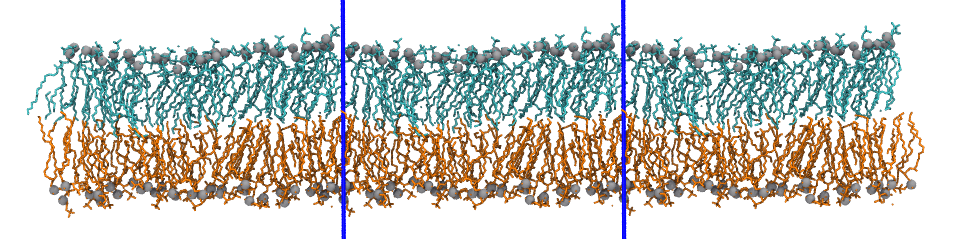}
\caption{DPPC 256 - Equilibration}
\end{subfigure}
}
\\
\multicolumn{2}{c}{
\begin{subfigure}{0.7\textwidth}
\centering
\smallskip
\includegraphics[width=\linewidth]{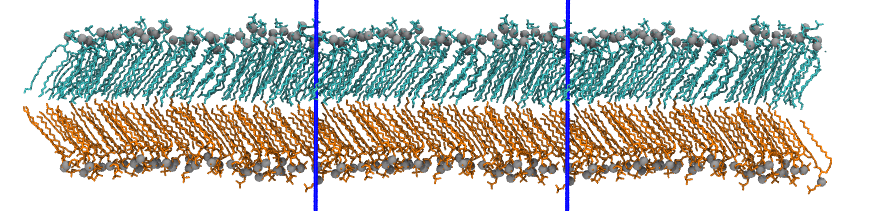}
\caption{DPPC 256 - Replication}
\end{subfigure}
}
\\
\multicolumn{2}{c}{
\begin{subfigure}{0.7\textwidth}
\centering
\smallskip
\includegraphics[width=\linewidth]{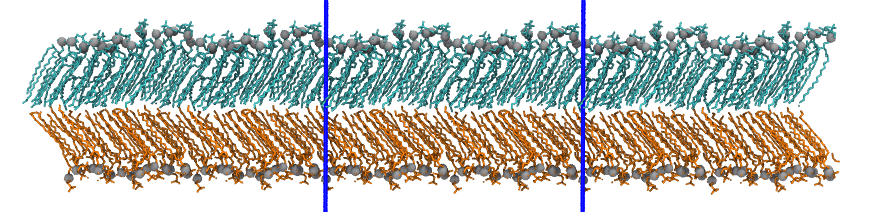}
\caption{DPPC 256 - Equilibration after replication}
\end{subfigure}
}
\end{tabular}

\caption{Screenshots of the DPPC bilayers constructed and used in the experiments on temperature treatments, either using CHARMM-GUI or by replication of a 64 DPPC bilayer simulated at 288~K. All systems are shown before and after a NVT and NPT short equilibration at 288~K, but always before the 50~ns thermalisation at 288~K.}
\label{si_fig_snapshot_construction}
\end{figure}

\begin{figure}[ht!]
\centering
\begin{tabular}[t]{cc}

\begin{subfigure}{0.35\textwidth}
\centering
\smallskip
\includegraphics[width=\linewidth]{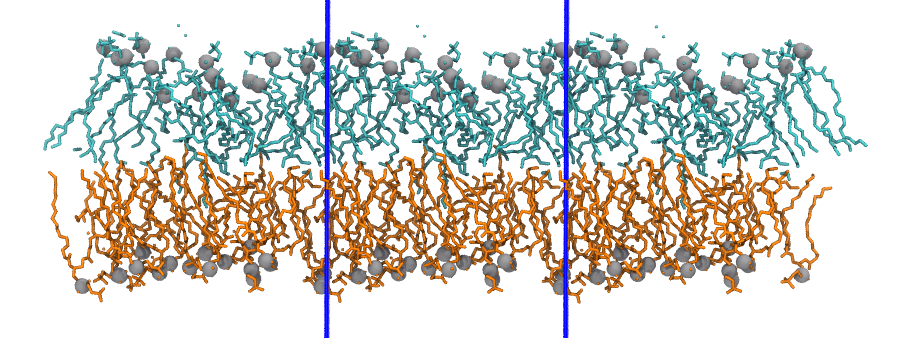}
\caption{DPPC 64}
\end{subfigure}
\\
\begin{subfigure}{0.7\textwidth}
\centering
\smallskip
\includegraphics[width=\linewidth]{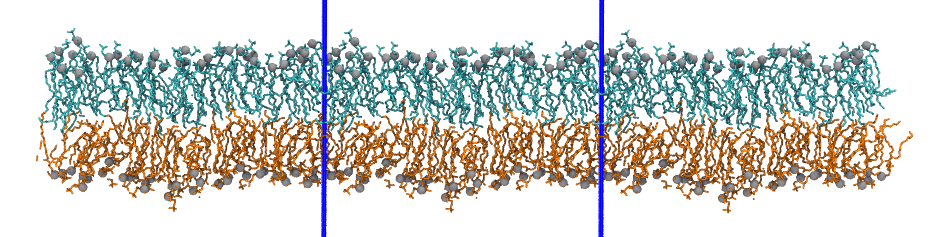}
\caption{DPPC 256}
\end{subfigure}
\\
\begin{subfigure}{0.7\textwidth}
\centering
\smallskip
\includegraphics[width=\linewidth]{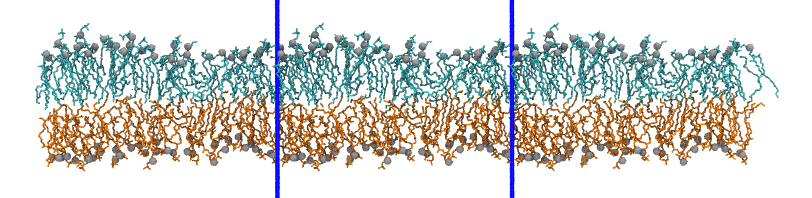}
\caption{DPPC 256 - Replication}
\end{subfigure}
\end{tabular}

\caption{Screenshots of the DPPC bilayers used in the experiments on temperature treatments after annealing at 358~K. In this work, all these systems were classified as being in the fluid phase.}
\label{si_fig_snapshot_fluid}
\end{figure}

In this section are presented the screenshots of the systems that were not used to prepare the figures in the \textit{Influence of the thermal history} section of the main text: the systems made of 256 DPPC molecules but obtained via CHARMM-GUI instead of replication of the small systems (Figure~\ref{si_fig_snapshot_non_replicated}), all the systems before the thermalisation (Figure~\ref{si_fig_snapshot_construction}) or in the fluid phase after annealing (Figure~\ref{si_fig_snapshot_fluid}). 

\begin{figure}[ht!]
\centering
\includegraphics[width=.9\linewidth]{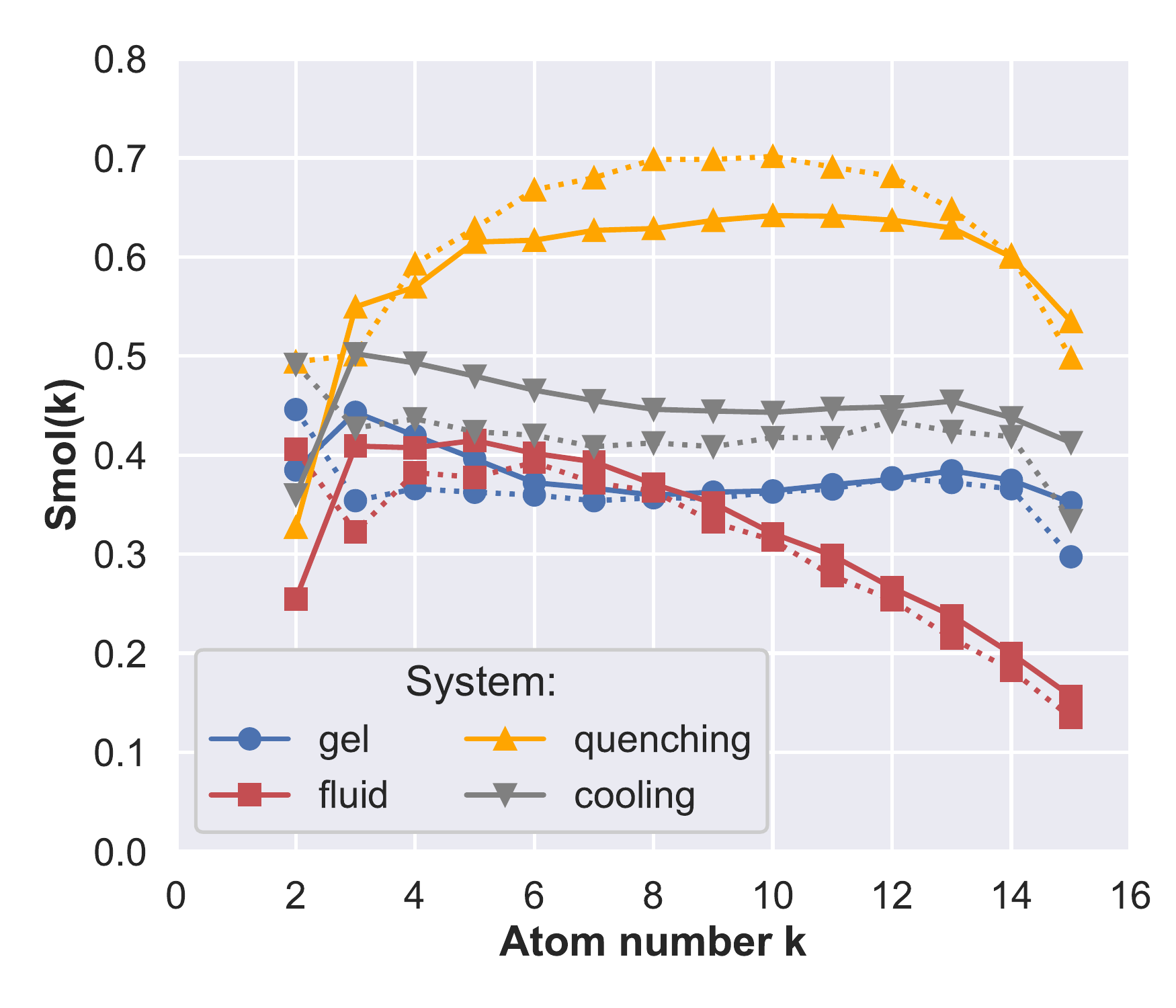}
\caption{Average order parameters of the carbon atoms from the DPPC tails measured in four 64 lipids bilayer systems: the initial system at 288~K in the gel phase (blue), the same system in the fluid phase after annealing at 358~K (red), and the result of both gentle cooling (gray) or quenching (orange) of the annealed system back to 288~K. While the system gently cooled goes back the configuration of the initial system in the gel phase, the quenched system reaches another phase: the disordered gel phase. Plain and dashed lines correspond respectively to the sn1 and sn2 tails of the lipids.}
\label{si_fig_temperature_smol}
\end{figure}

The increase of area per lipid $\darea$ of all the systems obtained during the different temperature treatments were gathered in Figure \ref{si_fig_area_ratio}. Similarly to the previous section, we measured the tail order parameters $S_{\mathrm{mol}}(k,n)$ for the different systems made of 64 DPPC molecules but obtained via different temperature treatment (Figure~\ref{si_fig_temperature_smol}). The tail order parameters confirmed that the gently cooled system was in a state similar to the one before annealing, noted gel here, but drastically different from the one obtained after quenching. The latter was found different from both gel and fluid phases, which highlight its nature of being of a new phase.

\newpage

\newpage

\newpage

\newpage
\section{Thermodynamics of tilted and disordered states}

In this section, we present the detailed results obtained and gathered to write the \textit{Thermodynamics of tilted and disordered states} section of the main text. 

\begin{figure}[ht!]
\centering
\begin{tabular}[t]{cc}

\begin{subfigure}{0.6\textwidth}
\centering
\smallskip
\includegraphics[width=\linewidth]{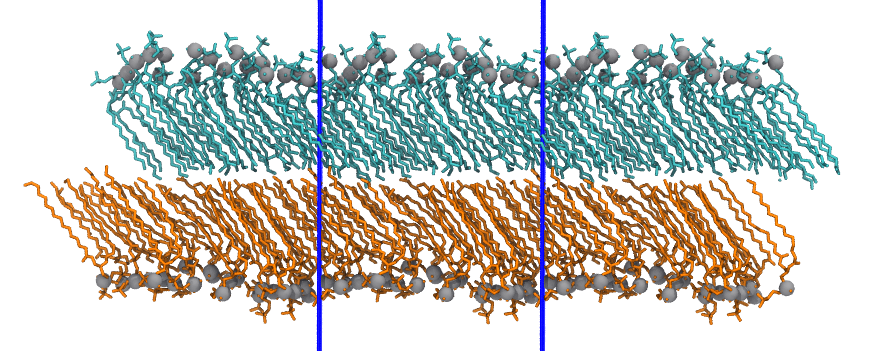}
\caption{From gently cooled system}
\end{subfigure}
\\
\begin{subfigure}{0.6\textwidth}
\centering
\smallskip
\includegraphics[width=\linewidth]{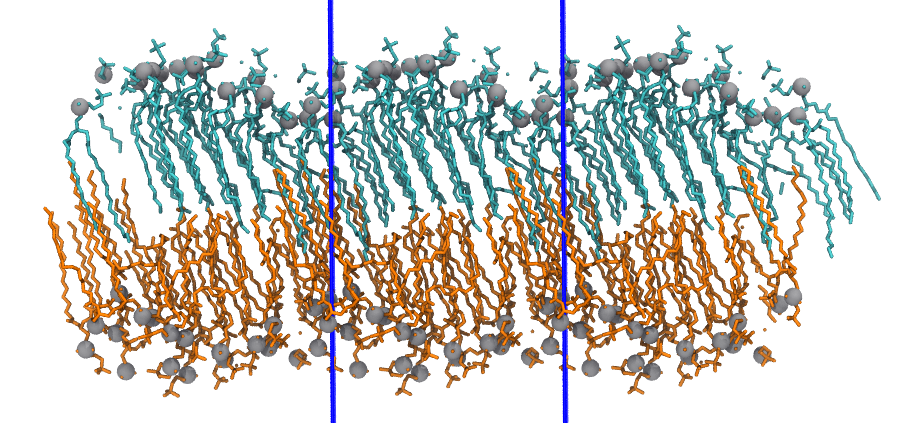}
\caption{From quenched system}
\end{subfigure}
\end{tabular}

\caption{Screenshots of the DPPC 64 bilayers obtained after energy minimisation using a conjugate gradient algorithm (a) from a gently cooling and found initially in a tilted gel phase, or (b) from a quenching and found initially in a disordered gel phase. Despite the energy minimisation, both systems remain in their initial phase.}
\label{si_fig_snapshot_is}   
\end{figure}

\begin{figure}[ht!]
\centering
\includegraphics[width=.9\linewidth]{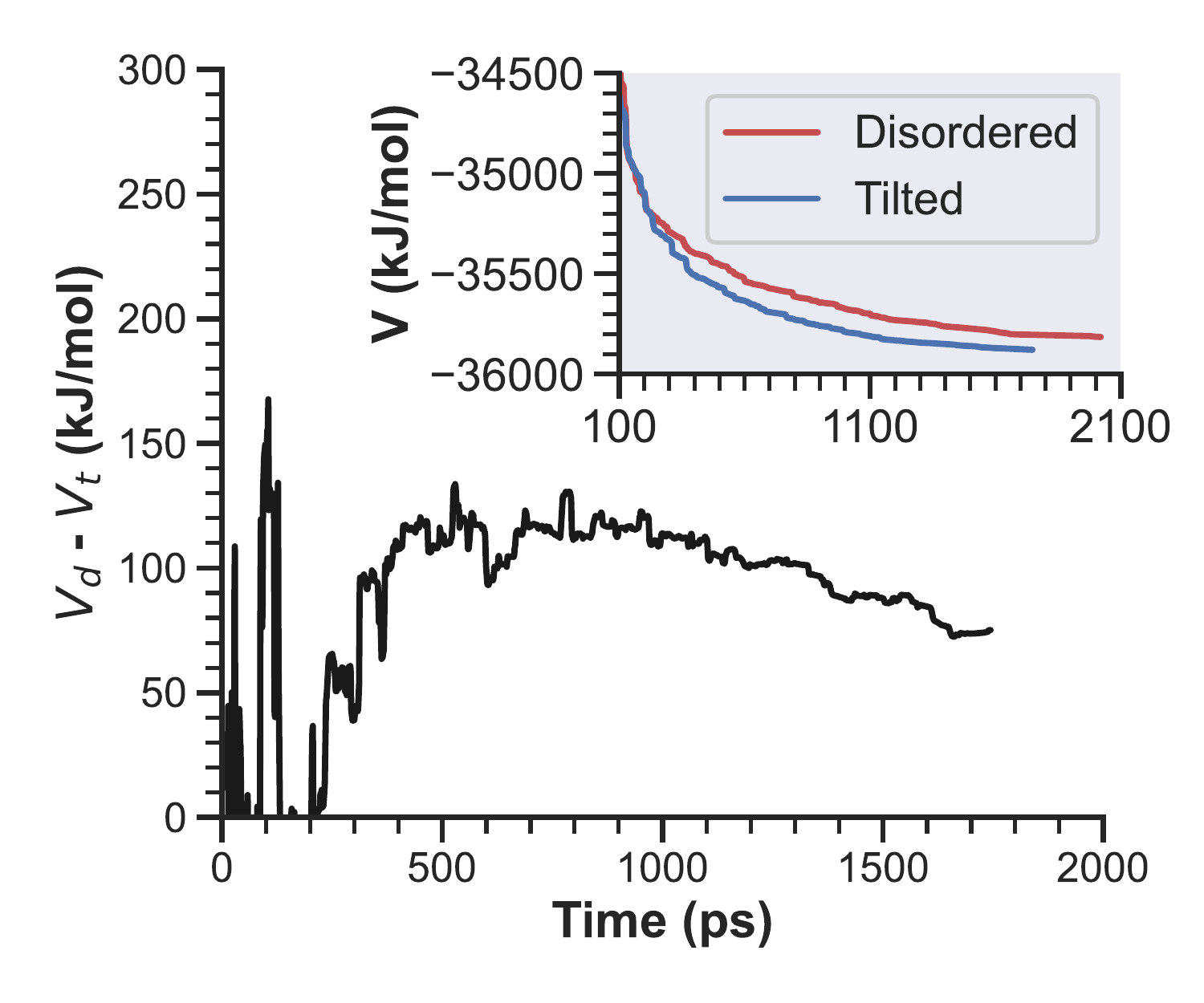}
\caption{Evolution of potential energies of the disordered system $V_d$ and of the tilted system $V_t$ during an energy minimisation simulation using a conjugate gradient integrator. The inset shows the evolution of the difference in potential energy between both system over time.}
\label{si_fig_potential_energy}
\end{figure}

As mentioned in the main text, we compute the energy minimization using the conjugate gradient method implemented in Gromacs. Starting from tilted and disordered configurations, we monitor the energy evolution with time as shown in Figure~\ref{si_fig_potential_energy} and we find that $V_d-V_t=64$kJ/mol. The corresponding inherent structures are depicted in Figure~\ref{si_fig_snapshot_is}.

\begin{figure}[ht!]
\centering
\includegraphics[width=.65\linewidth]{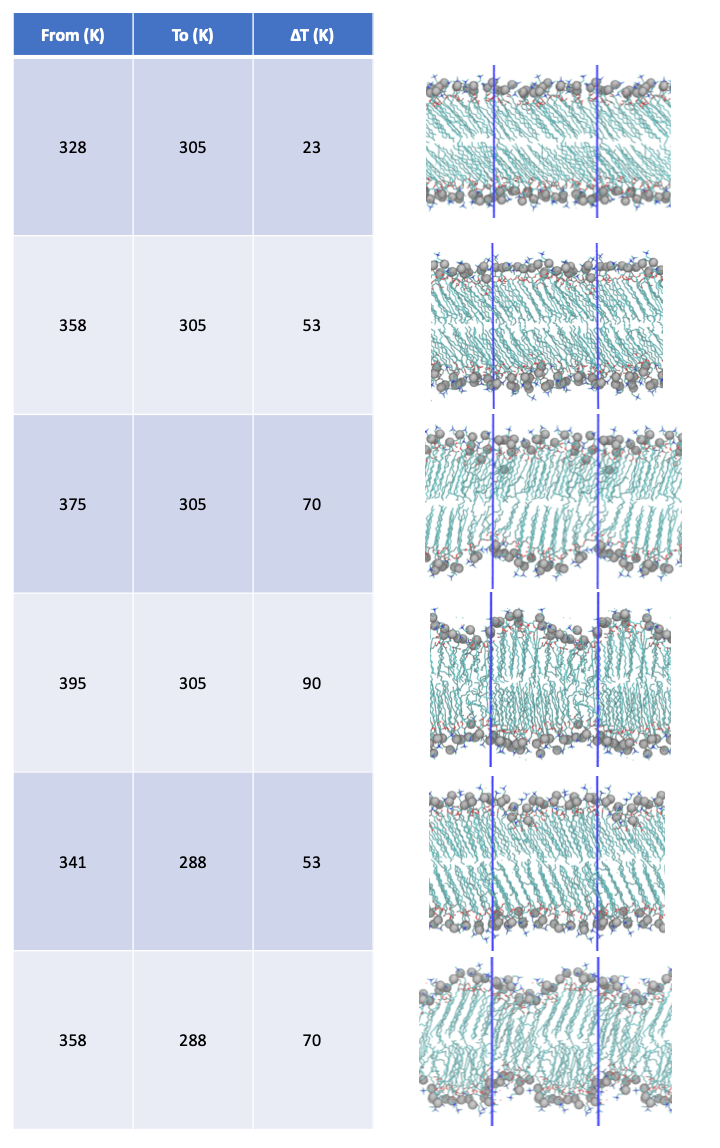}
\caption{Visualisation of the different DPPC 64 systems obtained \textit{via} quenching for different starting and final temperatures. To force the small system into a disordered phase, a $\Delta T$ of at least 70~K is required.}
\label{si_fig_quenching_dT}
\end{figure}

In order to compare the enthalpy of the systems in the tilted and disordered states, it was obviously critical to force a system in both phases using different temperature treatments. We picked the system at 64 DPPC molecules, where all phases were observed. However, while trying to obtain a disordered system at 305~K to compare the enthalpy of the disordered phase to the one of the real ripple phase, a quenching from 358 to 305~K ($\Delta T$ = 53) lead to a tilted gel phase. We made several attempts, and found that, regardless of the final temperature after quenching, a difference of temperature $\Delta T$ of at least 70~K is required to force the system in the disordered phase. The difference attempts and their results are summarised in Figure~\ref{si_fig_quenching_dT}.

\begin{figure}[ht!]
\centering
\includegraphics[width=.9\linewidth]{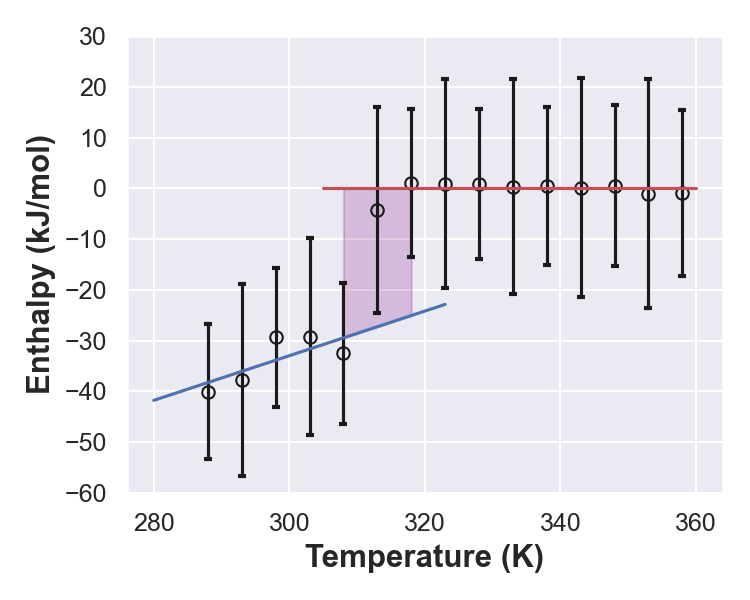}
\caption{Measurement of the enthalpy of a 64 DPPC system at different temperatures ranging from 288 to 358~K to compute the $\Delta H$ of the gel-fluid transition. The baseline calculated on the systems from 318 to 358~K was subtracted from all points, and the $\Delta H$ was calculated from the gray zone corresponding to the jump from one phase to another. Here $\Delta H$ = 27.3~kJ/mol.}
\label{si_fig_enthalpy_transition}
\end{figure}

The enthalpy of the (tilted) gel to fluid phase transition was measured by collecting the total enthalpy of a 64 DPPC molecules system after simulations at temperature ranging from 283 to 358~K (cf. Figure~\ref{si_fig_enthalpy_transition}). After subtraction of the baseline (here set to the enthalpy of the fluid systems) and dividing by the total number of lipids, the transition enthalpy was measured by integrating the enthalpy between the gel line and the fluid line, over the range where the transition occurs (here 308 to 318~K). By doing so, we measured a difference of enthalpy $\Delta H$ = 27.3~kJ/mol.

\newpage

\newpage

\newpage

\newpage
\section{A simple mechanical model that could explain the observed instability}

We briefly introduce a one-dimensional minimal model displaying an interesting thickness modulation instability, that supports the view that the corrugation observed in the simulations could be explained by a competition between hydration energy, tilt elasticity and partial melting of the lipids.  

\begin{figure}
\centering
  \resizebox{0.85\textwidth}{!}{\includegraphics{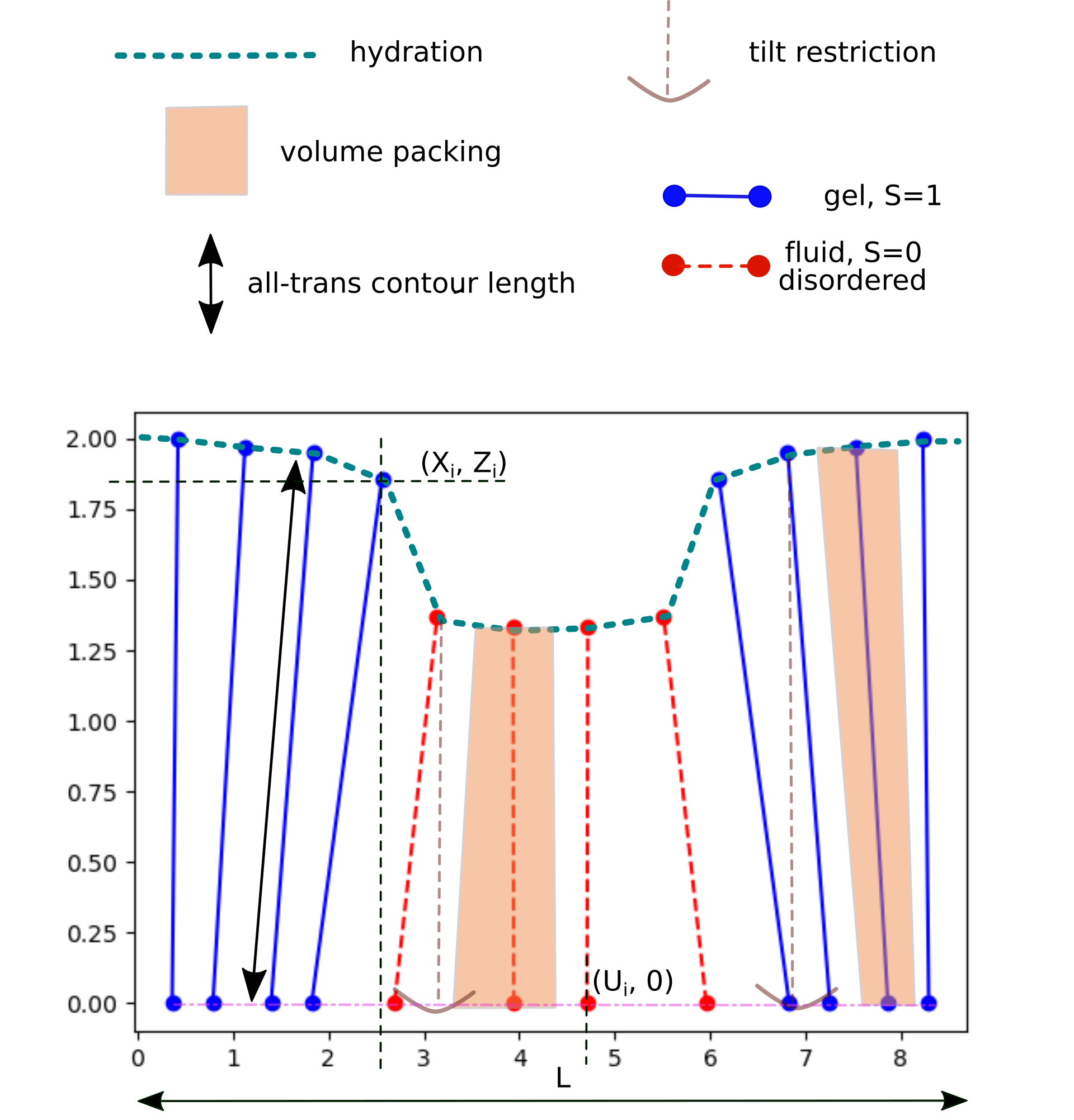}}
  \caption{Sketch of the mechanical model. Blue solid rod: gel state, red dashed rod: fluid or disordered state. Also represented are the four main contributions to the energy cost (hydration, packing, contour length and tilt). }
  \label{si_fig_mechanicalModel2}
\end{figure}

System \textbf{parameterization}

The system consists in a chain comprising a number $N\sim 10$ rods, each one representing a single lipid. Each rod is given by its two end points~(Figure~\ref{si_fig_mechanicalModel2}). The lower end-point has coordinates $(U_i,0)$ and remains at the mid-plane position. Only one leaflet is considered in the model, the other leaflet being assumed to follow the same behavior by symmetry. In addition, each rod possesses an internal state reminiscent from the Doniach two-states approach of the melting transition~\cite{1978_Doniach, Heimburg_BiophysicsMembrane}.  Our variables are

\begin{itemize}
\item the rods upper positions defined as $(X_i = L x_i, Z_i)$, $i \in [0,N-1]$; $x_i\in [0,1]$; $L$-periodicity assumed for $X_i$, 1-periodicity for $x_i$. 
\item the rods lower positions $(U_i = L u_i, 0)$; $u_i\in [0,1]$; $L$-periodicity assumed for $U_i$, 1-periodicity for $u_i$. 
\item the rod state $S_i = 0,1$, where 0 is \textit{gel}, 1 is \textit{fluid}, \textit{melted} or \textit{disordered}.
\end{itemize}

We also define the \textit{periodic difference} $a\dotminus b = (a - b -\lfloor a- b+\frac{1}{2}\rfloor)$ casually obtained by the python \texttt{a-b -np.floor(a-b+0.5)} macro. The periodic difference is an implementation of the periodic boundary conditions minimal image convention. Shift indices \textit{modulo} N are denoted with brackets, \textit{e.g.} $[i+1] \equiv (i+1) [N]$, with $[i+1]\in [0,N-1]$.

The mechanical model obeys the following \textbf{rules}
\begin{itemize}
\item
  the system tries to optimize its exposure to the solvent. Each bead attempts to reach separately the optimal   "area" per lipid (here length per lipid) $A_0$. The cost associated with it is related to the contour length of the rod upper beads
  \begin{equation}
    \mathcal{H}_{\mathrm{solvent}}=C_1 \sum_{i=1}^{N} \left[ L^2(x_{[i+1]}\dotminus x_i)^2+(Z_{[i+1]}-Z_i)^2-A_0^2 \right]^2
  \end{equation}

\item
  the "volume" (here area) of a lipid rod cannot depart too much from a target value $V_0$, whichever state the lipid occupies. This is mainly an incompressibility constraint imposed by the hydrophobic tails. If one takes the 1D lipid area the same as a trapezoid (approximate but hopefully accurate expression), the cost is 
  \begin{equation}
    \frac{1}{4}LZ_i(x_{[i+1]}\dotminus x_{[i-1]}+u_{[i+1]}\dotminus u_{[i-1]})
  \end{equation}
  then the corresponding cost function reads
  \begin{equation}
     \mathcal{H}_{\mathrm{packing}}=C_2\sum_{i=1}^{N} \left[L (x_{[i+1]}\dotminus x_{[i-1]}+u_{[i+1]}\dotminus u_{[i-1]})Z_i -4V_0\right]^2
  \end{equation}
  
\item the lipids in the gel state behave as stiff rods with well-defined contour length $L_c$, mimicking an all-\textit{trans} chain conformation. Therefore a state dependent energy term is introduced
  \begin{equation}
    \mathcal{H}_{\mathrm{all-trans}}=C_3\sum_{i=1}^{N} S_i \left[L^2 (x_i \dotminus u_i)^2+ Z_i^2 - L_c^2\right]^2
  \end{equation}
  
\item 
  the lipids in the fluid state have a restricted tilt, with an elastic coefficient of the order of magnitude of the bending modulus. Meanwhile, tilt restriction is necessary to frustrate the lipids in the gel state. Without it, the area per lipid in the gel state could reach the optimal value $A_0$ at the expense of an excessive tilt, obviously not seen experimentally nor numerically. Therefore, we introduce irrespective of the state, a tilt elasticity term
  \begin{equation}
   \mathcal{H}_{\mathrm{tilt\,restriction}}= C_4\sum_{i=1}^{N} \left[(x_{i} \dotminus u_{i})^2\right]^2
  \end{equation}

\item
  finally, though not strictly required, a repulsion term between end-tails improves the look of the resulting solutions.
  
  \begin{equation}
   \mathcal{H}_{\mathrm{cosmetic}}= C_5\sum_{i=1}^{N} (u_{[i+1]} \dotminus u_{i})^2
  \end{equation}

\item
  each disordered, or melted lipid, brings an additional thermodynamic free-energy contribution, irrespective of the position and orientation.
  \begin{equation}
   \mathcal{H}_{\mathrm{melting}}= \varepsilon(T)\sum_{i=1}^{N} (1-S_i) 
  \end{equation}
 State mismatch energy contributions could also easily be introduced.
 
\item
 a global chain tension term can be introduced as usual
  \begin{equation}
   \mathcal{H}_{\mathrm{tension}}= -F_e L
  \end{equation}
\end{itemize}

The \textbf{control parameters} below can be set arbitrarily. Our choice here is

\begin{itemize}
\item $A_0$ optimal interfacial area per lipid, close its fluid phase value, here 0.8
\item $L_c$ optimal contour length in \textit{all-trans} configuration, only in the gel phase, here 2.
\item $V_0$ optimal chain packing volume, cohesion forces, here 0.8*1.5 = 1.2
\item $F_e$ optional traction or tension force, here 0
\item $\varepsilon(T)$ Gibbs free-energy of melting, sends the fluid phase above the gel phase, depends linearly on temperature, vanishing around $T_m$, here 0.08 per bead.
\item Elastic parameters
  \begin{itemize}
  \item $C_1$ conjugated to the real area, here 1.
  \item $C_2$ conjugated to the packing volume; here 1.
  \item $C_3$ conjugated to the contour length, only in gel state, here 1.
  \item $C_4$ conjugated to tilt, related to Deserno-Terzi~\cite{2017_MertTerzi_Deserno} or Brown~\cite{2011_Watson_Brown} coupling, here 0.05
  \item $C_5$ conjugated to the separation between the lipid end-tails, \textbf{optional},  improves the chain structure appearance, here 0.05
  \end{itemize}
\end{itemize}

Once the model is set, one seeks for a minimum of the total energy. Minimization is done by gradient descent. Analytical gradients $\mathcal{H}_{,{x_i}}$,  $\mathcal{H}_{,{u_i}}$, $\mathcal{H}_{,{Z_i}}$ can be obtained and evaluated directly. A repeated iteration $(x_i;u_i;Z_i) \to (x_i- \delta t\cdot \mathcal{H}_{,{x_i}}; u_i-\delta t\cdot \mathcal{H}_{,{u_i}}; Z_i-\delta t\cdot \mathcal{H}_{,{Z_i}})$ leads to a minimum of energy $\mathcal{H}$ after a number of \textit{ca} $2\times 10^5$ iterations, $\delta t \sim 10^{-4}$,  a process similar to the relaxation stage in Molecular Dynamics. 

The gradient with respect to the chain scaling parameter $L$, provides a Irving-Kirkwood-Buff kind of virial expression, that matches the external applied force $F_e$. Local mechanical equilibrium is achieved automatically so long as $\mathcal{H}_{,{x_i}}$ and $\mathcal{H}_{,{u_i}}$ vanish separately. Forces oriented parallel to the $z$ direction are disregarded.  For completeness,  $\mathcal{H}_{,L}$ reads
\begin{eqnarray}
  \mathcal{H}_{,L} &=& \sum_{i=1}^{N}\left\lbrace%
  4C_1 L \left[ L^2 (x_{[i+1]}\dotminus x_i)^2 + (Z_{[i+1]}-Z_i)^2 -A_0^2)\right](x_{[i+1]}\dotminus x_i)^2\right.\nonumber\\
  & & \;\;+\, 2 C_2 Z_i\left[ LZ_i(x_{[i+1]}\dotminus x_{[i-1]} +u_{[i+1]}\dotminus u_{[i-1]})-4V_0\right]\nonumber\\
  & &\;\;\times\, (x_{[i+1]}\dotminus x_{[i-1]} +u_{[i+1]}\dotminus u_{[i-1]})\nonumber\\
  & & \left. + 4 C_4 L S_i \left[L^2(x_i \dotminus u_i)^2-L_c^2\right] (x_i \dotminus u_i)^2  + 2C_5 L (x_i\dotminus u_i)^2 \right\rbrace 
\end{eqnarray}
\medskip

\begin{figure}
\centering
\begin{tabular}{cc}
  \resizebox{0.5\textwidth}{!}{\includegraphics{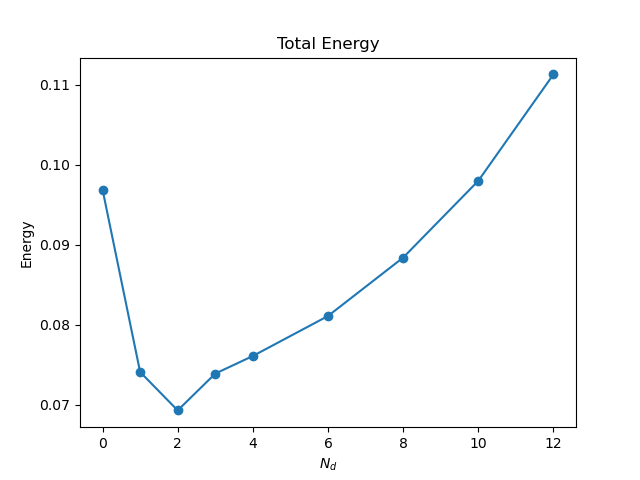}}&
  \resizebox{0.5\textwidth}{!}{\includegraphics{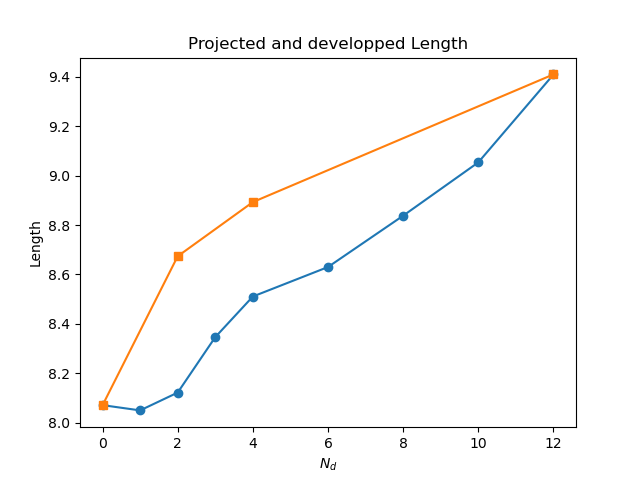}}
\end{tabular}
  \caption{Top, energy of a 12 beads chain as a function of the number of contiguous melted lipids. Bottom,  projected (circles) and contour (square) length as a function of the number of melted lipids}
\label{si_fig_energyAndLength}
\end{figure}

\begin{figure}
  \centering
  \begin{tabular}{cc}
     \resizebox{0.4\textwidth}{!}{\includegraphics{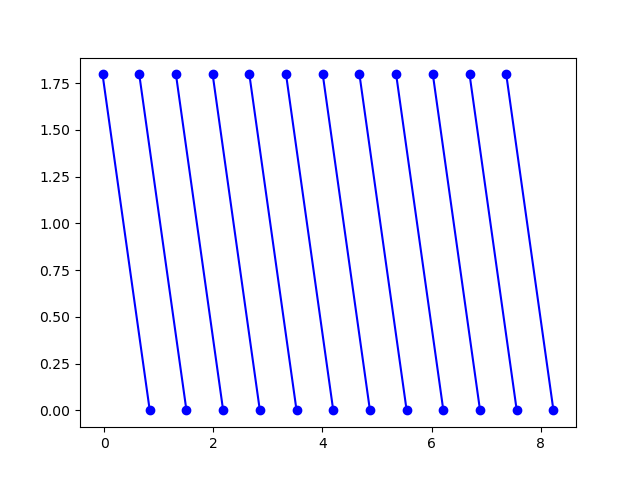}}&
     \resizebox{0.4\textwidth}{!}{\includegraphics{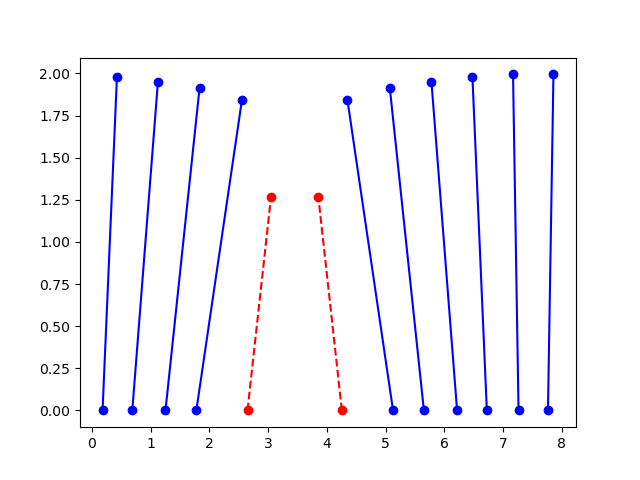}}\\
     \resizebox{0.4\textwidth}{!}{\includegraphics{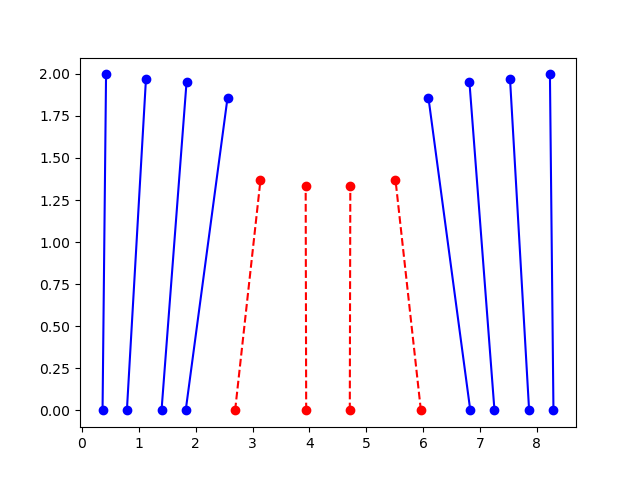}}&
     \resizebox{0.4\textwidth}{!}{\includegraphics{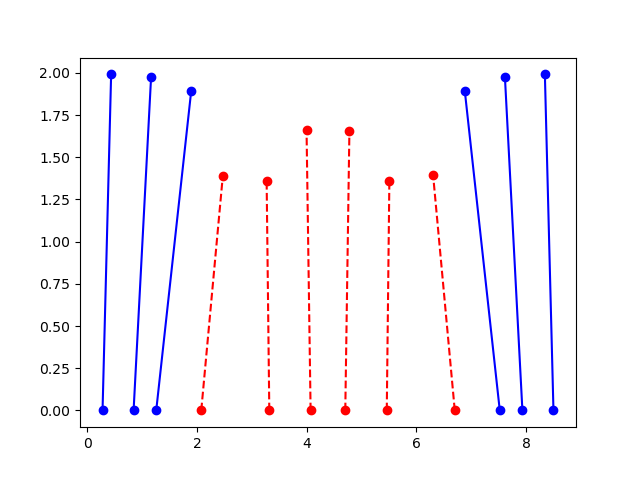}}\\
     \resizebox{0.4\textwidth}{!}{\includegraphics{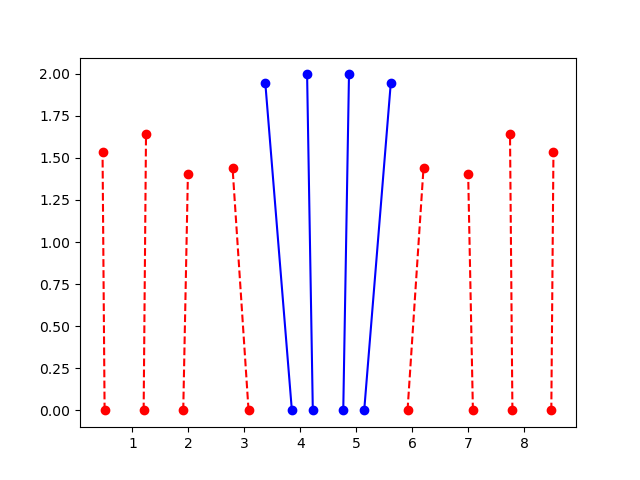}}&
     \resizebox{0.4\textwidth}{!}{\includegraphics{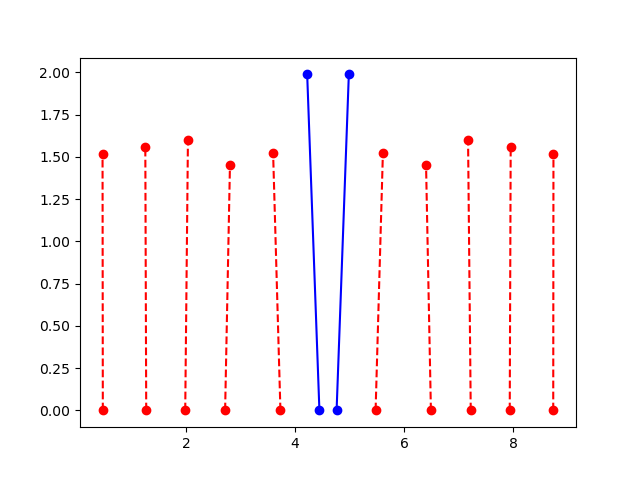}}\\
     \resizebox{0.4\textwidth}{!}{\includegraphics{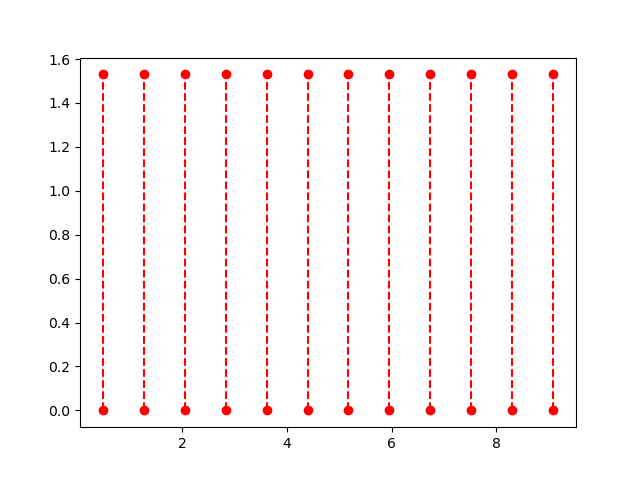}}
    & \\
    \end{tabular}
    \caption{From left to right and top to bottom, minimal energy configurations for states $(N_g,N_f)=$ (12,0),(10,2),(8,4),(6,6),(4,8), (2,10),(0,12).}
    \label{si_fig_minimalConfig} 
\end{figure}

Let us summarize the main \textbf{results} of the 1d model. We investigated pairs $(N_g,N_f)$ of $N_g$ consecutive gel lipids and $N_f$ fluid lipids (Figure~\ref{si_fig_minimalConfig}).  The structure of the  $(12,0)$ all-gel minimum was found to be an \textbf{homogeneous tilted phase}. The area per lipid was $0.67 < A_0=0.8$, see Figure~\ref{si_fig_energyAndLength}. The structure of the $(0,12)$ all-fluid minimum was a thin \textbf{non tilted phase}. The area per lipid was $0.8 \simeq  A_0$. The energy was \textit{on purpose} higher than the all-gel phase. Figure~\ref{si_fig_energyAndLength} shows that a minimum of energy is obtained for $(10,2)$. More fluid lipids leads to smoothly increasing energy until $(0,12)$. The lipid tilt is immediately released when the fluid state appears. The minimum in membrane thickness is significantly smaller than the all-fluid state (see vertical scale), consistent with partial interdigitation and simulation results. The instability is accompanied by a significant increase in contour length (analogue of interfacial area and hydration) while the projected area stays almost unchanged (Figure~\ref{si_fig_energyAndLength}). 
Interestingly, the local thickness of the groove in the (10,2) minimal configuration is significantly smaller than the average thickness of the fluid configuration (0,12), possibly pointing to some degree of interdigitation, and reminiscent from the observed numerical $L^d_{\beta}$ state. We note however that parameters have not yet been really optimized with respect to experimental data (area per lipid in fluid and gel phase, tilt, membrane thickness in fluid and gel phase, enthalpy of melting per lipid. As it is, this current model does not predict any lower critical lateral size for the apparition of the instability. It seems that the conformation $(4,2,4,2)$ has lower energy than $(10,2)$ which suggests that the ripple period could be very short. A gel-fluid neighbour mismatch penalty might help with enlarging the lateral corrugation size.

\bibliographystyle{elsarticle-num-names}
\bibliography{references.bib}